\documentclass[a4paper,10pt]{article}

\usepackage[latin1]{inputenc}
\usepackage[T1]{fontenc}
\usepackage[english]{babel}
\usepackage{amsmath, amssymb, amsfonts}
\usepackage{graphicx}
\usepackage{footnote}
\usepackage{float}
\usepackage{subcaption}
\usepackage{physics}
\usepackage{mathrsfs}
\usepackage{calligra}
\usepackage{url}
\usepackage{dsfont}
\usepackage[colorlinks]{hyperref}

\usepackage{amsthm}

\theoremstyle{remark}
\numberwithin{equation}{section}

\hypersetup{
    colorlinks=true,
    linkcolor=blue,
    filecolor=magenta,  
    urlcolor=[rgb]{0.00,0.00,0.50},    
    citecolor=red,
}

\advance \textheight by \topskip
\def\be{\begin{equation}}
\def\ee{\end{equation}}
\def\ba{\begin{eqnarray}}
\def\ea{\end{eqnarray}}

\def\Sp{{\mathrm{Sp}}}
\def\Mp{{\mathrm{Mp}}}

\def\Z{\mathbb Z}

\def\R{\mathbb R}

\def\C{\mathbb C}
\def\H{\mathbb H}

\def\D{\mathbb D}
\def\RP{\mathbb {RP}} 

\providecommand{\ii}{\mathrm{i}}

\usepackage{xcolor}

\begin{document}

%
%%%%%%%%%%%%%%%%%%%%%%%%%%%%%
\title{{\Large {\bf   \textcolor[rgb]{0.10,0.00,0.10}{Projective Time, Cayley Transformations and the Schwarzian Geometry of the Free Particle--Oscillator Correspondence}\\
[2pt]
 }}}

\author{{\bf Andrey Alcala 
 and  Mikhail S. Plyushchay 
} 
 \\
[8pt]
{\small \textit{Departamento de F\'{\i}sica,
Universidad de Santiago de Chile,
Av. Victor Jara 3493, Santiago, Chile}}\\
[4pt]
 \sl{\small{E-mails:   
\textcolor{blue}{andrey.alcala@usach.cl},
\textcolor{blue}{mikhail.plyushchay@usach.cl}
}}
}
\date{}
\maketitle

\begin{abstract}
We investigate the relation between the one--dimensional free particle and the harmonic
oscillator from a unified viewpoint based on projective geometry, Cayley transformations, and
the Schwarzian derivative.  Treating time as a projective coordinate on $\RP^1$ clarifies the
$SL(2,\R)\cong Sp(2,\R)$ conformal sector of the Schr\"odinger--Jacobi symmetry and provides a
common framework for two seemingly different correspondences: the Cayley--Niederer (lens)
map between the time--dependent Schr\"odinger equations and the conformal bridge
transformation relating the stationary problems.  We formulate these relations as canonical
transformations on the extended phase space and as their metaplectic lifts, identifying the
quantum Cayley map with the Bargmann transform.  General time reparametrisations induce
oscillator--type terms governed universally by the Schwarzian cocycle, connecting the present
construction to broader appearances of Schwarzian dynamics.
\end{abstract}

%%%%%%%%%%%%%%%%%%%%%%%%%%%%%%%%%%%%%%%%%%%%%%%%%%%%%%%%%%
\section{Introduction}\label{Introduction}

The one--dimensional free particle and the harmonic oscillator are paradigmatic examples of
classical and quantum dynamics.  They sit at the opposite ends of the intuitive spectrum:
\emph{freedom} versus \emph{confinement}.  Yet precisely because of their simplicity, these
systems recur---often implicitly---across theoretical physics and mathematics.  The free
particle underlies heat kernels and diffusion, semiclassical propagation and scattering, and
serves as a universal local model in perturbative constructions.  It is also a natural ``seed''
for integrable hierarchies: through inverse scattering and Darboux--Crum dressing one
generates reflectionless one--dimensional Schr\"odinger operators whose isospectral flows are
governed by the KdV equation and its hierarchy 
\cite{KayMoses}--\cite{SolBelokolos}. Under Wick rotation the free
Schr\"odinger equation turns into the heat equation, and the Cole--Hopf transformation (closely
related, at the linear level, to Darboux--type dressing) linearizes the Burgers equation \cite{Burgers1948,Hopf1950,Cole1951}.

The harmonic oscillator is the canonical normal form near stable equilibria, the backbone of
quantization around classical vacua, and the basic building block of quantum fields and strings.
Both systems also appear in less direct guises: via coherent and squeezed states, metaplectic
(Bogoliubov) transformations, Gaussian wave packets, and the geometric structures that
control their propagation and phase 
\cite{Heller1975}--\cite{CohKhare}.

At the same time, the two systems differ in a fundamental way that is already visible at the
spectral level.  The free particle has a continuous spectrum, while the oscillator has a purely
discrete one.  This immediately suggests that no standard Darboux--Crum (isospectral or
almost--isospectral) mechanism can relate them in a direct way.  Nevertheless, it has long
been known that they are tightly connected by symmetry.  In the free particle, the
(non--relativistic) conformal symmetry acts as a \emph{dynamical} symmetry, generated by the
$\mathfrak{sl}(2,\R)$ triple of time translation, dilation, and special conformal transformation,
together with the Heisenberg algebra of space translations and Galilean boosts; together
they form the Schr\"odinger (Jacobi) algebra $\mathfrak{h}_3\rtimes\mathfrak{sl}(2,\R)$  \cite{LevyLeblond1963}--\cite{BerndtSchmidt1998}. 

For the oscillator, the same algebraic skeleton reappears in a different realization, adapted to
confinement: time evolution becomes naturally compact (Newton--Hooke viewpoint), and the
$\mathfrak{sl}(2,\R)$ sector is realized through quadratic phase--space generators and their
metaplectic lift \cite{BacryLevyLeblond1968}--\cite{DuvalHorvathy2009}.

A central geometric theme is that the $\mathfrak{sl}(2,\R)$ symmetry is inherently
\emph{projective}: it acts on time by M\"obius transformations of $\RP^1$, and its global
structure is best understood after \emph{projectivizing} time.  From this viewpoint, the
Schwarzian derivative appears as the canonical projective object controlling reparametrisations,
cocycles, and induced quadratic terms \cite{OTbook,Osgood,OTams}.
 This is not merely a formal analogy.  The Schwarzian
derivative and its projective cocycle play a distinguished role in modern physics, most notably
in the low--energy sector of the Sachdev--Ye--Kitaev model and in two--dimensional
Jackiw--Teitelboim gravity, where the effective boundary dynamics is governed by the
Schwarzian action and its ${\rm Diff}(S^1)/SL(2,\R)$ structure 
\cite{SachdevYe1993}--\cite{SYKreview1}.

Historically, the projective viewpoint was already present in classical works of Cayley on
fractional linear transformations and projective invariants, where the Schwarzian derivative
enters as a fundamental differential invariant \cite{Cayley1846,Cayley1880}.  Much later, Niederer discovered a
remarkable map relating the free particle and the harmonic oscillator at the level of the
time--dependent Schr\"odinger equation (TDSE) \cite{Niederer1972,Niederer1973,Takagi},
 and Arnol'd developed a broader
geometric framework for such correspondences between differential equations \cite{ArnoldODE,ACGLR}.

In parallel, a different question arises if one looks at the stationary Schr\"odinger equation
(SSE): can one relate \emph{spectral} problems associated with freedom and confinement?
Although a direct Darboux--Crum link is excluded by the radically different spectral types,
a relation \emph{does} exist.  The conformal bridge transformation (CBT) establishes a map
between the free particle and oscillator at the level of the stationary problem, and admits
natural generalisations to a variety of backgrounds and symmetry--extended settings
\cite{CBT1}--\cite{JarLS}.

A priori, it is not obvious why a TDSE map (an evolution problem, valid for arbitrary initial
data) and an SSE map (a spectral problem) should be governed by a common organizing
principle.  One expects a relationship only indirectly, through the fact that propagators and
evolution kernels can be built from spectral data.  The main purpose of this article is to show
that, in the present context, there \emph{is} a unifying structure, and to identify it explicitly.
We argue that the appropriate ``roof'' is provided by a projective--geometric reformulation of
the $\mathfrak{sl}(2,\R)$ symmetry together with a complexified canonical transformation
generated by a rotated Cayley matrix, and by its quantum (metaplectic) counterpart.  This
framework clarifies, in a single language,
(i) the Cayley--Niederer map relating the free particle and oscillator at the level of the TDSE,
(ii) the CBT  relating their stationary problems,
and (iii) the appearance of the Schwarzian derivative as the universal projective object that
induces quadratic (oscillator--type) terms under general time reparametrisations.
In addition, the same viewpoint yields a transparent interpretation of the Bargmann transform
as a canonical (and, at the quantum level, metaplectic/unitary) passage between the
Schr\"odinger and Bargmann--Fock realisations of the Heisenberg algebra 
\cite{Bargmann1961}--\cite{BHall}.

The paper is organized as follows.
In Sec.~\ref{CTCCT} we introduce the complexified canonical transformation in phase space
defined by the Cayley matrix, and recall its interpretation in hyperbolic geometry as the map
between the upper half--plane and unit disc models, together with the role of the Cayley
parameter on $\RP^1$.
In Sec.~\ref{Free particle section} we review the free particle dynamical integrals and the
Schr\"odinger/Jacobi symmetry, emphasizing the projective (M\"obius) action on time.
In Sec.~\ref{QuantCayley} we construct the quantum Cayley map and discuss its metaplectic
realization, relating it to the Bargmann transform and to complex symplectic structures.
In Sec.~\ref{sec:QCayley_CBT} we present the quantum Cayley transform underpinning the CBT
 between the free particle and oscillator stationary problems.
In Sec.~\ref{SymmetriesHO} we discuss the classical symmetry structure of the oscillator and
formulate the Cayley--Niederer map in a way adapted to the compact (Newton--Hooke)
realization of the $\mathfrak{sl}(2,\R)$ sector.
In Sec.~\ref{SchrodingerProjective} we develop a unified treatment of coordinate and time
transformations of the TDSE and SSE, highlighting the natural appearance of $(\pm 1/2)$--weight
densities and the metaplectic (half--density) factors.
In Sec.~\ref{sec:free_to_osc_extended} we analyse the extended free particle and show how a
Schwarzian term generated by time reparametrization induces an oscillator--type potential.
In Sec.~\ref{SecGenF} we treat general reparametrisations and present a factorized metaplectic
form controlled by the Ermakov--Pinney amplitude.
Several technical developments and complementary viewpoints are collected in the Appendices:
Appendix~\ref{app:kahler_uplift} discusses compatible complex structures on $(\R^2,\omega)$ and
their parametrization by the upper half--plane; Appendix~\ref{app:metaplectic} reviews the
metaplectic representation; Appendix~\ref{app:Jordan-analytic} explains the role of Jordan
states from analyticity in the energy; Appendix~\ref{app:sp2r_abcd} connects $\Sp(2,\R)$
factorisations with ABCD optics; Appendix~\ref{app:demo_psi0zeta} provides a technical
derivation of a squeezed--vacuum Gaussian formula; Appendix~\ref{app:Sch_free_coherent}
summarizes Schr\"odinger--group Gaussian packets; Appendix~\ref{Appendix Schwarzian} collects
Schwarzian identities (including the cocycle property); and
Appendix~\ref{app:exact_vs_semiclassical_quantisation} comments on exact versus semiclassical
quantization of canonical transformations.

\section{Cayley transform and complexified canonical transformation}\label{CTCCT}

In this Section we introduce the \emph{complexified canonical transformation} in the
two--dimensional phase space defined by the $2\times2$ unitary Cayley matrix $C$.
This matrix preserves the symplectic form and, in this sense, provides the classical
phase--space counterpart of the familiar Cayley map in complex analysis and hyperbolic
geometry.  In the following sections it will serve as the basic ingredient for the
projectivization of time that underlies the $\mathrm{SL}(2,\R)$ (conformal) symmetry of the
free particle and of the harmonic oscillator, and it will clarify the origin and structure
of the conformal bridge and of the Cayley--Niederer maps between these systems at
the level of the stationary and time--dependent Schr\"odinger equations.  Finally, the same
viewpoint will lead to a natural interpretation of the Bargmann transform as a canonical
(and, at the quantum level, unitary) passage between the Schr\"odinger and Bargmann--Fock
realizations of the Heisenberg algebra.

\medskip
\noindent\textbf{Real symplectic basis and linear canonical transformations.}
Consider a two-dimensional phase space $\R^2$ with a real basis
\be\label{realbasis}
\xi_\alpha=(q,p)^{\mathsf T},\qquad
\{\xi_\alpha,\xi_\beta\}=\Omega_{\alpha\beta},\qquad
\Omega=i\sigma_2=
\begin{pmatrix}
0 & 1\\
-1 & 0
\end{pmatrix}
=:J.
\ee
Here $\Omega$ is the standard Poisson tensor (the inverse symplectic matrix);
with the convention $\omega=dp\wedge dq$ one has $\omega(u,v)=u^{\top}\breve{\Omega}v$
and $\Omega=\breve{\Omega}^{-1}$.
Linear canonical transformations are precisely those real linear maps
$\xi\mapsto \xi'=M\xi$ that preserve it, i.e.
\be\label{Sp2R}
M\in Sp(2,\R)\;\;\Longleftrightarrow\;\;
M^{\mathsf T}JM=J
\;\;\Longleftrightarrow\;\;
\det M=1,
\ee
so that $Sp(2,\R)\cong SL(2,\R)$ in one degree of freedom.

A phase-space function $F(q,p)$ generates a one-parameter Hamiltonian canonical
flow when applied to any function $f(q,p)$ by the adjoint action of the Poisson bracket,
\be\label{classflow}
\Phi(F;\tau):\ f(q,p)\ \mapsto\ f(q,p;\tau)=\exp(\tau X_F)f(q,p)
=\sum_{n=0}^{\infty}\frac{\tau^n}{n!}\{F,f\}_n,
\ee
where $X_F$ is the Hamiltonian vector field
\(
X_F=\frac{\partial F}{\partial \xi_\alpha}\,\Omega_{\alpha\beta}\,
\frac{\partial}{\partial \xi_\beta},
\), and
\(
\{F,f\}_0:=f,\) \( \{F,f\}_1:=\{F,f\},\)
\( \{F,f\}_n:=\{F,\{F,f\}_{n-1}\}\).
On the canonical coordinates themselves one has $\xi(\tau)=\exp(\tau X_F)\xi$,
and for quadratic $F$ the flow is linear:
\be\label{linearflow}
\xi_\alpha(\tau)=\bigl(M_F(\tau)\xi\bigr)_\alpha,\qquad
M_F(\tau)=e^{\tau A_F}\in Sp(2,\R).
\ee

\medskip
\noindent\textbf{Quadratic functions.}
The quadratic functions
\be\label{quadratics}
H_+=\frac12\,(p^2+q^2),\qquad
D=qp,\qquad
H_-=\frac12\,(p^2-q^2)
\ee
generate the Lie algebra $\mathfrak{sp}(2,\R)\cong\mathfrak{sl}(2,\R)$ via
\be\label{SL2R}
\{D,H_\pm\}=2H_\mp,\qquad
\{H_+,H_-\}=2D.
\ee
Their one-parameter Hamiltonian flows are represented by the $2\times 2$ matrices
\[
M_{H_+}(\tau)=
\begin{pmatrix}
\cos\tau & -\sin\tau\\
\sin\tau & \cos\tau
\end{pmatrix},\quad
M_D(\tau)=
\begin{pmatrix}
e^{-\tau} & 0\\
0 & e^{\tau}
\end{pmatrix},\quad
M_{H_-}(\tau)=
\begin{pmatrix}
\cosh\tau & -\sinh\tau\\
-\sinh\tau & \cosh\tau
\end{pmatrix},
\]
which satisfy the symplectic condition \eqref{Sp2R}.
The corresponding real $2\times 2$ generators $A_F$ in \eqref{linearflow},
are
\be\label{A-generators}
A_{H_+}=-J=-i\sigma_2,\qquad
A_D=-\sigma_3,\qquad
A_{H_-}=-\sigma_1,
\ee
in terms of standard Pauli matrices.
These generators, when taken with a minus sign, 
obey the  commutation relations of the form \eqref{SL2R}, 
and each $A_F$ lies in $\mathfrak{sp}(2,\R)$,  \(
A_F^{\mathsf T}J+JA_F=0.
\)

\medskip
\noindent\textbf{Parabolic (shear) basis.}
It is often convenient to switch to the alternative quadratic basis
\[
H_0:=\frac12\,p^2=\frac12\,(H_++H_-),\qquad
K:=\frac12\,q^2=\frac12\,(H_+-H_-),
\]
so that $(H_0,K,D)$ also spans $\mathfrak{sp}(2,\R)$.
The Poisson algebra in this basis reads
\[
\{D,H_0\}=2H_0,\qquad
\{D,K\}=-2K,\qquad
\{H_0,K\}=-D,
\]
and the corresponding Hamiltonian flows are the elementary shears
\[
M_{H_0}(\tau)=
\begin{pmatrix}
1 & -\tau\\
0 & 1
\end{pmatrix}
=\exp(-\tau\,\sigma_+),\qquad
M_{K}(\tau)=
\begin{pmatrix}
1 & 0\\
\tau & 1
\end{pmatrix}
=\exp(\tau\,\sigma_-),
\]
with the nilpotent generators
\(
\sigma_\pm:=\frac12(\sigma_1\pm i\sigma_2)\,.\)
In particular, $H_0$ generates the classical ``drift'' (free propagation)
$q(\tau)=q-\tau p$, $p(\tau)=p$, while $K$ generates the ``thin-lens'' shear
$q(\tau)=q$, $p(\tau)=p+\tau q$; these are precisely the two elementary
parabolic subgroups that will be used later for factorisations in $Sp(2,\R)$.
At the quantum level, the $Sp(2,\R)$ factorisations lift to the metaplectic
double cover $\mathrm{Mp}(2,\R)$ and therefore to operator (propagator)
factorisations; in particular, the harmonic and inverted-oscillator evolution
operators, as well as $U(\pi/4)=\exp(\frac{\pi}{4}\hat H_-)$,
see below,  are metaplectic
representatives of the corresponding classical matrices.

%%%%%%%%%%%%%%%%%%%%%%%%%%%%%%%%%%%%%%%%%%%%%%%%%%%%%%%%%%%%%%%%%%%%%%%%%%%%%%%%%%%%%%%%%%%%%%%%%%%%%%%%%%%%%%%%%%%%%%%%%%%%%%%%%%%%%%%%%%%%%%%%%%%%%%%%%%%%%%%%%%%%%%%%%%%%%%%%%%%%%%%%%%%%%%%%%%%%%%%%%%%%%%%%%%%%%%%%%%%%%%%%%%%
\medskip
\noindent\textbf{Complex canonical basis and the Cayley matrix.}
Let us  introduce  another, \emph{complex} canonical basis
\be\label{complexbasis}
\zeta _\alpha = (a^+, -ia^-)^{\mathsf T}=(C\xi)_\alpha\,,\qquad
\{\zeta _\alpha, \zeta _\beta\} = \Omega_{\alpha\beta},
\ee
where
%\[
$a^\pm = \frac{1}{\sqrt{2}}(q \mp ip).$
%\]
The two symplectic bases are linearly related  by the complex-valued matrix
\begin{equation}\label{xitxi}
C = \frac{1}{\sqrt{2}}
\begin{pmatrix}
1 & -i \\
-i & 1
\end{pmatrix} =\frac{1}{\sqrt{2}}( \mathbf{1}-i\sigma_1)=\exp\left(-i\frac{\pi}{4}\sigma_1\right)\,. 
\end{equation}
This is a specific   \emph{complex} symplectic, $C\in Sp(2,\C)\cong SL(2,\C)$, 
matrix:  it is unitary, $C^\dagger C= \mathbf{1}\,\Rightarrow\, C\in SL(2,\C)\cap SU(2)$, symmetric, $C^{\mathsf T}=C$, and
satisfies the relation
$
CJC=J\,.
$
 Being a particular case of the \( M_{H_-}(\tau) \) flow, 
 \begin{equation}
C = M_{H_-}\left(i{\pi}/{4} \right),\label{TMH-}
\end{equation}
it commutes with \( M_{H_-}(\tau) \). It is the Cayley matrix,  
whose  fourth power  is the central element $-I$ of $SU(2)$, 
\be
C^4 = -\mathbf{1}\,,\qquad C^8 = \mathbf{1}\,. \label{C8=1}
\ee

The one-parameter Hamiltonian \( SL(2,\mathbb{R}) \)--flows \( M_F(\tau) \) in the real basis \( \xi_\alpha \) and their matrix generators \( A_F \) are transformed by \( C \)-conjugation  into the \( SU(1,1) \) flows in the complex basis \( \zeta _\alpha \),
$\zeta (\tau)=\widetilde{M}_F\zeta $,
\(
\widetilde{M}_F(\tau) = C M_F(\tau) C^{-1},\) 
\( \widetilde{A}_F = C A_F C^{-1}\,,
\)
and one has
\be\label{Relation-Flows}
\widetilde{M}_{H_+}(\tau) = M_D(i\tau),\qquad
\widetilde{M}_D(\tau) = M_{H_+}(i\tau),\qquad
\widetilde{M}_{H_-}(\tau) = M_{H_-}\,,
\ee
and 
\(
\widetilde{A}_{H_+} = -i\sigma_3,\) 
\( \widetilde{A}_D = \sigma_2,\)
\(\widetilde{A}_{H_-} = {A}_{H_-}= -\sigma_1.
\)
In correspondence with  (\ref{TMH-}), the flows \( M_{H_-}(\tau) \) and \( \widetilde{M}_{H_-}(\tau) \) are the same in both bases, while \( \widetilde{M}_{H_+}(\tau) \) and \( \widetilde{M}_D(\tau) \) are related to \( M_D(\tau) \) and \( M_{H_+}(\tau) \) via the Wick rotation \( \tau \rightarrow i\tau \).
This corresponds to the map
\be
iD = iqp=\frac{1}{2}\left(a^-{}^2-a^+{}^2\right)\, \mapsto\, \widetilde{H}_+ = a^+ a^-
\label{iqpH+}
\ee
under the complexified canonical transformation  \( \xi_\alpha \mapsto \zeta _\alpha \).

Note here that
with any quadratic function $F(\xi)$, one can associate 
a symmetric matrix $K_F=K^{\mathsf T}_F$ by means of its Hessian,
\(
F=\frac{1}{2}\xi^{\mathsf T}K_F\xi\,,\) 
\( (K_F)_{\alpha\beta}=\frac{\partial^2 F}{\partial \xi_\alpha\partial\xi_\beta}\,.
\)
In terms of $K_F$, the $SL(2,\R)$ matrix generators $A_F$
are presented as
$
A_F=-J K_F\,,
$
and under the change of the basis (\ref{xitxi}), 
$ 
K_F\mapsto  \widetilde{K}_F=C^{-1}K_FC^{-1}\,.
$

\medskip
\noindent\textbf{Cayley conjugation: from $SL(2,\R)$ to $SU(1,1)$.}
A general element of the linear $SU(1,1)$ transformation  given by composition
of transformations $\widetilde{M}_{H_+}$,  $\widetilde{M}_{H_-}$ and $\widetilde{M}_{D}$
is 
\[
 \widetilde{S}=\begin{pmatrix}
 \alpha & \beta \\
\bar{\beta} & \bar{\alpha}
\end{pmatrix}=CSC^{-1}\in SU(1,1)\,,\qquad  S=
\begin{pmatrix}
a & b \\
c & d
\end{pmatrix}\in SL(2,\R)\,,
\]
 where $ad-bc=|\alpha|^2-|\beta|^2=1$.
Real  $a\,,b\,,c\,,d$, and complex $\alpha\,,\beta$, parameters
are related as
\(
\alpha=\frac{1}{2}\left((a+d)+i(b-c)\right)\,,\)
\( \beta =\frac{1}{2}\left((b+c)+i(a-d)\right)\,,\)	
\(
a=\Re \alpha+\Im \beta\,,\)
\(d=\Re \alpha-\Im \beta\,,\)
\(b=\Re \beta+\Im \alpha\,,\)
\(c=\Re \beta-\Im \alpha\,.\)

The $SU(1,1)$ map $\zeta\mapsto \zeta '=\widetilde{S}\,\zeta $ is
a  Bogoliubov transformation. 
In terms of $(a^-,a^+)^{\mathsf T}$, 
it takes a usual form 
\be\label{Bogol}
\begin{pmatrix}
a^-{}' \\
a^+{}'
\end{pmatrix}=B \begin{pmatrix}
a^- \\
a^+
\end{pmatrix}\,,\qquad
B= \begin{pmatrix}
u &v\\
\bar{u}&\bar{v}
\end{pmatrix}\,,\quad  u=\bar{\alpha}\,,\,\, v=i\bar{\beta}\,.
\ee
\vskip0.1cm
In what follows, we will see  how the  described classical  relations corresponding to the complexified canonical map underlie the CBT between the quantum free particle and harmonic oscillator systems, as well as the Bargmann transform that relates Schr\"odinger (coordinate) and Fock-Bargmann (holomorphic) representations of the Heisenberg group.

\vskip0.1cm

Matrix $M_{H_-}(\tau)$ of the  hyperbolic flow 
can be presented as a 
rotated through  $\pi/4$ dilation flow that is consistent with 
Cartan decomposition of $SL(2,\R)$,
\be\label{MH-MD}
M_{H_-}(\tau)=M_{H_+}(\pi/4)M_D(\tau) M_{H_+}(-\pi/4)\,.
\ee
This corresponds to the relation 
between  
$D$ and $H_-$, 
\[
H_-=\frac{1}{2}(p^2-q^2)=-D'=-q'p'\,,\qquad \text{where}\quad
\xi'_\alpha=(M_{H_+}(\pi/4)\xi)_\alpha\,,
\]
i.e. one of the generators of classical Bogoliubov transformations
(squeezing), $H_-$,  
is the same as $-D$ (dilation)  seen from the rotated by $\pi/4$ 
phase space coordinates. 
Relations \eqref{MH-MD} and  (\ref{TMH-}) allow 
 to present the basis-changing 
Cayley matrix $C$ in the  form 
\[
C=M_{H_+}(\pi/4)M_D(i\pi/4) M_{H_+}(-\pi/4)\,.
\]

In this way, the complex-valued symplectic matrix $C\in  SL(2,\C)\cap SU(2)$,
being a particular element of the compact real form of $SL(2,\C)$, 
establishes a unitary equivalence between the  two non-compact real forms of $ SL(2,\C)$,
$SL(2,\R)\cong  Sp(2,\R)$ and $SU(1,1)$.

\medskip
\noindent\textbf{Hyperbolic geometry: $\H_+$, $\D$, and the Cayley map.}
As is well known, the groups  $SL(2,\R)$ and $SU(1,1)$~\footnote{More exactly, their projective versions $PSL(2,\R)=SL(2,\R)/\mathcal{ Z}(SL(2,\R))=SL(2,\R)/\{I,-I\}$
and $PSU(1,1)=SU(1,1)/\mathcal{Z}(SU(1,1))=SU(1,1)/\{I,-I\}$.}
are isometry groups of the upper half-plane, $\H_+$,  and 
unit disc, $\D$,  Poincar\'e models of hyperbolic (Lobachevsky)
plane: 
\be\label{H+metic}
\H_+:=\{z=x+iy\in \C; \, y>0\}\,,\qquad
ds^2=\frac{dx^2+dy^2}{y^2}\,,\qquad
\partial \H_+=\RP^1\,,
\ee
\be\label{Dmetric}
\D:=\{w\in \C; \, |w|<1\}\,,\qquad
ds^2=4\frac{dw d\bar{w}}{(1-|w|^2)^2}\,,\qquad
\partial \D=S^1\,.
\ee
The upper-half plane $\H_+$  (together with its ideal boundary 
$\RP^1=\R\cup\{\infty\}$) is mapped  conformally into the unit disc $\D$ (and its boundary $S^1$)
by the same Cayley matrix (\ref{xitxi}), $C=\big(\begin{smallmatrix}
a & b \\
c & d\end{smallmatrix} \big)$, $a=d=\frac{1}{\sqrt{2}}$
$b=c=-\frac{i}{\sqrt{2}}$, 
but that now acts  as a fractional linear (M\"obius) transformation, $C\cdot z:=\frac{az+b}{cz+d}$\, , 
\be
z\mapsto {C}\cdot z=i\frac{z-i}{z+i}=\frac{z-i}{-iz+1}:=w\,\label{wiz}.
\ee
In accordance  with this, the isometry groups 
$SL(2,\R)$ and $SU(1,1)$ act 
on $\H_+$ and $\D$ projectively,
by fractional linear transformation,
\(
ds^2(z')=ds^2(z)\,,\)  
\( z'=S\cdot  z\,,\) \( S\in SL(2,\R)\,,\)
\(
ds^2(w')=ds^2(w)\,,\) 
\( w'=\widetilde{S}\cdot w\,,\) \( \widetilde{S}\in SU(1,1)\,.\)
Note that the two standard models of the hyperbolic plane,  (\ref{H+metic}) and (\ref{Dmetric}), also
appear naturally in geometric quantization as the moduli of compatible complex
structures (equivalently, holomorphic polarisations) on the symplectic plane; see
App.~\ref{app:kahler_uplift}. The upper half-plane $\H_+$ and the unit disc $\D$ structures also appear in  coherent squeezed states, see Sec.~\ref{sec:QCayley_CBT}.

Also note that the more common Cayley map
\be\label{w0Cay}
C_0\cdot z=\frac{z-\ii}{z+\ii}:=w_0\,,\quad
C_0=%(1+\ii)
\exp\!\Big(-\ii\frac{\pi}{3}(\mathbf n\cdot\boldsymbol{\sigma})\Big)\in SU(2),
\quad
\mathbf n=\frac{1}{\sqrt3}(1,1,1)\,,
\ee
differs from (\ref{wiz})  only by a rigid rotation of the unit disc,
$
w=\ii\,w_0=e^{\ii\pi/2}w_0,
$
i.e.\ by a $+\pi/2$ rotation. 
Unlike our matrix (\ref{xitxi}), 
$C_0$ 
produces a complexified linear canonical transformation $(q,p)^{\mathsf T}\mapsto e^{-i\pi/4}(a^+,a^-)^{\mathsf T} $  and satisfies 
$C_0^3=-\mathbf{1}$, $C_0^6=\mathbf{1}$, cf. (\ref{C8=1}).

\medskip
\noindent\textbf{Ideal boundary, Cayley parameter, and projectivized time.}
Restricting the Cayley map (\ref{wiz}) to the ideal boundary of $\H_{+}$ (i.e.\ to $z=t\in\mathbb{R}\cup\{\infty\}$),
one obtains a conformal identification of $\RP^{1}=\mathbb{R}\cup\{\infty\}$ with the unit circle $S^{1}$:
\begin{equation}\label{eq:cayley_RP1_to_S1}
w=w(t)=C\cdot t= i\,\frac{t-i}{t+i}\in S^{1}.
\end{equation}
In particular,
\(
t\in\{-1,0,1,\infty\}\ \longmapsto\ w\in\{-1,-i,1,i\}\subset S^{1},
\)
so that $t=\infty$ corresponds to the point $w(\infty)=i$.

A convenient parametrization of (a chart of) 
$\RP^{1}$ is given by the Cayley parameter\footnote{The coordinate $t=\tan\theta$ (equivalently, $t=\tan(\varphi/2)$ with $\varphi=2\theta$) is classical.
We call $t$ a \emph{Cayley parameter} because it is the real boundary coordinate induced by the Cayley fractional-linear map
$w=\frac{t-i}{t+i}$ (up to a rigid rotation), which identifies $\mathbb{RP}^1=\mathbb{R}\cup\{\infty\}$ with the unit circle $S^1$.
This terminology is consistent with the standard (matrix) Cayley transform introduced by Cayley~\cite{Cayley1846}.
 %This terminology is also consistent with Cayley's \emph{matrix} Cayley transform, which provides rational coordinates on rotations/orthogonal matrices~\cite{Cayley1846}.
For historical overviews of the Schwarzian in conformal/projective settings see, e.g.,~\cite{OTbook,Osgood,OTams}.}
\be\label{Caypar}
t=\tan\theta,\qquad -\frac{\pi}{2}<\theta<\frac{\pi}{2}\qquad (U_{0}),
\ee
which yields the simple angular form
\begin{equation}\label{eq:w_theta_phi}
w(\theta)= i\,\frac{\tan\theta-i}{\tan\theta+i}
          =-\,i\,e^{2i\theta}
          =-\,i\,e^{i\varphi},
\qquad \varphi:=2\theta\ (\mathrm{mod}\ 2\pi).
\end{equation}
In this description, the limits $\theta\to\pm\frac{\pi}{2}$ both correspond to $t\to\pm\infty$, i.e.\ to the same point
$\infty\in\RP^{1}$. One may therefore either identify the endpoints $\theta=\pm\frac{\pi}{2}$, or (as we will
prefer later) cover $\RP^{1}$ by two standard charts. Concretely, passing from $U_{0}$ to the complementary chart
$U_{\infty}$ amounts to using the coordinate
\(
t\mapsto \tilde t=-\frac{1}{t}\) \( (0\mapsto\infty),\) 
which corresponds to the shifts
\(
\theta\ \mapsto\ \theta+\frac{\pi}{2},\)
hence
\(\varphi\ \mapsto\ \varphi+\pi,\)
and therefore
\(w\ \mapsto\ -w.\)
So, here $U_0=\RP^1\setminus   \{\infty\}$,  and $U_\infty=\RP^1\setminus  
 \{0\}$. 
 Thus, on the circle $S^1$ the chart transition induces the antipodal map $w\mapsto -w$.
Equivalently, viewing $S^1$ as the space of \emph{unit vectors} in $\mathbb{R}^2$, the natural
projection $S^1\to\mathbb{RP}^1$ identifies antipodal points $\pm w$ as the same (unoriented) line,
and $w\mapsto -w$ is the corresponding deck involution.
This is consistent with  the geometric picture of $\RP^{1}$ as the space of (unoriented) lines through the origin in
$\R^{2}$. In this sense, $\RP^{1}$ is the moduli space (parameter space) of such lines, i.e. of
one-dimensional linear subspaces of $\R^{2}$. Choosing a unit vector on each line produces a
double cover $S^{1}\to\RP^{1}$,
 with the two
antipodal points $\pm w$ corresponding to the same underlying line.

This Cayley reparametrization will be used below in the discussion of the conformal symmetry
of the free particle and  
harmonic-oscillator systems.  Introducing a new time variable $\tau$ by the same
Cayley parameter
\(
t:=\tan\tau,
\)
a translation in $\tau$ becomes a fractional-linear  transformation of $t$:
\begin{equation}\label{eq:tau_translation_mobius}
\tau\mapsto\tau+\alpha
\qquad\Longrightarrow\qquad
t\mapsto t'=\tan(\tau+\alpha)=\frac{\cos\alpha\,t+\sin\alpha}{-\sin\alpha\,t+\cos\alpha}\,.
\end{equation}
Equivalently, on the boundary circle $S^{1}$ one has
$w=-\,i\,e^{2i\tau},$
and therefore $\tau\mapsto\tau+\alpha$ acts as a rigid rotation,
\(
w\mapsto w'=-\,i\,e^{2i(\tau+\alpha)}=e^{2i\alpha}\,w.
\)

Thus translations in $\tau$ correspond to the compact rotation subgroup acting on $S^{1}$.
In this language, the Niederer transformation \cite{Niederer1973}  
is precisely the Cayley transform of the time
variable accompanied by the corresponding transformation of the coordinate variable treated as a
$(-\tfrac12)$-weight density (cf.\ the Liouville transformation law for wave functions in the
stationary Schr\"odinger equation, Sec.~\ref{SchrodingerProjective}). 
We stress that the global implementation of conformal transformations requires some care; this will be addressed below by working on
$\RP^1$ (two charts) and,
when appropriate, promoting time to a dynamical variable in a reparametrization-invariant formulation.

%%%%%%%%%%%%%%%%%%%%%%%%%%%%%%%%%%%%%%%%%%%%%%%%%%%%%%%%%%%%%%%%%%%%%%%%%%%%%%%%%%%%%%%%
\section{Free particle: M\"obius and Schr\"odinger   symmetries}\label{Free particle section}

In this Section we recall the symmetry structure of the one--dimensional free particle
and fix our  notation.  We first list its continuous and discrete
classical symmetries and identify the corresponding integrals of motion, distinguishing
the ordinary (kinematical) ones from the dynamical generators.  We then pass to an
extended, reparametrization--invariant formulation in which time is promoted to a
dynamical variable and the theory is described by a first--class constraint.  This
extended viewpoint makes transparent how the $\mathrm{SL}(2,\R)$ (M\"obius) action on time
and the associated conformal boost can be realized by well defined canonical generators.
Finally, we projectivize time, \(t\in\R\to\RP^{1}\), discussing several equivalent ways to
do this; this resolves the global (non--completeness) problem of conformal boosts and
naturally leads to the density/half--density transformation laws that will be needed in
the quantum theory.

\medskip
\noindent\textbf{Classical continuous symmetries.}
A non-relativistic free particle described by the action
\be
S[x(t)] = \tfrac{1}{2} m \int_{t_i}^{t_f} \dot{x}^{2}\, dt\,= \int L \, dt\,,  
\label{freeac}
\ee
possesses  continuous symmetries:
\begin{subequations}\label{eq:schr-sym}
\begin{alignat}{2}
\mathcal{T}_{a}^{t}:\ & t \mapsto t' = t + a,\; x \to x' = x,
\qquad && \text{time translation}; \label{t-translation}\\[4pt]
\mathcal{T}_{b}^{x}:\ & x \mapsto x' = x + b,\; t \to t' = t,
\qquad && \text{space translation}; \label{s-translation}\\[4pt]
\mathcal{G}_{v}:\ & x \mapsto x' = x + v t,\; t \to t' = t,
\qquad && \text{Galilean boost}; \label{G-boost}\\[6pt]
\mathcal{D}_{\alpha}:\ & x \mapsto x' = e^{\alpha}x,\; t \mapsto t' = e^{2\alpha}t,
\qquad && \text{dilation}; \label{dilation}\\[6pt]
\mathcal{B}_{c}:\ & t \mapsto t' = \dfrac{t}{1 + c t},\; x \mapsto x' = \dfrac{x}{1 + c t},
\qquad && \text{conformal boost}; \label{c-boost}
\end{alignat}
\end{subequations}
%%%%%%%%%%%%%%%%%%%%%%%%%%%%%%%%%%%%%%%%%%%%%%%%%%%%%%%%%%%
Under transformations (\ref{t-translation}), (\ref{s-translation}), (\ref{dilation})
action does not change: $S \mapsto S' = S$. 
Under 
(\ref{G-boost}), one has $S \mapsto S' = S + \frac{m}{2}\,\big( 2 v x + v^{2} t \big) \Big|_{t_i}^{t_f}$, while
under (\ref{c-boost}), $ S \mapsto S' = S - \frac{m}{2}\, c\, \frac{x^{2}}{1 + c t} \Big|_{t_i}^{t_f}$.
\vskip0.1cm 

Conformal boost transformations (\ref{c-boost})
are not globally well defined; they
require a compactification (projectivization) 
of time:\; \( t \in \mathbb{R} \to \RP^{1} = \mathbb{R} \cup \{\infty\} \), see below.

\medskip
\noindent\textbf{Discrete symmetries and their relation to boosts.}
Action (\ref{freeac})  is also characterized by discrete symmetries :

$\bullet$ \( \mathcal{R}^{t} :\; t \mapsto -t \) , time reflection,
$\bullet$ \( \mathcal{R}^{x} :\; x \mapsto -x \) , space reflection,
$\bullet$ Time inversion\footnote{ Although $t \mapsto -1/t$ is an element of the continuous $\rm{SL}(2,\R)$ symmetry, it is often useful to single it out
as a discrete involution with special geometric meaning (exchange of charts on 
$\mathbb{R}\mathbb{P}^1$, see below).}:
\[
t \mapsto t' = \mathcal{S}(t) = -\frac{1}{t} \; , \qquad
x' = \mathcal{S}(x) = -\frac{x}{t} \;\; \Rightarrow \;\;
S \mapsto S' = S + \frac{m}{2}\!\left( - \frac{x^{2}}{t} \right) \Big|_{t_i}^{t_f}\,.
\]
The last  discrete symmetry also requires the projectivization of time, 
\( t \in \mathbb{R} \to \mathbb{R}\mathbb{P}^{1} \) ;
 then \( \mathcal{S} : t = 0 \leftrightarrow t = \infty \),
and 
\( \mathcal{S}^{2}(t) = t,\quad \mathcal{S}^{2}(x) = -x = \mathcal{R}^{x}(x) \;\; \Rightarrow \;\; \mathcal{S}^{4} = \text{Id}\,. \)

Galilean and conformal boosts can be generated by  the space and time translations  composed with discrete symmetries $\mathcal{S}$ and $\mathcal{R}^{x}$,
$$
 \mathcal{G}_{v}= \mathcal{R}^{x} \circ \mathcal{S} \circ \mathcal{T}^{x}_{-v} \circ 
 \mathcal{S} \,,\qquad  \mathcal{B}_{c}=\mathcal{R}^{x} \circ \mathcal{S} \circ \mathcal{T}^{t}_{-c} \circ \mathcal{S}\,.
 $$
 
 \medskip
\noindent\textbf{Integrals of motion: kinematical and dynamical generators.}
From solution of equation of motion,
$
x(t) = x(0) + \dot{x}(0) \, t = x(0) + \frac{1}{m}p \, t,
$
we identify the integrals, 
\be
P=m\dot{x}(0)=p \; , \qquad G = m x(0) = m x - p t\,,\label{pG}
\ee
\be
H_{0} = \frac{1}{2 m} p^{2} \,,\quad D = \frac{1}{m} G p = x p - 2 H_{0} t\,,
\quad
K = \frac{1}{2 m} G^{2} = \frac{1}{2} m x^{2} - D t - H_{0} t^{2}\,.\label{H0DK}
\ee
The integrals $G,$ $D,$ $K$  are dynamical,    explicitly depending on time,
$\frac{dG}{dt} = \frac{\partial G}{\partial t} + \{ G , H_{0} \} = 0\,.$

\medskip
\noindent\textbf{Schr\"odinger (Jacobi) algebra and Poisson brackets.}
Dynamical integrals  transform solutions of equations of motion into solutions, and generate Lie algebra 
of the (centrally extended) Schr\"odinger group (also known as the Jacobi group)
$
\mathfrak{h}_3 \rtimes \mathfrak{sp}(2,\mathbb{R})\,
$
given by  nonzero Poisson brackets
\be
\{ D , H_{0} \} = 2 H_{0} \; , \qquad \{ D , K \} = - 2 K \; , \qquad \{ K , H_{0} \} = D \; ,
\label{Dh0K}
\ee
\[
\{ D , P \} = P \; , \qquad \{ D , G \} = - G \; ,
\qquad
\{ H_{0} , G \} = - P \; , \qquad \{ K , P \} = G\,,
\]
with mass $m$ playing the role of the central charge of $1D$ Heisenberg algebra \( \mathfrak{h}_{3} \), 
\( \{ G,\,P \} = m \).

The linear fractional  (M\"obius) transformations
$PSL(2,\mathbb{R})$ act on \( t \in \mathbb{RP}^{1} \), 
\[
t \to t' = \frac{a t + b}{c t + d} \; , \quad a , b , c , d \in \mathbb{R} \; , \quad
a d - b c = 1\,.
\]
Their infinitesimal form and corresponding generators are
\[
\delta t = \delta \alpha \; , \quad l_{-1} = \partial_{t}\,;\qquad
\delta t = \delta \beta \, t \; , \quad l_{0} = t \, \partial_{t}\,; \qquad
\delta t = \delta \gamma \, t^{2} \; , \quad l_{+1} = t^{2} \, \partial_{t}\,.
\]
Vector fields \( l_{0} , l_{\pm 1} \) generate the $\mathfrak{sl}(2,\R)$ algebra in the standard form
(\ref{l0l+l-sl}) (equivalent to (\ref{Dh0K}) under the identification $l_{-1}\equiv H_{0}$, $l_{0}\equiv -\tfrac{1}{2}D$, $l_{+1}\equiv K$),
\be
 \; \{ \, l_{0} , l_{\pm 1} \, \} = \pm \, l_{\pm 1}\,,\qquad 
\{ \, l_{-1} , l_{+1} \, \} = 2 \, l_{0}\,,\label{l0l+l-sl}
\ee
and represent  the
\( (n = -1 , 0 , 1) \) slice of the Witt algebra
\be
\{ l_{n} , l_{m} \} = ( m - n ) \, l_{n+m}\,,\qquad
n , m \in \mathbb{Z}\,.\label{l+l-l0}
\ee

Infinitesimal transformations of \( x \) under
continuous symmetries are 
$
\delta x = \bar{\delta} x + \dot{x} \, \delta t \, ,
$
where
$
\bar{\delta} x = \varepsilon \{ x , F \}\,,
$
 $ F = ( P , G , H_{0} , D , K ).$
Explicitly, 
\[
F=P:\quad  \mathcal{T}^{x}_{\varepsilon} (x) = \bar{\delta} x = \varepsilon\,,\qquad
\delta t = 0\,,
\]
\[
F=G:\quad \mathcal{G}_{\varepsilon} (x) = - \varepsilon \, t \, , \qquad 
\varepsilon = - \delta v\,,
\quad \delta t = 0\,,
\]
\[
F=H_0:\quad \mathcal{T}^{t}_{\varepsilon} (x) = 0\,,
\qquad \delta t = \delta \alpha=-\varepsilon,
\]
\[
F=D:\,\, \mathcal{D}_{\varepsilon} (x) = \bar{\delta} x + \delta \beta \, \dot{x} \, t
= \varepsilon ( x - 2 \dot{x} t ) + \delta \beta \, \dot{x} \, t= \varepsilon \, x\,,
\,\, \varepsilon = \tfrac{1}{2} \delta \beta\,,\quad \delta t = \delta \beta\, t\,,
\]
\[
F=K:\quad 
\mathcal{K}_{\varepsilon} (x) = - \varepsilon ( x - \dot{x} t ) t + 
\delta \gamma \, \dot{x} \, t^{2}=\delta \gamma \, x \, t \,, \quad  \varepsilon = - \delta \gamma\,,\quad 
\delta t = \delta \gamma \, t^{2}\,.
\]

\medskip
\noindent\textbf{Extended formulation: reparametrization invariance and constraint.}
Let us promote \( t \) to be a dynamical
variable,  \(t \to t(s) ,\; x \to x(s)\), and present action (\ref{freeac}) 
in the  reparametrization-invariant form,
\be
S[x(s),t(s)] =\int \tfrac{1}{2}m  \frac{(\dot{x})^{2}}{\dot{t}} \, ds\,, \label{freeacext}
\ee
where \(\; \dot{x} = dx/ds \; , \; \dot{t} = dt/ds \; , \; \dot{t} > 0 \; .\)
Canonical momenta are
$p_{x} = \frac{\partial L}{\partial \dot{x}} = m \frac{\dot{x}}{\dot{t}}$, 
$p_{t} = \frac{\partial L}{\partial \dot{t}} = - \frac{m}{2} 
\Big( \frac{\dot{x}}{\dot{t}} \Big)^{2}$.
Reparametrization invariance of \( S \)  corresponds to the change
$s \to \alpha(s)$, $\dot{\alpha} > 0\,,$
and is generated by the first-class constraint 
\be
\mathcal{C}_{\rm rep} := p_{t} + \frac{1}{2m} p_{x}^{2} \approx 0\,,
\label{repacon}
\ee
that, according to Dirac,  at the quantum level transforms into the  time-dependent Schr\"odinger equation for 1D free particle.

The first order action corresponding to (\ref {freeacext}) is
\(
S = \int \big( \dot{t}\, p_{t} + \dot{x}\, p_{x} - e\, 
\mathcal{C}_{\rm rep} \big)\, ds\,,
\)
where $
e = e(s)
$
is  Lagrange multiplier (einbein,  lapse function). 
The integrals  have the same form (\ref{pG}), (\ref{H0DK}) with $p=p_x$,
and they  weakly  Poisson-commute with constraint \eqref{repacon} and Hamiltonian
\(
\mathcal{H} = e\, \mathcal{C}_{\rm rep}\approx 0\,.\)

\medskip
\noindent\textbf{$sl(2,\R)$ generators on extended phase space and projectability.}
Vector fields generating conformal
symmetry transformations of \( t \) and \( x \) are
\begin{eqnarray}\label{J-}
&J_{-} = \partial_{t} \; ,  \qquad \delta t = \varepsilon \, , \quad \delta x = 0\,,&\\
&J_{0} = t \partial_{t} + \tfrac{1}{2} x \partial_{x} \; , \qquad
\delta t = \varepsilon t \, , \quad \delta x = \tfrac{1}{2} \varepsilon x\,,& \label{J0}
\\
&J_{+} = t^{2} \partial_{t} + t x \partial_{x} \; , \qquad
\delta t = \varepsilon t^{2} \, , \quad \delta x = \varepsilon t x\,.&
\label{J+}
\end{eqnarray}
 They satisfy the \( sl(2,\mathbb{R}) \) algebra  in the form  
(\ref{l0l+l-sl}).  The phase space functions 
\begin{eqnarray}
&\mathcal{J}_{-} = p_{t} = \mathcal{C}_{\rm rep} - H_{0} \, ,& \label{4a}\\
&\mathcal{J}_{0} = t p_{t} + \tfrac{1}{2} x p_{x} = \tfrac{1}{2} D + t\, \mathcal{C}_{\rm rep} \, ,& \label{4b}\\
&\mathcal{J}_{+} = t^{2} p_{t} + t x p_{x} - \tfrac{1}{2} m x^{2} = - K + t^{2} \mathcal{C}_{\rm rep} \, & \label{4c}
\end{eqnarray}
``corrected" by the constraint (\ref{repacon})
correspond to (\ref{J-})--(\ref{J+})
realized as the  Hamiltonian
vector fields 
\(
X_{F} \;=\; \partial_{\mu} F \; \Omega_{\mu\nu} \; \frac{\partial}{\partial z_{\nu}} \, 
\)
in the extended phase space $z_\mu=(x,t, p_x,p_t)$,
$\{ z_{\mu} , z_{\nu} \} = \Omega_{\mu\nu}$.
The presence of the terms proportional
to the constraint $\mathcal{C}_{\rm rep}$ in (\ref{4a})--(\ref{4c}) 
reflects the fact that any two
phase space functions that differ by
the term $\lambda(t,x)\, \mathcal{C}_{\rm rep}$ generate the same
transformation on the physical
phase space; they belong to the
same equivalence class.
Reparametrizing the world-line
$( s \to s + \varepsilon(s) )$ is a pure gauge transformation,  it
moves along the orbits of  \(\mathcal{C}_{\rm rep}\). 
The pieces $t\, \mathcal{C}_{\rm rep}$ and $t^{2}\mathcal{C}_{\rm rep}$ in (\ref{4b}), (\ref{4c})
subtract precisely those gauge
directions that would otherwise
contaminate action of $D$ or $K$
on $(t,x)$.
Hence the ``mixing'' is not an
ad-hoc trick but a standard feature
of gauge systems: to realize a
space-time symmetry on the extended
configuration/phase space, one must
pick a generator that is projectable
onto the physical phase space, and
that inevitably differs from the na\"\i ve
Noether charge by a term proportional to
the first-class constraint.
The term $-\tfrac{1}{2} m x^{2}$ in (\ref{4c}) is harmless;
it Poisson-commutes with both
$x$ and $t$.

To have a truly global (everywhere-defined, complete) action of the
conformal boost as a diffeomorphism
generated by $J_{+} = t^{2}\,\partial_{t} + t x\,\partial_{x}$
on the extended configuration/phase
space, we have to do the
projective compactification of the
$t$ variable.

\medskip
\noindent\textbf{M\"obius action, conformal boosts, and half--density weight.}
Under the linear fractional  transformation, $t\rightarrow T=\frac{at+b}{ct+d}$, 
$ad-bc=1$, the spatial coordinate transforms as $x\rightarrow X=\frac{x}{ct+d}$ in accordance with 
(\ref{t-translation}), (\ref{dilation}), (\ref{c-boost}). 
 So, the quantity  $x(t)(dt)^{-1/2}$ is invariant under M\"obius $SL(2,\R)$ transformations, 
 \be\label{x=weight-1/2}
 x(t)(dt)^{-1/2}= X(T)(dT)^{-1/2}\,,\quad \Longleftrightarrow \quad X(T)=x(t)\left(\frac{dT}{dt}\right)^{1/2}\,.
 \ee
 Geometrically, $(dt)^{1/2}$ is a half-density of weight $+\frac{1}{2}$, and $x(t)$ transforms as the time-density of weight $-\frac{1}{2}$, i.~e. as a section of the ($-\frac{1}{2}$) power of the density bundle  over projective time, or,  equivalently,   
 as a 
$(-\frac{1}{2}$)-density under time reparametrizations.

\medskip
\noindent\textbf{Projectivisation of time: two--chart description of $\RP^{1}$.}
To projectivize (compactify) time, we introduce two charts.

$\bullet$ Chart U${}_0$ (affine, ``standard'', $t=0\in$  U${}_0$) :
\[
t = \frac{T_{1}}{T_{2}} \, , \qquad x = \frac{X}{T_{2}} \, , \qquad \text{with } T_{2} \neq 0 .
\]
A fiber coordinate $X$ is assumed
to be a scalar under linear $SL(2,\mathbb{R})$ transformations, while $T_1\rightarrow aT_1+bT_2$, $T_2\rightarrow cT_1+dT_2$, 
and so, $t$ undergoes a linear fractional  transformation. 

Na\"ively,  the change  $t \to 1/t$ would  correspond to the matrix $\left(\begin{smallmatrix}0&1\\[2pt]1&0\end{smallmatrix}\right)$ with $\det=-1$,  
 which is orientation reversing, an element of
$PGL(2,\mathbb{R})$ but not $PSL(2,\mathbb{R})$.
To preserve orientation and stay in
$PSL(2,\mathbb{R})$, we take instead

$\bullet$ Chart U${}_\infty$  ($t=\infty\in$  U${}_\infty$):
\[
\widetilde{t} = - \frac{1}{t} \, , \qquad \widetilde{x} = - \frac{x}{t} \, ,
\]
obtained from   U${}_0$ by the M\"obius map with matrix
$
\left(\begin{smallmatrix}0&1\\[2pt]-1&0\end{smallmatrix}\right)
 \in SL(2,\mathbb{R}) \, .
$
In homogeneous variables,  the transition from chart U${}_0$ to U${}_\infty$ 
is the linear change,
\(
(\, \widetilde{T}_{1} , \widetilde{T}_{2} , \widetilde{X} \,) = (\, - T_{2} , T_{1} , - X \,) \,,
\)
and so, 
\[
\widetilde{t} = \frac{\widetilde{T}_{1}}{\widetilde{T}_{2}} = - \frac{T_{2}}{T_{1}} = - \frac{1}{t} \, ,
\qquad
\widetilde{x} = \frac{\widetilde{X}}{\widetilde{T}_{2}} = - \frac{X}{T_{1}} = - \frac{x}{t} \, , 
\qquad \text{with } T_{1} \neq 0 .
\]
Working in two charts
makes the finite $SL(2,\mathbb{R})$ flows of conformal symmetry 
global : the ``blow-up"  is just a chart
switch.

In chart U${}_0$, Lagrangian takes the form 
\[
\frac{m}{2}\frac{\dot{x}^{2}}{\dot{t}} = \frac{m}{2} \frac{\mathcal{A}_2^{2}}{\mathcal{B}} \cdot \frac{1}{T_{2}^{2}} \, , \qquad
\text{where } \quad \mathcal{A}_2 = \dot{X}\, T_{2} - X \dot{T}_{2} \, , \quad
\mathcal{B} = \dot{T}_{1}\, T_{2} - T_{1} \dot{T}_{2} \, .
\]
With conjugate momenta  $P_{T_{1}}$, $P_{T_{2}}$, $P_X$, 
we obtain two first-class constraints
\[
\Phi_{\mathbb{RP}^{1}} := T_{1} P_{T_{1}} + T_{2} P_{T_{2}} + X P_{X} \approx 0 \, ,\quad
\Phi_{\mathrm{rep}} := T_{2} P_{T_{1}} + \frac{1}{2m} T_{2}^{2} P_{x}^{2} \approx 0 \, ,
\quad
\{ \Phi_{\mathbb{RP}^{1}} , \Phi_{\mathrm{rep}} \} = 0 .
\]
Constraint $\Phi_{\mathbb{RP}^{1}}$ generates gauge
transformation
\(
\Phi_{\mathbb{RP}^{1}} :\; ( T_{1} , T_{2} , X ) \to \lambda ( T_{1} , T_{2} , X ) \, , \)
\( \lambda \neq 0 \, .
\)
Hence, simultaneously
rescaled by the same constant $\lambda$
coordinates $( T_{1} , T_{2} , X )$ belong to the same gauge
orbit. The phase space functions 
\be
t = \frac{T_{1}}{T_{2}} \, , \qquad x = \frac{X}{T_{2}} \, ,
\qquad
p_{t} := T_{2} P_{T_{1}} \, , \qquad p_{x} := T_{2} P_{X}\label{xtpxptTX}
\ee
Poisson-commute with $\Phi_{\mathbb{RP}^{1}}$, 
hence are gauge-invariant under this constraint, and we have the canonical transformation
\[
( T_{1} , P_{T_{1}} ;\; T_{2} , P_{T_{2}} ;\; X , P_{X} )
\;\to\;
\big( t = T_{1}/T_{2} ,\; p_{t} ;\; \ln T_{2} ,\; \Phi_{\mathbb{RP}^{1}} ;\; x = X/T_{2} ,\; p_{x} \big).
\]

Constraint $\Phi_{\mathrm{rep}}$ generates reparametrizations
and with (\ref{xtpxptTX}) equivalently  is presented in the form (\ref{repacon}),
\(
\Phi_{\mathrm{rep}} = p_{t} + \frac{1}{2m} p_{x}^{2} \approx 0 \, .
\)
Generators of Galilean boosts,
dilations and conformal boosts
expressed in terms of $x$, $p_{x}$ and  $t$ given by (\ref{xtpxptTX}),
$
G = m x - t p_{x} \, ,$
$
D = \frac{1}{m} p_{x} G \, , $
$K = \frac{1}{2m} G^{2} \, ,
$
Poisson-commute with both
constraints.

Introducing the gauge
\(
\chi_{\mathbb{RP}^{1}} := T_{2} - 1 \approx 0
\)
for $\Phi_{\mathbb{RP}^{1}} \approx 0$, and reducing the
system to the surface of the second-class constraints,
$( \Phi_{\mathbb{RP}^{1}} \approx 0 ,\; \chi_{\mathbb{RP}^{1}} \approx 0 ),$
we return to the reparametrization--invariant formulation based on
 action \eqref{freeacext} with $t = T_{1}$, $x = X$.

In chart U${}_\infty$, we have
$
\frac{\dot{x}^{2}}{\dot{t}} = \frac{\mathcal{A}_1^{2}}{\mathcal{B}} \cdot \frac{1}{T_{1}^{2}} \,,
$
where $\mathcal{B}$ has the same form as in chart U${}_0$, while
$\mathcal{A}_1$ is obtained from $\mathcal{A}_2$ by changing  $T_2$ for $T_1$. The difference of Lagrangians on intersection of the charts 
U${}_0\cap$U${}_\infty$ is a total derivative,
\(
L_0-L_\infty=\frac{d}{ds}\left(\frac{m}{2} \frac{X^2}{T_1T_2}\right)\,,
\)
confirming that the transition from one chart to another is a canonical transformation.

In chart U${}_\infty$,  the two first-class constraints
take the same form 
with the change of phase space
variables $T_{1}, P_{T_{1}}, T_{2}, P_{T_{2}}, X, P_{X}$
for their tilted analogs,
with
\(
\widetilde{P}_{T_{1}} = - P_{T_{2}} \, , \) 
\( \widetilde{P}_{T_{2}} = P_{T_{1}} \, , \)
\( \widetilde{P}_{X} = - P_{X}\).
Fixing in chart U${}_\infty$ the gauge $\tilde{T}_{2} = 1$,
gives local coordinates
$
\widetilde{t} = \widetilde{T}_{1},$  $\widetilde{x} = \widetilde{X}$.

\medskip
\noindent\textbf{Projectivisation as a circle: angular coordinate and Cayley parameter.}
Alternatively, the projectivization of time can be achieved based on an equivalent
interpretation of $\RP^1$ as a circle $S^1$ with antipodal identification of the
points on it. For this, we assign $t=\tan\theta$ with $-\pi/2<\theta<\pi/2$, cf. (\ref {Caypar}). 
This corresponds to the domain spanned by all lines through the origin in the
$(T_1,T_2)$ plane except the vertical one (chart $U_0$). Chart $U_\infty$ is
obtained by $\theta\mapsto\theta+\pi/2$ and $\widetilde{t}=-1/t=-\cot\theta$
with the domain spanned in the $(T_1,T_2)$ plane by all lines through the origin
except the horizontal one. Setting $\phi:=2\theta\in(-\pi,\pi)$, then passing
from $U_0$ to $U_\infty$ is $\phi\mapsto\phi+\pi$, i.e.\ antipodal identification
on $S^1$. Thus, $\RP^1\simeq S^1/\{\phi\sim \phi+\pi\}$. Equivalently, one could
consider $\phi\in(-\pi,\pi]$, i.e.\ including $+\pi$ and excluding $-\pi$ values,
and declaring $\phi=\pi\,\leftrightarrow\, t=\infty$. This is purely a convention
to avoid double counting at the seam. However, we prefer to use the
stereographic/Cayley coordinate
\be
t=\tan\frac{\phi}{2}\,,\qquad \phi\in(-\pi,\pi)\,,
\label{t=tanphi}
\ee
by implying the antipodal identification on $S^1$ realized via the usage of the
second chart $\phi\mapsto\phi+\pi$. Note that
\be
\frac{dt}{d\phi}=\frac{1}{2\cos^2(\phi/2)}\,,\qquad
\frac{d\phi}{dt}=2\cos^2(\phi/2)=\frac{2}{1+t^2}\,.
\label{dt-dphi}
\ee
In this way, translation in $\phi$, $\phi\mapsto\phi+\alpha$, corresponds to a
linear fractional transform of $t$,
\[
t\mapsto \frac{\cos{\frac{\alpha}{2}}\,t +\sin{\frac{\alpha}{2}}}
{-\sin{\frac{\alpha}{2}}\,t+\cos{\frac{\alpha}{2}}}\,,
\]
cf. (\ref {eq:tau_translation_mobius}),
that is an element of the compact (elliptic) one-parameter subgroup of
$SL(2,\R)$. The translation for $\alpha=\pi$ corresponds to the transition
between the charts $U_0$ and $U_\infty$, $t\mapsto -1/t$.

\medskip
\noindent\textbf{Quantum free particle and Schr\"odinger covariance.}
With (\ref{t=tanphi}), the Lagrangian from (\ref{freeacext}) becomes
\be
L_\phi=m\frac{\cos^2(\phi/2)\dot{x}^2}{\dot{\phi}}\,,
\ee
and the reparametrization (Hamiltonian) constraint takes the form
\be
\mathcal{C}_\phi\equiv p_\phi+\frac{p_x^2}{4m\cos^2(\phi/2)}\approx 0\,.
\ee
This is the free-particle constraint written in a compact time coordinate $\phi$.
Dirac quantization amounts to imposing $\widehat{\mathcal C}_\phi\Psi=0$ on
physical states $\Psi(x,\phi)$, with $\hat p_\phi=-i\hbar\partial_\phi$ and
$\hat p_x=-i\hbar\partial_x$. Hence
\be
i\hbar\,\partial_\phi\Psi(x,\phi)
=\frac{\hat p_x^{\,2}}{4m\cos^2(\phi/2)}\,\Psi(x,\phi)
=-\frac{\hbar^2}{4m\cos^2(\phi/2)}\,\partial_x^2\Psi(x,\phi)\,.
\label{phi-Schro}
\ee
Returning to the affine time $t$ via (\ref{t=tanphi}), i.e.\ using \eqref{dt-dphi}
and $\Psi(x,\phi)=\psi(x,t(\phi))$ so that $\partial_\phi\Psi=(dt/d\phi)\,\partial_t\psi$,
one obtains after cancelling the common factor $dt/d\phi$ the standard free-particle
time-dependent Schr\"odinger equation
\be
i\hbar\,\partial_t\psi(x,t)=-\frac{\hbar^2}{2m}\,\partial_x^2\psi(x,t)\,.
\label{free-Schro}
\ee
In particular, $\phi\to\pm\pi$ corresponds to $t\to\pm\infty$.

%%%%%%%%%%%%%%
%%%%%%%%%%%%%%%%%%%%%%%%%%%%%%%%%%%%%%%%%%%%%%%%%%%%%%%%%%%%%%%%%%%%%%%%%%%%%%%%%%%%%%%%%%%%%%%%%%%%%%%%%%%%%%%%

%%%%%%%%%%%%%%
%%%%%%%%%%%%%%%%%%%%%%%%%%%%%%%%%%%%%%%%%%%%%%%%%%%%%%%%%%%%%%%%%%%%%%%%%%%%%%%%%%%%%%%%%%%%%%%%%%%%%%%%%%%%%%%%

\section{Quantum Cayley map: Bargmann transform and the complex-metaplectic operator}\label{QuantCayley}

In this Section we describe the quantum counterpart of the complexified canonical
transformation generated by the Cayley matrix $C$ introduced in Section~\ref{CTCCT}.
On the one hand, this transformation is implemented (in a canonical choice of
normalization) by the unitary Bargmann transform, which may be viewed as a unitary
change of polarization from the Schr\"odinger realization on $L^{2}(\R,dq)$ to the
Bargmann--Fock space of holomorphic functions.  On the other hand, the same canonical
map admits an operator realization inside the (complexified) metaplectic group as an
exponential of a quadratic generator; this makes transparent the relation between the
Cayley matrix and a metaplectic similarity action on $(\hat q,\hat p)$.  We also explain
how the remaining overall scalar factor is fixed once the operator is interpreted as the
intertwiner between the Schr\"odinger and Bargmann--Fock realisations of the Heisenberg
algebra.

\medskip
\noindent\textbf{Bargmann transform as a unitary change of polarization.}
In what follows we work in units $\hbar=1$.
On $L^2(\mathbb{R},dq)$ we use
$\hat q\,\psi(q)=q\,\psi(q)$,  $\hat p\,\psi(q)=-i\,\partial_q\psi(q)$, so that
$i\hat p=\partial_q$, and assume the choice of units with dimensionless $q$.

Consider the Cayley matrix (\ref {xitxi}), 
\(
C=\frac{1}{\sqrt{2}}(1-i\sigma_1)\in SL(2,\C)\cap SU(2).
\)
It generates the complex linear canonical transformation
$(q,p)^{\mathsf T}\mapsto (u,v)^{\mathsf T}$,
\be\label{Cuv}
(u,v)^{\mathsf T}=C(q,p)^{\mathsf T},\qquad
u=\frac{1}{\sqrt{2}}(q-ip)=a^+,\qquad
v=\frac{1}{\sqrt{2}}(p-iq)=-ia^-,
\ee
where $a^\pm=\frac{1}{\sqrt{2}}(q\mp ip)$ is the standard complex canonical pair.

We look for the generating function of the first type for this canonical transformation
in the form
$F(q,u)=\frac{1}{2}\alpha q^2+\beta qu+\frac{1}{2}\gamma u^2.$
From the relations $p=\partial_q F$ and $v=-\partial_u F$ we obtain
$\alpha=-i$, $\beta=i\sqrt{2}$, $\gamma=-i$, and hence
\(
F(q,u)=i\left(-\frac{1}{2}q^2+\sqrt{2}\,qu-\frac{1}{2}u^2\right).
\)

For a Schwartz function $f(q)$, we define the quantum analog of this classical (complex)
canonical transformation by the integral operator
\[
(\mathcal{B}f)(z)=k\int_{\R}\exp\!\big(-iF(q,z)\big)\,f(q)\,dq,
\]
where we identify the new canonical coordinate $u$ with the complex variable $z\in\C$.
The constant $k$ is fixed by requiring that the lowest Hermite function
\(
\psi_0(q)=\pi^{-1/4}\exp(-q^2/2)
\)
is mapped to the constant function $1$. This gives $k=\pi^{-1/4}$, and therefore
\be\label{BTransform}
(\mathcal{B}f)(z)=\int_{\R}K(z,q)\,f(q)\,dq,\qquad
K(z,q)=\pi^{-1/4}\exp\!\left(-\frac{1}{2}z^2+\sqrt{2}\,zq-\frac{1}{2}q^2\right).
\ee

\medskip
\noindent\textbf{Bargmann--Fock Hilbert space and unitarity.}
The function $K(z,q)$ is nothing but the integral kernel of the (unitary) Bargmann transform
from the Hilbert space $L^2(\R)$ to the Bargmann--Fock Hilbert space $\mathcal{F}$ of entire
holomorphic functions with
\[
\|f\|^2=\int_{\C}|f(z)|^2\,d\mu(z)<\infty,\qquad
d\mu(z)=\pi^{-1}e^{-|z|^2}\,d^2z,\qquad
d^2z=d(\Re z)\,d(\Im z),
\]
and inner product
\[
\langle f|g\rangle=\int_{\C}\overline{f(z)}\,g(z)\,d\mu(z).
\]
\noindent
Since $\mathcal{B}:L^2(\mathbb R)\to\mathcal F$ is unitary, its inverse is the adjoint with respect
to $\langle\cdot,\cdot\rangle_{L^2}$ and $\langle\cdot,\cdot\rangle_{\mathcal F}$, and is given by
\[
(\mathcal{B}^{-1}F)(q)=\int_{\C}\overline{K(z,q)}\,F(z)\,d\mu(z),\quad
\overline{K(z,q)}=\pi^{-1/4}\exp\!\left(-\frac{1}{2}\bar z^{\,2}+\sqrt{2}\,\bar z\,q-\frac{1}{2}q^2\right).
\]

Thus, the complexified linear canonical transformation defined by the Cayley matrix $C$
is implemented at the quantum level by the unitary Bargmann transform between the
Schr\"odinger (coordinate) and Fock--Bargmann representations, i.e.\ the 
unitary isomorphism
$\mathcal{B}:L^2(\mathbb{R})\to\mathcal{F}$ \cite{Takhtajan,BHall}.
  
\medskip
\noindent\textbf{Intertwining of the Heisenberg generators.}
In the Bargmann--Fock representation one has the standard holomorphic realization
\(
\mathcal{B}\,\hat a^+\,\mathcal{B}^{-1}=z,\) 
\( \mathcal{B}\,\hat a^-\,\mathcal{B}^{-1}=\partial_z,\)
\( [\partial_z,z]=1,
\)
so that
\[
\mathcal{B}\,\hat q\,\mathcal{B}^{-1}=\frac{1}{\sqrt{2}}(z+\partial_z),\qquad
\mathcal{B}\,(i\hat p)\,\mathcal{B}^{-1}=\frac{1}{\sqrt{2}}(\partial_z-z).
\]
Equivalently, the canonical pair $(u,v)=(a^+,-ia^-)$ is represented as
$u\mapsto z$ and $v\mapsto -i\,\partial_z$ on $\mathcal{F}$.

Note that if in \eqref{BTransform} one restricts $z\in\C$ to a real variable $y\in\R$ (that would contradict the complex nature of the $u$ variable),
the resulting integral operator is no longer unitary: unitarity here is tied to the
holomorphic (complex) polarization and the Gaussian measure $d\mu(z)$.

The quantum analog of \eqref{Cuv} can also be realized by an (non-unitary) complex-metaplectic (see App. \ref{app:metaplectic}) operator,  acting by similarity on $(\hat q,\hat p)$; this operator also provides, after pairing with the holomorphic polarization, the unitary Bargmann transform described above.

\medskip
\noindent\textbf{Complex-metaplectic operator realization.}
We now provide an operator realization of the same complexified canonical transformation
\eqref{Cuv} generated by the Cayley matrix $C$, and explain how its overall normalization is
fixed once it is interpreted as an intertwiner between the Schr\"odinger and Bargmann--Fock
realisations of the Heisenberg algebra.

\medskip
\noindent\textbf{From the inverted oscillator flow to the Cayley matrix.}
Consider the quadratic Hamiltonian (of the inverted harmonic oscillator)
\be\label{Hminus_def}
\hat H_-=\frac12(\hat p^{\,2}-\hat q^{\,2})
=-\frac12\Big((\hat a^+)^2+(\hat a^-)^2\Big),
\qquad
\hat a^\pm=\frac{1}{\sqrt2}(\hat q\mp i\hat p).
\ee
In correspondence with classical relation
(\ref{classflow}) for $F=H_-$, define the one-parameter (generally non-unitary) similarity flow
\be\label{U_s_def}
U(s)=\exp\!\big(s\,\hat H_-\big).
\ee
Using $[\hat q,\hat p]=i$, one finds
$[\hat H_-,\hat q]=-\,i\hat p,$
$ [\hat H_-,\hat p]=-\,i\hat q,$
and therefore the Heisenberg-picture operators
$\hat q(s)=U(s)\hat q\,U(s)^{-1}$, $\hat p(s)=U(s)\hat p\,U(s)^{-1}$ satisfy
\[
\frac{d}{ds}\binom{\hat q(s)}{\hat p(s)}=-\,i\sigma_1\binom{\hat q(s)}{\hat p(s)}
\quad\Longrightarrow\quad
\binom{\hat q(s)}{\hat p(s)}=\exp(-is\sigma_1)\binom{\hat q}{\hat p}.
\]
At the special value $s=\pi/4$ one gets
\(
\exp\!\left(-i\frac{\pi}{4}\sigma_1\right)=\frac{1}{\sqrt2}\big(1-i\sigma_1\big)=C,
\)
cf. classical relation (\ref{xitxi}), and 
hence
\be\label{adjoint_C}
U(\pi/4)\,\binom{\hat q}{\hat p}\,U(\pi/4)^{-1}
=
C\,\binom{\hat q}{\hat p}.
\ee
Equivalently,
\(
U(\pi/4)\,\hat q\,U(\pi/4)^{-1}
=\hat a^+,
\)
\(
U(\pi/4)\,\hat p\,U(\pi/4)^{-1}
=-\,i\hat a^-,
\)
which is precisely the quantum counterpart of \eqref{Cuv}.
In the usual ``time-evolution'' notation $e^{-it\hat H_-}$, the choice
$U(\pi/4)=e^{(\pi/4)\hat H_-}$ corresponds to the pure imaginary time $t=i\pi/4$.

\medskip
\noindent\textbf{Disentangling (factorized) form.}
Introduce $\hat H_+=\hat a^+\hat a^-+\frac12$ and the $\mathfrak{su}(1,1)$ generators
\be\label{K+K-K0}
K_+=\frac12(\hat a^+)^2,\qquad K_-=\frac12(\hat a^-)^2,\qquad K_0=\frac12\hat H_+,
\ee
so that $\hat H_- =-(K_++K_-)$. Then one has the standard (analytic continuation of the)
$\mathrm{SU}(1,1)$ disentangling identity, and at $s=\pi/4$ it yields the factorization,
see \eqref{app:eq:su11}, 
\be\label{Uc_factorised}
U(\pi/4)
=
\exp\!\Big(-\frac12(\hat a^+)^2\Big)\,
\exp\!\Big(\frac{1}{2} \ln 2\,\hat H_+\Big)\,
\exp\!\Big(-\frac12(\hat a^-)^2\Big)\,.
\ee
\noindent

\medskip
\noindent\textbf{Overall constant and its fixation by the Bargmann--Fock normalization.}
Let us now introduce
\be\label{Uc_with_c}
U_C=c\,U(\pi/4)=c\,\exp\!\Big(\frac{\pi}{4}\hat H_-\Big),
\qquad c\in\C^\times .
\ee
As an operator acting by similarity on $(\hat q,\hat p)$, the scalar $c\neq0$ is invisible:
$U_C \hat A U_C^{-1}=U(\pi/4)\hat A U(\pi/4)^{-1}$ for any operator $\hat A$.
Thus, \emph{the canonical transformation alone does not fix $c$}. 
\medskip
\noindent
The scalar is fixed only after one specifies an \emph{intertwining problem} and a
corresponding \emph{normalization convention}.  In particular, the unitary Bargmann transform
$\mathcal{B}:L^2(\R,dq)\to\mathcal F$ is fixed by the Bargmann--Fock Hilbert-space
normalization, while the non-unitary Schr\"odinger-space operator $U(\pi/4)$ admits
different convenient scalings on polynomial (rigged) cores.

\paragraph{Bargmann--Fock normalization ($\mathcal B\psi_0=1$).}
The Bargmann--Fock space $\mathcal F$ is the holomorphic realization in which
\(
\hat a^+\mapsto z,\) \( \hat a^-\mapsto \partial_z,\)
\( [\partial_z,z]=1,
\)
and therefore the corresponding generalized eigenbra $\langle z|$ of $\hat a^+$,
\(
\langle z|\,\hat a^+=z\,\langle z|,
\)
\(
\langle z|0\rangle=1,
\)
produces a holomorphic wavefunction from a Schr\"odinger state $|\psi\rangle$ by
\be\label{B_as_matrix_element}
(\mathcal{B}\psi)(z):=\langle z|\psi\rangle.
\ee

\medskip
\noindent\textbf{Dilations and the half-density factor.}
In the coordinate representation this becomes the integral transform \eqref{BTransform}, since
the kernel $K(z,q)$ is the solution of the eigenvalue equation
\(
\hat a^+\,K(z,q)=z\,K(z,q),\) \(  \hat a^+=\frac{1}{\sqrt2}\big(q-\partial_q\big),
\)
namely
\[
K(z,q)=A(z)\,\exp\!\Big(-\frac12 q^2+\sqrt2\,zq\Big).
\]
Holomorphicity in $z$ and compatibility with the Fock--Bargmann normalization fix
$A(z)=A_0\,e^{-z^2/2}$, which reproduces the Gaussian factor in \eqref{BTransform}.
Equivalently, the full exponent in $K(z,q)$ equals $-iF(q,z)$, where $F(q,u)$ is the
type-I generating function of the classical complexified canonical transformation \eqref{Cuv}
(with the identification $u\equiv z$).

The exponential phase alone does \emph{not} determine a
unitary quantum operator. A correct quantization of a linear canonical transformation
(metaplectic operator) requires a square-root Jacobian prefactor (equivalently, an action
on half-densities), which is the quantum remnant of ordering choices (and of the metaplectic correction). In the Bargmann case, this prefactor is fixed by holomorphicity and
by Hilbert-space normalization.  With the standard convention that the oscillator vacuum
\(
\psi_0(q)=\pi^{-1/4}e^{-q^2/2}
\)
is mapped to the constant function $1$, one gets $A_0=\pi^{-1/4}$.
This fixes the usual \emph{unitary} Bargmann transform $\mathcal{B}:L^2(\R,dq)\to\mathcal F$.

\paragraph{Schr\"odinger-side normalization ($U_C\chi_n=\psi_n$) and the factor $2^{1/4}$.}
We now adopt an alternative, Schr\"odinger-side normalization for $U_C$, which in general differs from the Bargmann--Fock (unitary) convention above. 
The operator $U(\pi/4)=\exp(\frac{\pi}{4}\hat H_-)$ acting on
$q$-space is \emph{non-unitary} on $L^2(\R,dq)$, and its overall scalar is not fixed by the
similarity action on $(\hat q,\hat p)$.
A different and very natural choice is to fix the Schr\"odinger-side normalization by requiring that,
when $U_C$ is applied to the polynomial (Jordan) states $\chi_n(q):=q^n/\sqrt{n!}$, 
see \eqref{app:MonomialsMainText}, one has
\[
U_C\,\chi_0=\psi_0,
\qquad\text{equivalently}\qquad
U_C\,\chi_n=\psi_n \ \ (n\ge0),
\]
where $\{\psi_n\}$ are the normalized harmonic-oscillator eigenfunctions.

To see what this implies for the scalar, one may use the factorization
\[
U(\pi/4)=\exp\!\Big(-\frac12 \hat q^{\,2}\Big)\,
         \exp\!\Big(\frac12 \hat H_0\Big)\,
         \exp\!\Big(\frac{i}{2}\ln 2\,\hat D\Big),
\qquad
\hat H_0=\frac12\hat p^{\,2},\quad
\hat D=\frac12(\hat q\hat p+\hat p\hat q)\,,
\]
see Appendix \ref{app:sp2r_abcd}.
In the coordinate representation, $\hat p=-i\partial_q$ and
\(
i\hat D=q\,\partial_q+\frac12 .
\)
Consequently, for any real $\alpha$ one has the exact dilation formula
\be\label{dilation_formula_halfdensity}
\exp(i\alpha \hat D)\,f(q)
=\exp\!\Big(\alpha\Big(q\partial_q+\frac12\Big)\Big)\,f(q)
=e^{\alpha/2}\,f(e^\alpha q),
\ee
and the presence of the constant $+\tfrac12$ term is precisely what produces the
half-density (Jacobian) factor $e^{\alpha/2}$.
Moreover, \eqref{dilation_formula_halfdensity} implies unitarity of dilations on $L^2(\R,dq)$:
\[
\int_\R\bigl|\exp(i\alpha \hat D)\,f(q)\bigr|^2\,dq
=\int_\R e^{\alpha}\,|f(e^\alpha q)|^2\,dq
=\int_\R |f(u)|^2\,du.
\]
Applying \eqref{dilation_formula_halfdensity} to the last factor with
$\alpha=\frac12\ln 2$ yields
\[
\exp\!\Big(\frac{i}{2}\ln 2\,\hat D\Big)\,f(q)
=2^{1/4}\,f(\sqrt2\,q),
\qquad\text{in particular}\qquad
\exp\!\Big(\frac{i}{2}\ln 2\,\hat D\Big)\,1=2^{1/4}.
\]
Since $\exp(\frac12 \hat H_0)\,1=1$, we obtain
\(
U(\pi/4)\,1=2^{1/4}\,e^{-q^2/2}=(2\pi)^{1/4}\,\psi_0(q),
\)
and therefore the condition $U_C\,\chi_0=\psi_0$ fixes uniquely
\[
U_C:=(2\pi)^{-1/4}\,U(\pi/4)=(2\pi)^{-1/4}\,\exp\!\Big(\frac{\pi}{4}\hat H_-\Big).
\]
In this way, the discrepancy between the Bargmann-kernel constant $\pi^{-1/4}$ and the
Schr\"odinger-side constant $(2\pi)^{-1/4}$ is completely explained:
\(
(2\pi)^{-1/4}=2^{-1/4}\,\pi^{-1/4},
\)
and the factor $2^{\pm1/4}$ is traced back directly to the constant $+\tfrac12$ term in
$i\hat D=q\partial_q+\tfrac12$ (the metaplectic ``half-form'' contribution).
If one were to omit this term (i.e.\ replace $i\hat D$ by $q\partial_q$), then the
prefactor $e^{\alpha/2}$ in \eqref{dilation_formula_halfdensity} would disappear, but
dilations would no longer be unitary on $L^2(\R)$:
$\|\exp(\alpha q\partial_q)f\|_{L^2}^2=e^{-\alpha}\|f\|_{L^2}^2$.

\medskip
\noindent\textbf{A commutative diagram on the polynomial core.}
On the polynomial core, set $e_n(z):=z^n/\sqrt{n!}$ and define the coefficient-wise
``analytic continuation'' map $\mathcal{A} :\chi_n\mapsto e_n$.
(Since $\chi_n(q)=q^n/\sqrt{n!}$ are polynomials, this identification is unambiguous on
$\operatorname{span}\{\chi_n\}$, but it does \emph{not} define a canonical map on all of
$L^2(\R)$.)
Since $(\mathcal B\psi_n)(z)=e_n(z)$ and $U_C\chi_n=\psi_n$, one has on
$\operatorname{span}\{\chi_n\}$ the algebraic identity
\[
\mathcal{B}\circ U_C= \mathcal{A},
\qquad\text{equivalently}\qquad
U_C=\mathcal{B}^\dagger\circ \mathcal{A},
\]
summarized by the commutative diagram
\be\label{cdiagram}
\boxed{
\begin{array}{ccc}
\operatorname{span}\{\chi_n\} & \xrightarrow{\quad U_C\quad} &
\operatorname{span}\{\psi_n\}\subset L^2(\R,dq) \\
\ \downarrow \mathcal{A} &  & \downarrow \mathcal{B} \\
\operatorname{span}\{e_n\}\subset\mathcal F & \xrightarrow{\ \ \mathrm{id}\ \ } &
\operatorname{span}\{e_n\}\subset\mathcal F
\end{array}
}
\ee
(which is to be understood on this dense common core).
%%%%%%%%%%%%%%%%%%%%%%%%%%%%%%%%%%%%%%%%%%%%%%%%%%%%%%%%%%%%

\medskip
\noindent\textbf{Conformal-bridge interpretation.}
The same operator $U(\pi/4)=e^{(\pi/4)\hat H_-}$ can be interpreted as the evolution generated by the inverted harmonic oscillator at the imaginary time $t=i\pi/4$, and it serves as the quantum ``conformal bridge'':
it maps polynomial Jordan states of the free particle at $E=0$ to harmonic-oscillator
eigenstates, plane waves to oscillator coherent states, and $\psi_{0}$ to a squeezed state.
We record this here only for completeness; a detailed discussion is presented in the next Section.

Note also that the conformal-bridge transformation is underpinned by the parallel realisations of the generators of the $\mathfrak{sl}(2,\mathbb{R})$ conformal algebra in the Schr\"odinger representation on $L^{2}(\mathbb{R},dq)$ and in the Bargmann--Fock representation on $\mathcal F$.
This parallels the classical relation \eqref{Relation-Flows} between the corresponding flows in real and complex symplectic bases, and is summarized by the following scheme:
\be\label{Schr--BF}
\boxed{
\begin{array}{c|c}
\textstyle L^{2}(\mathbb{R},dq) & \textstyle \mathcal{F}\\ \hline
\textstyle i\hat{D}=q\,\partial_{q}+\frac{1}{2}
& \textstyle \hat{H}_{+}=z\,\partial_{z}+\frac{1}{2}\\[1.2ex]
\textstyle \hat{H}_{+}=\frac{1}{2}\bigl(-\partial_{q}^{2}+q^{2}\bigr)
& \textstyle i\hat{D}=\frac{1}{2}\bigl(-\partial_{z}^{2}+z^{2}\bigr)\\[1.2ex]
\textstyle \hat{H}_{-}=-\frac{1}{2}\bigl(\partial_{q}^{2}+q^{2}\bigr)
& \textstyle \hat{H}_{-}=-\frac{1}{2}\bigl(\partial_{z}^{2}+z^{2}\bigr)
\end{array}
}
\ee

%%%%%%%%%%%%%%%%%%%%%%%%%%%%%%%%%%%%%%%%%%%%%%%%%%%%%%%%%%%%%%%%%%%%%%
\section{Quantum Cayley transform and conformal bridge transformation}
\label{sec:QCayley_CBT}
In the previous Section  we constructed the quantum Cayley operator
\[
U(\pi/4)=\exp\!\Big(\frac{\pi}{4}\,\hat H_-\Big),
\qquad
\hat H_-=\frac12(\hat p^{\,2}-\hat q^{\,2})
=-\frac12\big((\hat a^+)^2+(\hat a^-)^2\big),
\]
and fixed the overall scalar by the Schr\"odinger-side normalization condition
\be\label{eq:CBT_UC_def}
U_C:=(2\pi)^{-1/4}\,U(\pi/4),
\qquad
U_C\,\chi_n=\psi_n
\ee
(which determines $U_C$ uniquely up to an irrelevant global phase).  In the present Section
we use $U_C$ as the basic intertwiner implementing, at the quantum level, the complexified
canonical transformation generated by the Cayley matrix $C$.  This provides a convenient
and conceptually transparent realization of the conformal bridge viewpoint \cite{CBT1,CBT5,CBT8}, 
and it allows
one to relate oscillator coherent/squeezed states to their free-particle images (or
preimages) in a uniform language.

%----------------------------------------------------------------------
\medskip
\noindent\textbf{From $E=0$ Jordan states to oscillator eigenstates.}
The free-particle Hamiltonian is $\hat H_0=\frac12\hat p^{\,2}$.
At $E=0$ one has the polynomial Jordan chain $\{\chi_n\}_{n\ge0}$, see
App.~\ref{app:Jordan-analytic}.  In particular, the dilation generator
\(
\hat D:=\frac12(\hat q\hat p+\hat p\hat q)
\,\Longrightarrow\)
\(i\hat D=q\,\partial_q+\frac12
\)
acts diagonally on $\chi_n$:
\be\label{eq:CBT_iD_on_chi}
i\hat D\,\chi_n=\Big(n+\frac12\Big)\chi_n.
\ee

The key conformal-bridge statement  is
the similarity transformation
\be\label{eq:CBT_conjugation_iD_to_Hplus}
U_C\,(i\hat D)\,U_C^{-1}=\hat H_+,
\ee
where $\hat H_+=\hat a^+\hat a^-+\frac12$ is the harmonic-oscillator Hamiltonian,
that follows from $U_C\,(\hat{q},i\hat{p})\,U_C^{-1}=
(\hat a^+, \hat a^-)$.
Combining \eqref{eq:CBT_iD_on_chi} with \eqref{eq:CBT_conjugation_iD_to_Hplus} yields
\[
\hat H_+\,(U_C\chi_n)
=
U_C\,(i\hat D)\chi_n
=
\Big(n+\frac12\Big)\,U_C\chi_n\,,
\]
so $U_C\chi_n$ is an eigenstate of $\hat H_+$ with eigenvalue $n+\tfrac12$. 
With the normalization \eqref{eq:CBT_UC_def} this becomes the explicit bridge (\ref{cdiagram}), 
\be\label{eq:CBT_UC_chi_to_psi}
U_C\,\chi_n=\psi_n,\qquad n=0,1,2,\ldots .
\ee
Thus, at the level of stationary problems, the Cayley operator trades the
non-compact spectral problem for $i\hat D$ (on the free side at $E=0$) for the
compact oscillator spectrum of $\hat H_+$.

\medskip
\noindent\textbf{From plane waves ($E>0$) to Glauber coherent states.}
\label{subsec:plane_to_coherent}
Consider free-particle momentum eigenfunctions (as distributions)
\[
\phi_k(q):=\frac{1}{\sqrt{2\pi}}\,e^{ikq},
\qquad
\hat p\,\phi_k = k\,\phi_k,
\qquad
E=\frac{k^2}{2}>0.
\]
The Cayley similarity implements the complexified canonical transformation
\[
U(\pi/4)\,\hat q\,U(\pi/4)^{-1}=\hat a^+,
\qquad
U(\pi/4)\,\hat p\,U(\pi/4)^{-1}=-i\,\hat a^-,
\]
hence the same holds for $U_C$.
Applying this to $\hat p\,\phi_k=k\phi_k$ gives
\(
\hat a^-\,\big(U_C\phi_k\big)
=
i k\,\big(U_C\phi_k\big).
\)
Therefore $U_C\phi_k$ is (up to an overall $k$-dependent phase/constant) the
harmonic-oscillator \emph{Glauber coherent state} with eigenvalue $\alpha=ik$:
\be\label{eq:CBT_plane_to_coherent}
U_C\,\phi_k \ \propto\  \psi_{\alpha}(q)\big|_{\alpha=ik},
\qquad
\hat a^-\,\psi_\alpha=\alpha\,\psi_\alpha.
\ee
For convenience we recall the standard normalized coordinate-space form
\be\label{eq:CBT_coherent_wavefunction}
\psi_\alpha(q)
=
\pi^{-1/4}\exp\!\left(
-\frac12\bigl(q-\sqrt2\,\Re\alpha\bigr)^2
+i\sqrt2\,(\Im\alpha)\,q
-\frac{i}{2}\,\Re\alpha\,\Im\alpha
\right),
\ee
which for purely imaginary $\alpha=ik$ reduces to a Gaussian with a plane-wave phase.

%%%%%%%%%%%%%%%%%%%%%%%%%%%%%%%%%%%%%%%%%%%%%%%%%%%%%%%%%%%%%%%%%%%%%%%%%%%%%%%%%%%%%%%%%

\medskip
\noindent\textbf{Free-particle preimages of oscillator coherent squeezed states.}
\label{subsec:free_preimages_squeezed}
We use the $\mathfrak{su}(1,1)$ generators $(K_+,K_-,K_0)$ defined in
Eq.~\eqref{K+K-K0}.  For $\alpha,\zeta\in\C$ we introduce the standard
displacement and single--mode squeezing operators
\be
D(\alpha):=\exp(\alpha\,\hat a^{+}-\bar\alpha\,\hat a^{-}),
\qquad
S(\zeta):=\exp(\zeta\,K_{+}-\bar\zeta\,K_{-}),
\label{eq:DandS_def_23}
\ee
and the normalized oscillator coherent  squeezed states (Perelomov/Jacobi-group coherent states)
\be
\ket{\alpha,\zeta}:=D(\alpha)\,S(\zeta)\ket{0}.
\label{eq:sqcoh_def_23}
\ee
Writing $\zeta=r\,e^{\ii\phi}$, we set
\be
\kappa:=e^{\ii\phi}\tanh r,
\qquad |\kappa|<1\quad (r\in\R),
\label{eq:kappa_def_23}
\ee
and employ the standard $\mathfrak{su}(1,1)$ disentangling (valid for $|\kappa|<1$),
\be
S(\zeta)
=\exp(\kappa\,K_{+})\,
\exp\!\bigl(\ln(1-|\kappa|^{2})\,K_{0}\bigr)\,
\exp(-\bar\kappa\,K_{-})\,,
\label{eq:S_disentangle_23}
\ee
see Appendix 	\ref{app:sp2r_abcd}.
Since $K_{-}\ket{0}=0$ and $K_{0}\ket{0}=\frac14\ket{0}$, one gets the squeezed vacuum in the
number basis as
\be
S(\zeta)\ket{0}
=(1-|\kappa|^{2})^{1/4}\sum_{n=0}^{\infty}
\frac{\sqrt{(2n)!}}{2^{n}\,n!}\,\kappa^{n}\,\ket{2n}.
\label{eq:sqvac_number_23}
\ee
In the coordinate representation this is a Gaussian wavefunction, see Appendix \ref{app:demo_psi0zeta},
\be\label{psi0zeta}
\psi_{0,\zeta}(q):=\langle q|S(\zeta)|0\rangle
=\pi^{-1/4}(1-|\kappa|^{2})^{1/4}(1+\kappa)^{-1/2}
\exp\!\left(-\frac12\,\frac{1-\kappa}{1+\kappa}\,q^{2}\right)\,,
\ee
where the square root is taken on a fixed branch (different branch choices change only an
overall phase).  For $|\kappa|<1$ one has $\Re\,\bigl(\frac{1-\kappa}{1+\kappa}\bigr)>0$, so
$\psi_{0,\zeta}\in L^{2}(\R,dq)$.

\medskip
\noindent\textbf{Free-particle preimages under $U_C^{-1}$.}
Define the free-particle preimages by
\be
\Psi_{\alpha,\zeta}(q):=\bigl(U_C^{-1}\ket{\alpha,\zeta}\bigr)(q).
\label{eq:preimage_def_23}
\ee
A convenient parametrization of normalized free Gaussians is given in Appendix \ref{app:Sch_free_coherent}
(Robertson--Schr\"odinger saturation): for $\tau\in\H_{+}$ and real $(q_{0},p_{0})$,
\be
\psi_{q_{0},p_{0};\tau}(q)
=\left(\frac{\Im\tau}{\pi|\tau|^{2}}\right)^{1/4}
\exp\!\left(\ii p_{0}q-\frac{\ii}{2\tau}(q-q_{0})^{2}\right),
\qquad \tau\in\H_{+}.
\label{eq:free_gauss_tau_23}
\ee
The inverse Cayley map generated by $U_C^{-1}$ identifies the squeezing parameter $\kappa$
with the upper-half-plane parameter $\tau$ by the same Cayley transform used in (\ref {wiz}):
\be
\kappa \;=\; \ii\,\frac{\tau-\ii}{\tau+\ii},
\qquad\Longleftrightarrow\qquad
\tau \;=\; \ii\,\frac{1-\ii\kappa}{1+\ii\kappa}.
\label{eq:kappa_tau_cayley_23}
\ee
Moreover, the displacement parameter $\alpha$ fixes the real center $(q_{0},p_{0})$ in the
usual way (with $\hat a^{-}=\frac{1}{\sqrt2}(\hat q+\ii\hat p)$):
\(
q_{0}=\sqrt2\,\Re\alpha\,,\,\) \( p_{0}=\sqrt2\,\Im\alpha\).
With these identifications one has, up to an overall (physically irrelevant) phase,
\be
\Psi_{\alpha,\zeta}(q)\;=\;e^{\ii\Theta(\alpha,\zeta)}\,\psi_{q_{0},p_{0};\tau}(q),
\qquad
\tau=\ii\,\frac{1-\ii\kappa}{1+\ii\kappa}.
\label{eq:preimage_is_free_gaussian_23}
\ee
\medskip
\noindent\textbf{$L^{2}$-condition and analytic continuation.}
From \eqref{eq:free_gauss_tau_23} the preimage belongs to $L^{2}(\R,dq)$ if and only if
\be
\Im\tau>0
\qquad\Longleftrightarrow\qquad
|\kappa|<1.
\label{eq:L2_condition_23}
\ee
Thus all \emph{unitary} oscillator squeezed coherent states (real $r$ in \eqref{eq:kappa_def_23})
have square-integrable free preimages.  If one analytically continues the squeezing parameter
beyond the unit disk (so that $|\kappa|\ge 1$), then $\tau$ leaves $\H_{+}$ and the Gaussian
in \eqref{eq:free_gauss_tau_23} ceases to be normalizable; this reflects the fact that
$U_C^{-1}$ is an unbounded (non-unitary) similarity transform on $L^{2}(\R)$.

\medskip
\noindent\textbf{Bogoliubov viewpoint.}
Conjugation by $S(\zeta)$ implements the Bogoliubov transformation (\ref{Bogol}),
\be
S(\zeta)\,\hat a^{\mp}\,S(\zeta)^{-1}
=\hat a^{\mp}\cosh r-\;e^{\pm\ii\phi}\,\hat a^{\pm}\sinh r,
\label{eq:Bogoliubov_23}
\ee
i.e. the $SU(1,1)$ action generated by \eqref{eq:S_disentangle_23}.  Under the inverse Cayley
intertwiner $U_C^{-1}$ this becomes the corresponding linear canonical transformation on the
free-particle side, encoded by the M\"obius action on $\tau$ (equivalently, by the Cayley
relation \eqref{eq:kappa_tau_cayley_23}).

%%%%%%%%%%%%%%%%%%%%%%%%%%%%%%%%%%%%%%%%%%%%%%%%%%%%%%%%%%%%%%%%%%%%%%%%%%%%%%%%%%%%%%%%

\medskip
\noindent\textbf{$\cos$ and $\sin$ as free-particle images of cat states.}
Define the (unnormalized) even/odd cat combinations \cite{CatStates}
\[
\ket{\mathrm{cat}_\pm(\alpha)}:=\ket{\alpha}\pm\ket{-\alpha}.
\]
Choosing $\alpha=ik$ with $k\in\R$ and using \eqref{eq:CBT_plane_to_coherent}, one finds
\(
U_C^{-1}\ket{\mathrm{cat}_\pm(ik)}
\ \propto\
\phi_k \pm \phi_{-k}
\ \propto\
e^{ikq}\pm e^{-ikq}.
\)
Thus, up to constants and phases,
\be\label{eq:CBT_cos_sin}
U_C^{-1}\ket{\mathrm{cat}_+(ik)}\ \propto\ \cos(kq),
\qquad
U_C^{-1}\ket{\mathrm{cat}_-(ik)}\ \propto\ \sin(kq).
\ee
In this way the standard even/odd superpositions of oscillator coherent states
have natural free-particle counterparts given by the trigonometric eigenfunctions.

%%%%%%%%%%%%%%%%%%%%%%%%%%%%%%%%%%%%%%%%%%%%%%%%%%%%%%%%%%%%%%%%%%%%%%

%%%%%%%%%%%%%%%%%%%%%%%%%%%%%%%%%%%%%%%%%%%%%%%%%%%%%%%%%%%%%%%%%%%%%%%%%%%%%%%%%%%%%%%%%%%%%%%%%%%%%%%%%%%%%%%%%%%%%%%%%

\section{Classical symmetries of the harmonic oscillator and Cayley--Niederer  transform}\label{SymmetriesHO}

In this Section we turn to the one--dimensional harmonic oscillator and recall its classical
symmetry structure in a form adapted to later comparison with the free particle.  Besides
time translations generated by $H_{+}$, the oscillator possesses Newton--Hooke analogs of
spatial translations and Galilean boosts, realized as dynamical integrals of motion with
explicit $(2\pi/\omega)$--periodic time dependence.  Together with a distinguished set of
quadratic dynamical integrals they generate a $\mathfrak{h}_3 \rtimes \mathfrak{sp}(2,\R)$
Lie--algebraic structure, and we also indicate the $\omega\to0$ limit in which the free--particle
generators are recovered.  We then rewrite the quadratic dynamical integrals in an explicit
trigonometric form, emphasizing in particular the $\pi/\omega$ periodicity and the natural
compactification of time.  This prepares the ground for the  Cayley--Niederer map:
via an appropriate reparametrization (chart--wise on $\RP^{1}$) the oscillator evolution is
related to the free particle, and the corresponding time reparametrizations acquire a
manifest M\"obius form.  We conclude with the discrete symmetries of the oscillator.

\medskip
\noindent\textbf{Equations of motion and their solution.}
Consider the one--dimensional harmonic oscillator ($\omega>0$),
\be
L_{\rm osc}=\frac{m}{2}\left(\dot x^{2}-\omega^{2}x^{2}\right)\,,\qquad
H_{+}=\frac{p^{2}}{2m}+\frac{m\omega^{2}x^{2}}{2}\,,\qquad \{x,p\}=1\,.
\label{eq:HO_L_H}
\ee
The canonical equations are
\(
\dot x=\{x,H_{+}\}=\frac{p}{m}\,,\)
\( \dot p=\{p,H_{+}\}=-m\omega^{2}x\,,\)
and therefore $\ddot x+\omega^{2}x=0$. With initial data $x(0)=x_{0}$, $p(0)=p_{0}$,
\be
x(t)=x_{0}\cos\omega t+\frac{p_{0}}{m\omega}\sin\omega t\,,\qquad
p(t)=p_{0}\cos\omega t-m\omega x_{0}\sin\omega t\,.
\label{eq:HO_solution}
\ee

\medskip
\noindent\textbf{Newton--Hooke translations and boosts: dynamical integrals $P(t)$ and $B(t)$.}
Solving \eqref{eq:HO_solution} for $(x_{0},p_{0})$ yields the dynamical integrals
\be
P(t)=p\cos\omega t+m\omega x\sin\omega t\,,
\qquad
B(t)=mx\cos\omega t-\frac{p}{\omega}\sin\omega t\,,
\label{eq:HO_PB_dynamical}
\ee
which satisfy
\(
\frac{dN(t)}{dt}=\frac{\partial N(t)}{\partial t}+\{N(t),H_{+}\}=0\,,\)  $N(t)=P(t)\,,\,\,B(t)$.
Their Poisson bracket reproduces the Heisenberg algebra (with the mass $m$ as the central charge),
\(
\{B(t),P(t)\}=m\,.
\)
The Newton--Hooke relations with $H_{+}$ for any $t$ read
\be
\{H_{+},P(t)\}=-\,\omega^{2}B(t)\,,\qquad
\{H_{+},B(t)\}=P(t)\,.
\label{eq:HO_NewtonHooke_basic}
\ee
Thus, time translations generated by $H_{+}$ rotate the translation/boost doublet.

\medskip
\noindent\textbf{$sl(2,\R)$ dynamical integrals and the $\omega\to0$ limit.}
Guided by the free particle case, define 
\be
H_{0}^{\rm NH}(t)=\frac{1}{2m}\,P(t)^{2}\,,
\qquad
D^{\rm NH}(t)=\frac{1}{m}\,B(t)P(t)\,,
\qquad
K^{\rm NH}(t)=\frac{1}{2m}\,B(t)^{2}\,,
\label{eq:HO_H0DK_NH_def}
\ee
which are dynamical integrals.
Their Poisson bracket algebra 
is $sl(2,\R)$: 
\be
\{D^{\rm NH},H_{0}^{\rm NH}\}=2H_{0}^{\rm NH}\,,\qquad
\{D^{\rm NH},K^{\rm NH}\}=-2K^{\rm NH}\,,\qquad
\{K^{\rm NH},H_{0}^{\rm NH}\}=D^{\rm NH}\,,
\label{eq:HO_sl2_algebra_NH}
\ee
together with the action on $(P,B)$,
\be
\{D^{\rm NH},P\}=P\,,\qquad
\{D^{\rm NH},B\}=-B\,,\qquad
\{H_{0}^{\rm NH},B\}=-P\,,\qquad
\{K^{\rm NH},P\}=B\,.
\label{eq:HO_action_on_PB_NH}
\ee
The oscillator Hamiltonian $H_{+}$ is the compact (Newton--Hooke) combination
\be
H_{+}=H_{0}^{\rm NH}+\omega^{2}\,K^{\rm NH}\,,
\label{eq:Hplus_as_compact_comb_NH}
\ee
which indeed reduces  to $H_+=\frac{p^{2}}{2m} +\frac{m\omega^{2}x^{2}}{2}=const$ for any $t$.

\medskip
\noindent\textbf{Explicit trigonometric form; $\pi/\omega$ periodicity and $\RP^{1}$.}
On a fixed time slice $t=0$,
$B(0)=mx_0$,  $P(0)=p_0$, and one has
\(
H_{0}=\frac{p^{2}_0}{2m}\,,\)
\(
D^{\rm NH}(0)=x_0p_0:=D_{0}\,,\)
\( K^{\rm NH}(0)=\frac{1}{2}mx^{2}_0:=K_{0}\,.\)
From \eqref{eq:HO_PB_dynamical} and \eqref{eq:HO_H0DK_NH_def} one obtains
\be
H_{0}^{\rm NH}(t)=H_{0}\cos^{2}\omega t+\frac{\omega}{2}\,D_{0}\,\sin(2\omega t)
+{\omega^{2}}\,K_{0}\sin^{2}\omega t\,,
\label{eq:HO_H0_trig_NH}
\ee
\be
D^{\rm NH}(t)=D_{0}\cos(2\omega t)
+\left(\omega K_{0}-\frac{1}{\omega}H_{0}\right)\sin(2\omega t)\,,
\label{eq:HO_D_trig_NH}
\ee
\be
K^{\rm NH}(t)=K_{0}\cos^{2}\omega t-\frac{1}{2\omega}\,D_{0}\,\sin(2\omega t)
+\frac{1}{\omega^{2}}\,H_{0}\sin^{2}\omega t\,.
\label{eq:HO_K_trig_NH}
\ee
In the limit $\omega\to0$, 
$P(t)\to p,$ $B(t)\to mx-pt,$
$H_{0}=\frac{p^{2}}{2m}$, 
$D^{\rm NH}(t)\to xp-2H_{0}\,t,$
and $K^{\rm NH}(t)\to K_0-D_0t+H_0t^2$,
cf. (\ref{pG}) and (\ref{H0DK}).

Although $(x(t),p(t))$ are $2\pi/\omega$--periodic, the dynamical integrals
$D^{\rm NH}(t)$ and $K^{\rm NH}(t)$ depend on $2\omega t$ and therefore have period $\pi/\omega$.
From the viewpoint of the $sl(2,\R)$ generators, time is naturally a coordinate on a circle with
identified antipodes, i.e.\ on $\RP^{1}$.

A convenient Cayley parameter on this $\RP^{1}$ is
\be
z=e^{2\ii\omega t}\,,
\label{eq:HO_Cayley_z}
\ee
cf. (\ref{eq:w_theta_phi}), which is invariant under $t\mapsto t+\pi/\omega$ and makes the double--angle periodicity manifest.

\medskip
\noindent\textbf{Compact $sl(2,\R)$ evolution and ladder variables.}
Define the complementary combination
\be
J^{\rm NH}:=H_{0}^{\rm NH}-\omega^{2}\,K^{\rm NH}\,,
\label{eq:HO_J_def_NH}
\ee
so that $H_{+}=H_{0}^{\rm NH}+\omega^{2}K^{\rm NH}$ and $J^{\rm NH}$ are the two linear combinations of $(H_{0}^{\rm NH},K^{\rm NH})$, with $J^{\rm NH}$  corresponding to $H_-$ considered above.
Using \eqref{eq:HO_sl2_algebra_NH} one finds
\be
\dot D^{\rm NH}=\{D^{\rm NH},H_{+}\}=2J^{\rm NH}\,,\qquad
\dot J^{\rm NH}=\{J^{\rm NH},H_{+}\}=-2\omega^{2}D^{\rm NH}\,,
\label{eq:HO_DJ_system_NH}
\ee
and hence $\ddot D^{\rm NH}+(2\omega)^{2}D^{\rm NH}=0$.
Introduce the ladder combinations
\be
J_{\pm}^{\rm NH}:=J^{\rm NH}\pm \ii\omega D^{\rm NH}\,.
\label{eq:HO_Jpm_def_NH}
\ee
Then \eqref{eq:HO_DJ_system_NH} gives
\(\dot J_{\pm}^{\rm NH}=\{J_{\pm}^{\rm NH},H_{+}\}=\pm 2\ii\omega\,J_{\pm}^{\rm NH}\,,\)
and therefore
\be
J_{\pm}^{\rm NH}(t)=e^{\pm 2\ii\omega t}\,J_{\pm}^{\rm NH}(0)=z^{\pm 1}J_{\pm}^{\rm NH}(0)\; 
\,,
\label{eq:HO_Jpm_solution_NH}
\ee
with $z=e^{2\ii\omega t}$ from \eqref{eq:HO_Cayley_z}.

\medskip
\noindent\textbf{Cayley--Niederer map and the emergence of M\"obius transformations.}
To reveal the fractional--linear structure, introduce, coherently with (\ref{Caypar}) and (\ref{t=tanphi}),  the projective time variable
\be
\tau=\frac{1}{\omega}\tan(\omega t)\,,
\qquad\Longleftrightarrow\qquad
t=\frac{1}{\omega}\arctan(\omega\tau)
\quad\text{(chart-wise on }\RP^{1}\text{)}\,,
\label{eq:HO_tau_tan}
\ee
and the rescaled coordinate
\be
y=\frac{x}{\cos(\omega t)}=x\,\sqrt{\frac{d\tau}{dt}}\,.
\label{eq:HO_y_rescale}
\ee
The factor $\sqrt{d\tau/dt}$ shows that $x$ behaves as a half-density of weight $-1/2$ under time
reparametrizations in this construction, cf. (\ref {x=weight-1/2}).
A direct computation shows that $y(\tau)$ satisfies the free equation $d^{2}y/d\tau^{2}=0$.
Equivalently, the free and oscillator Lagrangians are related up to a total derivative:
\(
\frac{m}{2}\left(\frac{dy}{d\tau}\right)^{2}d\tau
=
\left[
\frac{m}{2}\left(\dot x^{2}-\omega^{2}x^{2}\right)
+\frac{d}{dt}\left(\frac{m\omega}{2}\,x^{2}\tan(\omega t)\right)
\right]dt\,\), cf. \eqref {eq:free_actiontilde}
 below. 
The map \eqref{eq:HO_tau_tan}--\eqref{eq:HO_y_rescale} is the (classical) 
Cayley--Niederer
transformation relating the harmonic oscillator to the free particle.

In the $(\tau,y)$ variables, the $sl(2,\R)$ transformations act in the standard projective way,
\be
\tau'=\frac{a\tau+b}{c\tau+d}\,,\qquad
y'=\frac{y}{c\tau+d}\,,\qquad ad-bc=1\,,
\label{eq:HO_Mobius_on_tau}
\ee
so that in terms of the physical time $t$ one obtains the fractional--linear law for $\tan(\omega t)$,
\be
\tan(\omega t')=\frac{a\,\tan(\omega t)+b\,\omega}{c\,\tan(\omega t)+d\,\omega}\,,
\label{eq:HO_Mobius_on_tan}
\ee
cf. (\ref {eq:tau_translation_mobius}).
This explains why, in contrast to the free particle case, the $sl(2,\R)$ transformations are not
manifestly M\"obius in $t$: the physical time $t$ is an angular coordinate on $\RP^{1}$, while
$\tau=\tan(\omega t)/\omega$ is an affine coordinate on the projective line.

For completeness, the induced transformation of $x$ can be written as
\be
x' = y' \cos(\omega t')
= x\,\frac{\cos(\omega t')}{\cos(\omega t)}\,
\frac{1}{\,d+\frac{c}{\omega}\tan(\omega t)\,}\,.
\label{eq:HO_x_transform}
\ee
A more detailed discussion of the Cayley--Niederer transformation including its quantum implementation
will be given below in Section  \ref{sec:free_to_osc_extended}.

\medskip
\noindent\textbf{Discrete symmetries.}
Besides the continuous symmetries above, the oscillator admits the standard discrete symmetries:

$\bullet$  Parity: $x\mapsto -x$, $p\mapsto -p$, $t\mapsto t$;\quad
$\bullet$ Time reversal: $t\mapsto -t$, $x\mapsto x$, $p\mapsto -p$.

\noindent
In addition, the free--particle conformal inversion is naturally realized for the oscillator
as an inversion of the \emph{projective} time variable:
\be
\omega\tau\mapsto \omega\tau'=-\,\frac{1}{\omega \tau}\,.
\label{eq:HO_projective_inversion}
\ee
Using $\omega\tau=\tan(\omega t)$, this corresponds (on $\RP^{1}$) to a shift by a quarter period in time $t$,
\be
\tan(\omega t')=-\cot(\omega t)=\tan\!\left(\omega t-\frac{\pi}{2}\right)
\qquad\Rightarrow\qquad
t'\equiv t-\frac{\pi}{2\omega}\ \ (\mathrm{mod}\ \tfrac{\pi}{\omega})\,,
\label{eq:HO_quarter_shift}
\ee
equivalently, on the Cayley circle \eqref{eq:HO_Cayley_z} it is the antipodal map $z\mapsto z'=-z$.
On phase space this can be represented as the canonical quarter--rotation
\(
(x,p)\ \mapsto\ \left(\frac{p}{m\omega},\ -\,m\omega x\right),
\)
under which
\be
H_{0}^{\rm NH}\ \longleftrightarrow\, \omega^{2}\,K^{\rm NH}\,,\qquad
D^{\rm NH}\ \longmapsto\ -D^{\rm NH}\,,
\label{eq:HO_inversion_on_generators}
\ee
$J^{\rm NH}\mapsto -J^{\rm NH}$, while the compact combination $H_{+}=H_{0}^{\rm NH}+\omega^{2}K^{\rm NH}$ is preserved.

%%%%%%%%%%%%%%%%%%%%%%%%%%%%%%%%%%%%%%%%%%%%%%%%%%%%%%%%%%%%%%%%%%%%%%%%%%%%%%%%%%%%%%%%%%%%%%%%%%%%%%%%%%%%%%%%%%%%%%%%%%%%%%%%%%%%%%%%%%%%%%%

\providecommand{\Schw}[2]{\left\{#1\,;\,#2\right\}} 

\section{Stationary and  time-dependent Schr\"odinger equations, and \texorpdfstring{$(\pm 1/2)$}{(+/-1/2)}-weight densities }\label{SchrodingerProjective}

Before we discuss the quantum version of the Cayley--Niederer transform for the
free--particle--harmonic oscillator correspondence, we recall here a few structural
facts about the stationary and time--dependent Schr\"odinger equations that are
naturally formulated in projective terms.  For the stationary equation in Liouville
(normal) form, the unique reparametrization that preserves the absence of the
first-derivative term is the $(-\tfrac12)$ (inverse half-density) law, and the
potential term $Q(x)$ transforms as a projective connection with an inhomogeneous
Schwarzian contribution.  For the TDSE, by contrast, the natural requirement is
unitarity of the Hilbert-space identification under spatial changes of variables,
which leads to a $(+\tfrac12)$ (half-density) law; a nontrivial map preserving the
standard flat kinetic term necessarily involves a correlated time reparametrization
(lens/Cayley--Niederer-type transform), in which the Schwarzian term is the
\emph{time} Schwarzian of $t=t(\tau)$.  We conclude by reconciling these two
weights, which arise from different goals (Liouville form versus unitarity).

\medskip
\noindent\textbf{Liouville transformation and the $(-\tfrac12)$ (inverse half-density) law.}
Consider the stationary 1D Schr\"odinger equation
\be
-\frac{\hbar^2}{2m}\,\psi''(x)+V(x)\psi(x)=E\,\psi(x),
\label{app:SSE}
\ee
which we rewrite in Liouville (normal) form
\be
\psi''(x)+Q(x)\psi(x)=0,
\qquad
Q(x):=\frac{2m}{\hbar^2}\bigl(E-V(x)\bigr).
\label{app:SSE_Liouville}
\ee
In the discussion below one may regard $Q(x)$ as an arbitrary function.

Let $y=y(x)$ be locally monotone, with inverse $x=x(y)$.  A mere change of variable produces
a first-derivative term; the unique rescaling that keeps the \emph{Liouville form} (no first derivative)
is the \emph{Liouville rescaling}
\be
\psi(x)=\bigl(y'(x)\bigr)^{-1/2}\,\phi(y),
\qquad y'=\frac{dy}{dx}.
\label{app:Liouville_rescale}
\ee
Equivalently, $\phi(y)=\bigl(x'(y)\bigr)^{-1/2}\psi(x(y))$.
This is precisely the invariance law for an \emph{inverse half-density}:
\be
\psi(x)\,(\mathrm{d}x)^{-1/2}=\phi(y)\,(\mathrm{d}y)^{-1/2}.
\label{app:inverse_half_density}
\ee
Indeed, since $\mathrm{d}y=y'(x)\mathrm{d}x$, one has $(\mathrm{d}y)^{-1/2}=(y')^{-1/2}(\mathrm{d}x)^{-1/2}$,
and \eqref{app:inverse_half_density} is equivalent to \eqref{app:Liouville_rescale}.
This $(-\tfrac12)$ weight should be understood as the density weight relevant to the \emph{Liouville-normal-form problem}
for the ODE,
cf. (\ref{x=weight-1/2}) and (\ref{eq:HO_y_rescale}).

\medskip
\noindent\textbf{Projective-connection transformation of $Q$: two equivalent forms.}
Substituting \eqref{app:Liouville_rescale} into \eqref{app:SSE_Liouville} yields again the Liouville form
\be
\phi_{yy}(y)+\widetilde Q(y)\,\phi(y)=0,
\label{app:Liouville_transformed_eq}
\ee
with the projective-connection transformation law
\be
\widetilde Q(y)=\bigl(x'(y)\bigr)^2\,Q\bigl(x(y)\bigr)+\frac12\,\Schw{x}{y}\,,
\label{app:Q_transform_SSE_xy}
\ee
where  $\Schw{x}{y}$ is the Schwarzian derivative \cite{OTbook,Osgood,OTams,PlyuSch},
see App. \ref{Appendix Schwarzian}.
Using the inverse-map identity \eqref{app:Schwarz_inverse}, the same formula can be rewritten in terms of
$\Schw{y}{x}$ \emph{with a common prefactor}:
\be
\widetilde Q(y)=\bigl(x'(y)\bigr)^2\left[\,Q\bigl(x(y)\bigr)-\frac12\,\Schw{y}{x}\Big|_{x=x(y)}\right]\,.
\label{app:Q_transform_SSE_yx}
\ee
Thus $Q$ is not a quadratic differential (which would transform as $\widetilde Q=(x')^2 Q$), but 
an affine connection-like object whose differences are quadratic differentials, 
with the inhomogeneous Schwarzian term encoding its projective-connection (cocycle) nature, see App. \ref{Appendix Schwarzian}.

\medskip
\noindent\textbf{Projective coordinate and the intrinsic meaning of $Q$.}
Let $\psi_1,\psi_2$ be two linearly independent solutions of \eqref{app:SSE_Liouville}.
Their ratio
\be
w(x):=\frac{\psi_1(x)}{\psi_2(x)}
\label{app:w_ratio}
\ee
is a projective coordinate (defined up to M\"obius transformations $w\mapsto (aw+b)/(cw+d)$), and one has the identity
\be
\Schw{w}{x}=2\,Q(x)\,. 
\label{app:w_Schwarz_Q}
\ee
Equivalently, the reduction-of-order integral
\be
u(x):=\int^{x}\frac{\mathrm{d}t}{\psi(t)^2}
\label{app:u_def}
\ee
is invariant under reparametrizations when $\psi$ obeys the inverse half-density law \eqref{app:inverse_half_density},
because
\be
\psi(x)^2=\bigl(y'(x)\bigr)^{-1}\phi(y)^2
\quad\Longrightarrow\quad
\frac{\mathrm{d}x}{\psi(x)^2}=\frac{\mathrm{d}y}{\phi(y)^2},
\label{app:u_invariance_step}
\ee
and therefore $\int^x \mathrm{d}t/\psi(t)^2=\int^{y(x)}\mathrm{d}\eta/\phi(\eta)^2$.
In this language the identity \eqref{app:w_Schwarz_Q} can be viewed as $\Schw{u}{x}=2Q(x)$.

\medskip
\noindent\textbf{TDSE: spatial reparametrizations at fixed time and unitarity.}
Consider the TDSE
\be
\ii\hbar\,\partial_t\Psi(t,x)
=
\left(-\frac{\hbar^2}{2m}\,\partial_x^2+V(t,x)\right)\Psi(t,x).
\label{app:TDSE}
\ee
Let $x=x(y)$ be a time-independent diffeomorphism with Jacobian
\(
J(y):=\frac{\mathrm{d}x}{\mathrm{d}y}>0.
\)

\medskip
\noindent\textbf{(a) Pure coordinate substitution.}
With $\widetilde\Psi(t,y):=\Psi\bigl(t,x(y)\bigr)$ one has
\be
\partial_x=\frac{1}{J}\partial_y,
\qquad
\partial_x^2=\frac{1}{J^2}\partial_y^2-\frac{J'}{J^3}\partial_y,\qquad J(y):=\frac{\mathrm{d}x}{\mathrm{d}y}>0.
\label{app:dx_to_dy}
\ee
and \eqref{app:TDSE} becomes a TDSE with a first-derivative term in the spatial operator.
Thus a general spatial reparametrization does \emph{not} preserve the standard flat form
$-(\hbar^2/2m)\partial_y^2+\widetilde V(t,y)$.

\medskip
\noindent\textbf{(b) Unitary half-density transform and the spatial Schwarzian term.}
The natural requirement for quantum mechanics is unitarity, i.e.\ preservation of
$\int |\Psi(t,x)|^2\,\mathrm{d}x$.
Since $\mathrm{d}x=J(y)\mathrm{d}y$, the unitary identification
$L^2(\mathrm{d}x)\cong L^2(\mathrm{d}y)$ is implemented by
\be
\Phi(t,y):=\sqrt{J(y)}\,\widetilde\Psi(t,y)
=
\sqrt{\frac{\mathrm{d}x}{\mathrm{d}y}}\,
\Psi\bigl(t,x(y)\bigr).
\label{app:unitary_half_density_map}
\ee
Equivalently,
\(
|\Psi(t,x)|^2\,\mathrm{d}x=|\Phi(t,y)|^2\,\mathrm{d}y,
\)
\(
\Psi(t,x)\,(\mathrm{d}x)^{1/2}=\Phi(t,y)\,(\mathrm{d}y)^{1/2}.
\) 
Thus, in the TDSE/Hilbert-space context, $\Psi$ behaves as a \emph{spatial} $(+1/2)$-density.

A direct substitution $\widetilde\Psi=J^{-1/2}\Phi$ into the pulled-back equation yields
\be
\ii\hbar\,\partial_t\Phi
=
-\frac{\hbar^2}{2m}\,\frac{1}{J(y)^2}\,\partial_y^2\Phi
+
\left(
V\bigl(t,x(y)\bigr)
-\frac{\hbar^2}{4m}\,\frac{1}{J(y)^2}\,\Schw{x}{y}
\right)\Phi.
\label{app:TDSE_spatial_general}
\ee
Two key features appear:
(i) the kinetic term acquires the variable coefficient $J^{-2}$ (a variable, 
``position-dependent  mass'' \cite{BravoPlyu});
(ii) a spatial Schwarzian correction contributes to the effective potential.
Therefore, preserving the \emph{flat} kinetic term with constant coefficient requires $J=\mathrm{const}$,
i.e.\ $x$ linear in $y$.
This is why the nontrivial free $\leftrightarrow$ oscillator map cannot be achieved by a purely spatial diffeomorphism:
it must involve time.

\medskip
\noindent\textbf{Lens (Cayley--Niederer-type) transform: time reparametrization and time-Schwarzian term.}
To preserve the \emph{standard TDSE form} with a flat kinetic term, one uses a correlated transformation
of time, space, and wave function.

Let
\be
t=t(\tau),\qquad \rho(\tau):=\frac{\mathrm{d}t}{\mathrm{d}\tau}>0,
\label{app:time_reparam}
\ee
and scale space linearly,
\be
x=\sqrt{\rho(\tau)}\,y.
\label{app:scaling_space}
\ee
(The spatial map is linear in $y$, hence its \emph{spatial} Schwarzian vanishes identically.)

Define the transformed wave function by
\be
\Psi\bigl(t(\tau),x\bigr)
=
\rho(\tau)^{-1/4}\,
\exp\!\left(\frac{\ii}{\hbar}\,S(\tau,y)\right)\,
\Phi(\tau,y),
\qquad x=\sqrt{\rho(\tau)}\,y\,,
\label{app:lens_transform}
\ee
with quadratic phase
\be
S(\tau,y):=\frac{m}{4}\,\frac{\rho'(\tau)}{\rho(\tau)}\,y^2,
\qquad ({}'=\mathrm{d}/\mathrm{d}\tau).
\label{app:phase_S}
\ee
The prefactor $\rho^{-1/4}$ is exactly the unitary half-density factor corresponding to
$\mathrm{d}x=\sqrt{\rho}\,\mathrm{d}y$:
\be
\int_{\R}|\Psi|^2\,\mathrm{d}x=\int_{\R}|\Phi|^2\,\mathrm{d}y.
\label{app:norm_invariance_lens}
\ee

\medskip
\noindent\textbf{Result: transformed TDSE and the \emph{time} Schwarzian.}
Substituting \eqref{app:lens_transform}--\eqref{app:phase_S} into \eqref{app:TDSE} yields again a standard TDSE,
\be
\ii\hbar\,\partial_\tau\Phi(\tau,y)
=
\left(
-\frac{\hbar^2}{2m}\,\partial_y^2
+
\widetilde V(\tau,y)
\right)\Phi(\tau,y),
\label{app:TDSE_transformed}
\ee
with transformed potential
\be
\widetilde V(\tau,y)
=
\rho(\tau)\,V\!\left(t(\tau),\sqrt{\rho(\tau)}\,y\right)
+
\frac{m}{4}\,\Schw{t}{\tau}\,y^2
\,. 
\label{app:V_tilde_general}
\ee
Here $\Schw{t}{\tau}$ is the Schwarzian derivative of $t=t(\tau)$,
\be
\Schw{t}{\tau}
=
\frac{t'''(\tau)}{t'(\tau)}-\frac{3}{2}\left(\frac{t''(\tau)}{t'(\tau)}\right)^2
=
\frac{\rho''(\tau)}{\rho(\tau)}-\frac{3}{2}\left(\frac{\rho'(\tau)}{\rho(\tau)}\right)^2.
\label{app:Schwarz_time}
\ee
This is the crucial structural point:
\emph{in the lens/Cayley-Niederer transform the Schwarzian term is the Schwarzian of the time reparametrization},
not of the spatial change (which is linear in $y$ and has zero spatial Schwarzian).
Using \eqref{app:Schwarz_inverse}, one may also express this in terms of the inverse map $\tau=\tau(t)$:
\(
\Schw{t}{\tau}=-\rho(\tau)^2\,\Schw{\tau}{t}\Big|_{t=t(\tau)}.
\)

\medskip
\noindent\textbf{Pulled-back TDSE and the origin of the chirp and half-density factors.}
To make explicit what the TDSE becomes \emph{before} the chirp/half-density rescaling in
\eqref{app:lens_transform}, define the pulled-back wave function
\[
\widetilde{\Psi}(\tau,y):=\Psi\bigl(t(\tau),x(\tau,y)\bigr),
\qquad x(\tau,y):=\sqrt{\rho(\tau)}\,y,
\qquad \rho(\tau)=\frac{dt}{d\tau}>0.
\]
A direct chain-rule computation gives
\(
\partial_t\Psi=\frac{1}{\rho}\left(\partial_\tau-\frac{\rho'}{2\rho}\,y\,\partial_y\right)\widetilde{\Psi},\) 
\(\partial_x=\frac{1}{\sqrt{\rho}}\,\partial_y,\)
\(\partial_x^2=\frac{1}{\rho}\,\partial_y^2 .\)
Substituting into the general TDSE \eqref{app:TDSE} (with arbitrary $V(t,x)$) yields,
equivalently,
\begin{equation*}
\frac{1}{\rho(\tau)}\left(
i\hbar\partial_\tau
+\frac{\hbar^2}{2m}\partial_y^2
-i\hbar\,\frac{\rho'}{2\rho}\,y\,\partial_y
-\rho(\tau)\,V\bigl(t(\tau),\sqrt{\rho(\tau)}\,y\bigr)
\right)\widetilde{\Psi}(\tau,y)=0,
\end{equation*}
or, after completing the square,
\begin{equation*}
\left(
i\hbar\partial_\tau
+\frac{\hbar^2}{2m}\left(\partial_y-\frac{i}{\hbar}\,\frac{m}{2}\,\frac{\rho'}{\rho}\,y\right)^2
+\frac{i\hbar}{4}\,\frac{\rho'}{\rho}
+\frac{m}{8}\left(\frac{\rho'}{\rho}\right)^2 y^2
-\rho\,V\bigl(t(\tau),\sqrt{\rho}\,y\bigr)
\right)\widetilde{\Psi}=0.
\end{equation*}
The unitary chirp factor $U_S(\tau)=\exp\!\bigl(\frac{i}{\hbar}S(\tau,y)\bigr)$ with $S$ as in \eqref{app:phase_S}
removes the elongated derivative by conjugation,
\(
U_S^{-1}\left(\partial_y-\frac{i}{\hbar}\,\frac{m}{2}\,\frac{\rho'}{\rho}\,y\right)U_S=\partial_y,\)
\(U_S^{-1}(i\hbar\partial_\tau)U_S=i\hbar\partial_\tau-\partial_\tau S,
\)
so that the induced $-\partial_\tau S$ term combines with the remaining quadratic contribution to produce precisely
the time-Schwarzian potential term in \eqref{app:V_tilde_general}, \eqref{app:Schwarz_time}. 
Finally, the half-density factor $\rho^{-1/4}$ in \eqref{app:lens_transform} cancels the residual pure-imaginary term,
since
\(
\rho^{1/4}(i\hbar\partial_\tau)\rho^{-1/4}
=
i\hbar\partial_\tau - \frac{i\hbar}{4}\,\frac{\rho'}{\rho}.
\)
This gives a concrete mechanism behind the required normalization and the associated Maslov/Fresnel phase bookkeeping when solutions are continued globally across caustics.

\medskip
\noindent\textbf{Cayley--Niederer map (free $\leftrightarrow$ oscillator).}
The classical Cayley--Niederer choice
\be
t=\tan\tau,\qquad \rho(\tau)=\frac{dt}{d\tau}=\sec^2\tau,\qquad x=\sec\tau\,y,
\label{app:Niederer_map}
\ee
cf. (\ref{eq:HO_tau_tan}),
gives
\(
\Schw{t}{\tau}=2,
\)
so for $V\equiv0$ one obtains a harmonic oscillator term
$\widetilde V=(m/2)\,y^2$ in \eqref{app:TDSE_transformed} (frequency $\omega=1$ in $\tau$-time).

\medskip
\noindent\textbf{Action principle viewpoint and unitarity of the lens transform.}
The TDSE \eqref{app:TDSE} follows from the action functional
\be
\mathcal{S}[\Psi,\Psi^*]
=
\int \mathrm{d}t\int \mathrm{d}x\,
\left[
\frac{\ii\hbar}{2}\left(\Psi^*\partial_t\Psi-\Psi\,\partial_t\Psi^*\right)
-\frac{\hbar^2}{2m}\,|\partial_x\Psi|^2
- V(t,x)\,|\Psi|^2
\right].
\label{app:TDSE_action}
\ee
Under the transformation \eqref{app:lens_transform} with $t=t(\tau)$ and $x=\sqrt{\rho}\,y$,
one has $\mathrm{d}t\,\mathrm{d}x=\rho(\tau)\,\mathrm{d}\tau\,\sqrt{\rho(\tau)}\,\mathrm{d}y$,
while the prefactor $\rho^{-1/4}$ and the phase \eqref{app:phase_S} are precisely chosen so that:
(i) the kinetic term retains the flat form $-(\hbar^2/2m)\partial_y^2$,
(ii) the term $\propto y\partial_y$ cancels,
and (iii) the potential transforms as in \eqref{app:V_tilde_general}.
As a result, $\mathcal{S}[\Psi,\Psi^*]$ is mapped to the same functional of $(\Phi,\Phi^*)$ (with $\widetilde V$),
up to a total $\tau$-derivative coming from the quadratic phase (a boundary term in the action).
Consequently, the map \eqref{app:lens_transform} is implemented by a (time-dependent) unitary operator
between Hilbert spaces, which is also reflected in the exact norm preservation \eqref{app:norm_invariance_lens}.

\medskip
\noindent\textbf{No contradiction between $(-\tfrac12)$ and $(+\tfrac12)$ weights.}
Let us stress here that the different ``weights'' arise from different \emph{goals}:

\begin{itemize}
\item For the stationary ODE \eqref{app:SSE_Liouville}, the goal is to preserve the \emph{Liouville normal form}
(no first derivative). This selects the inverse half-density law \eqref{app:Liouville_rescale},
equivalently \eqref{app:inverse_half_density}, and produces the projective-connection law
\eqref{app:Q_transform_SSE_xy}--\eqref{app:Q_transform_SSE_yx} with the \emph{spatial} Schwarzian.

\item For the TDSE as a Hilbert-space evolution, the natural requirement is \emph{unitarity}
(preservation of $\int |\Psi|^2 \mathrm{d}x$). This yields the half-density law
\eqref{app:unitary_half_density_map}.
To preserve the standard flat kinetic term in a nontrivial way (free $\leftrightarrow$ oscillator),
one must also reparametrize time and scale space as in \eqref{app:time_reparam}--\eqref{app:scaling_space},
leading to the \emph{time} Schwarzian term $\Schw{t}{\tau}$ in \eqref{app:V_tilde_general}.
\end{itemize}

%%%%%%%%%%%%%%%%%%%%%%%%%%%%%%%%%%%%%%%%%%%%%%%%%%%%%%%%%%%%%%%%%%%%%%

\section{Extended free particle and the Schwarzian-induced oscillator}
\label{sec:free_to_osc_extended}

\noindent
In this Section we relate the free particle and the harmonic oscillator
(at both the classical and quantum levels) in the \emph{extended} formulation,
where time is promoted to a dynamical variable and the dynamics is encoded by a
first--class constraint.
A general time reparametrization, combined with the accompanying space rescaling,
induces a canonical transformation on the extended phase space that maps the free
constraint to an oscillator--type constraint with a \emph{time--dependent} frequency.
The latter is governed by the \emph{time Schwarzian} of the reparametrization,
$\Omega^2(\tau)=\tfrac12 \omega^2\{F;\eta\}$, so that the oscillator potential is
``Schwarzian--induced''.

We then quantize this canonical map in the metaplectic/Fourier--integral framework.
In particular, we show how the most economical choice of generating function
implements the point transformation by elementary Fourier integrals (producing
delta distributions), and how the quantum free constraint is intertwined into the
time--dependent oscillator Schr\"odinger equation with potential
$\widetilde V(\tau,y)=\tfrac{m}{2}\Omega^2(\tau)y^2$.
Finally, we single out the constant--frequency case (the Cayley--Niederer map) as the
projective--time choice $t(\tau)=\omega^{-1}\tan(\omega\tau)$ and relate the free
and oscillator propagators; in this way the Mehler kernel and its Maslov/branch
structure emerge directly from the free kernel.

\medskip
\noindent\textbf{Classical canonical equivalence in the extended formulation.}
Let us return to the classical reparametrization-invariant formulation of the free particle described by action 
(\ref{freeacext}),
\be
S_{\text{free}}[x(s),t(s)]
=
\int ds\, \frac{m}{2}\,\frac{\dot x^{\,2}}{\dot t}\,,
\qquad
\dot{}\equiv \frac{d}{ds}\,,
\label{eq:free_action_repar}
\ee
in which time parameter $t$ is promoted to be the dynamical variable. 
As it was discussed in Section \ref{Free particle section}, Hamiltonian 
 formulation on the extended phase space
$(x,t,p_x,p_t)$ is characterized by the first-class constraint
\be
\mathcal C_{\text{free}}
=
p_t+\frac{p_x^{\,2}}{2m}\approx 0\,.
\label{eq:Phi_free_constraint}
\ee
Let us  perform the same time reparametrization and space rescaling as in
\eqref{app:time_reparam}--\eqref{app:scaling_space} : 
\be
t=\frac{1}{\omega}F(\omega\tau)\,,
\qquad
\rho(\tau):=\frac{dt}{d\tau}=F'(\eta)>0\,,
\qquad x=\sqrt{\rho(\tau)}\,y\,,
\label{eq:config_map_section7}
\ee
where $\omega>0$ is a constant of dimension of frequency, and so, $\eta=\omega\tau$ and $\omega t:=\lambda$ are dimensionless.
Action \eqref{eq:free_action_repar} takes then the form
\be
{S}_{\Omega}[y,\tau] 
=
\int ds\, \left(\frac{1}{2}m\,\frac{\dot y^{\,2}}{\dot \tau}
-\frac{1}{2}m\,\Omega^2(\tau)y^2
+\frac{d}{ds}S(\tau,y)\right)
\,,
\label{eq:free_actiontilde}
\ee
where 
\be\label{OmwgaFreq}
\Omega^2(\tau)=\frac{1}{2}\omega^2\Schw{F}{\eta}\,, \qquad \text{and}\qquad
S(\tau,y)=\frac{m\omega}{4}\frac{F''(\eta)}{F'(\eta)}\,y^2\,.
\ee
The first two terms in this reparametrization-invariant action 
correspond to Lagrangian of harmonic oscillator with time-dependent frequency.
Free particle constraint \eqref{eq:Phi_free_constraint} transforms now into
\be\label{PhiOmega}
\Phi_{\Omega}=P_\tau+H_{\Omega}\approx 0\,,\qquad
H_{\Omega}=\frac{1}{2m}P_y^2+\frac{1}{2}m\,\Omega^2(\tau)y^2\,.
\ee
Relations 
\begin{align}
&p_x
=\frac{1}{\sqrt{F'(\eta)}}\left(P_y+\frac{m\omega}{2}\mathcal P_F(\eta)\,y\right)\,,&
\label{eq:pq_inverse}\\[6pt]
&p_t
=\frac{1}{F'(\eta)}\left(
P_\tau-\frac{\omega}{2}\mathcal P_F(\eta)\,y P_y
+\frac{m\omega^2}{4}\left(\{F;\eta\}-\frac12 \left(\mathcal P_F(\eta)\right)^2\right) y^2
\right)\,&
\label{eq:pt_inverse}
\end{align}
provide a canonical lift of the point transformation \eqref{eq:config_map_section7},
where $\mathcal P_F(\eta)=\frac{F''(\eta)}{F'(\eta)}$ is pre-Schwarzian \eqref{app:preSchwarz_def}.
The inverse  relations that present $P_y$ and $P_\tau$ in terms of $(t,p_t,x,p_x)$ can be  obtained easily from  \eqref{eq:pq_inverse} and \eqref{eq:pt_inverse} via the exchange
\be\label{dualityT}
(t,p_t,x,p_x)\longleftrightarrow (\tau,P_\tau,y,P_y)\,,\qquad
F\longleftrightarrow R=F^{-1}\,,\qquad
\eta \longleftrightarrow \lambda\,,
\ee
and using identities 
\[
F'(\eta)=\frac{1}{R'(\lambda)}\,,\qquad \mathcal P_F(\eta)=-\frac{1}{R'(\lambda)}
\mathcal P_R(\lambda)\,,\qquad 
\Schw{F}{\eta}=-\frac{1}{(R'(\lambda))^2} \Schw{R}{\lambda}\,,
\]
with  $F'$ and $R'$ denoting the  derivatives in $\eta$ and $\lambda$, respectively,
and $\mathcal P_R(\lambda)=\frac{R''(\lambda)}{R'(\lambda)}$.

The relation between corresponding Liouville $1$-forms 
is 
\be\label{Liouville1form}
p_t dt+p_x dx=P_\tau d\tau +P_y dy + dS,
\ee
where  $S=S(\tau,y)$ 
is exactly the function from \eqref{OmwgaFreq}, which appears under the sign of total derivative  in
the transformed action \eqref{eq:free_actiontilde} as well as in quantum transformation
\eqref{app:lens_transform}.

A generating function  of the described canonical transformation 
\eqref{eq:config_map_section7}, \eqref{eq:pq_inverse}, \eqref{eq:pt_inverse}
is
\begin{align}
F_{22}(t,P_\tau;\,x,P_y)
=
\frac{1}{\omega}R(\lambda)\,P_\tau+\sqrt{R'(\lambda)}\,x\,P_y-\frac{m\omega}{4}\,
\mathcal P_R(\lambda)
\,x^2\,,
\label{eq:F22}
\end{align}
from which 
$p_t=\frac{\partial F_{22}}{\partial t},$ 
$\tau=\frac{\partial F_{22}}{\partial P_\tau},$ 
$p_x=\frac{\partial F_{22}}{\partial x},$ 
$y=\frac{\partial F_{22}}{\partial P_y}.$
Three other possible forms of generating functions can be obtained from
\eqref{eq:F22} and \eqref{Liouville1form} by applying a corresponding Legendre transformation.
Particularly, 
\begin{eqnarray}\label{eq:F23}
F_{23}(t,P_\tau;p_x,y)&=&F_{22}-xp_x-yP_y\nonumber\\
&=&
\frac{1}{\omega}R(\lambda)\,P_\tau-\frac{y}{\sqrt{R'(\lambda)}}\,p_x-
\frac{m\omega}{4}\,\frac{R''(\lambda)}{(R'(\lambda))^2}\,y^2\,.
\end{eqnarray}
In analogous way, one obtains $F_{32}(p_t,\tau;x,P_y)$ and $F_{33}(p_t,\tau;p_x,y)$,
whose explicit form can be written directly from \eqref{eq:F22} and \eqref{eq:F23}
by using the transformation law 
\[
F_{22}\longleftrightarrow -F_{33}\,,\qquad F_{23}\longleftrightarrow -F_{32}\,,
\]
under exchange \eqref{dualityT}. Note that the last term in \eqref {eq:F22}
and \eqref {eq:F23}, and its analogs in $F_{32}$ and $F_{33}$, is the function
$S$ from \eqref {OmwgaFreq}
expressed in corresponding phase space variables.

Under canonical transformation \eqref {eq:config_map_section7}, 
\eqref{eq:pq_inverse}, \eqref{eq:pt_inverse}, constraint 
\eqref{eq:Phi_free_constraint} takes the form  $\frac{1}{F'(\eta)}\Phi_\Omega\approx 0$,
that is equivalent to the first class constraint \eqref{PhiOmega}.
\vskip0.1cm
So, at the classical level, the extended formulation of the free particle described by reparametrization-invariant  action \eqref{eq:free_action_repar} is\emph{ canonically equivalent} to analogous  extended formulation \eqref{eq:free_actiontilde} of the harmonic oscillator  with time-dependent frequency in general case of transformation \eqref{eq:config_map_section7}. 
%%%%%%%%%%%%%%%%%%%%%%%%%%%%%%%%%%%%%%%%%%%%%%%%%%%%%%%%%%%%%%%%%%%%%%%%%%%%%%%%%%%%%%%%%%%%%%%%%%%%%%%%%%%%%
%%%%%%%%%%%%%%%%%%%%%%%%%%%%%%%%%%%%

\medskip
\paragraph{Quantum implementation and distributional collapse.}
Each of the four classical generating functions $F_{ij}$ gives rise, upon
quantization, to a Fourier--integral (metaplectic) operator with phase $F_{ij}$.
We fix the normalization so that each integration over a conjugate pair carries
the unitary factor $(2\pi\hbar)^{-1/2}$; in particular, a double integral carries
$(2\pi\hbar)^{-1}$. For the choice $F_{22}(t,P_\tau;x,P_y)$ given in \eqref{eq:F22}
we define $\mathcal U_{22}$ by
\be\label{eq:U22_PP_def}
(\mathcal U_{22}\Psi)(P_\tau,P_y)
=
\int_{\R}\!dt\int_{\R}\!dx\,
\frac{1}{2\pi\hbar}\,
\exp\!\left(\frac{i}{\hbar}F_{22}(t,P_\tau;x,P_y)\right)\Psi(t,x).
\ee
The unitary Fourier transform in a conjugate pair is taken as
\be\label{eq:unitary_FT_pair}
(\mathcal F_{x\to p_x}\phi)(p)
=
\frac{1}{\sqrt{2\pi\hbar}}
\int_{\R}e^{-\frac{i}{\hbar}p_xx}\phi(x)\,dx,
\qquad
\mathcal F^{-1}_{x\to p_x}=\mathcal F_{p_x\to x},
\ee
and similarly for the pairs $(t,p_t)$, $(\tau,P_\tau)$ and $(y,P_y)$.

\smallskip
\noindent
Passing to the $(\tau,y)$--representation by inverse Fourier transforms in $P_\tau$
and $P_y$, we write
\be\label{eq:U22_taurep}
(\widetilde{\mathcal U}_{22}\Psi)(\tau,y)
:=
\big(\mathcal F^{-1}_{P_\tau\to\tau}\mathcal F^{-1}_{P_y\to y}\,\mathcal U_{22}\Psi\big)(\tau,y)
=
\int_{\R}\!dt\int_{\R}\!dx\,
K_{22}(\tau,y;t,x)\,\Psi(t,x),
\ee
with kernel
\be\label{eq:K22_def}
K_{22}(\tau,y;t,x)
=
\int_{\R}\!dP_\tau\int_{\R}\!dP_y\,
\frac{1}{(2\pi\hbar)^2}\,
\exp\!\left[
\frac{i}{\hbar}\Big(F_{22}(t,P_\tau;x,P_y)-P_\tau\tau-P_y y\Big)
\right].
\ee
Here the phase is fixed by the classical generating function $F_{22}$,
while the overall amplitude (up to a constant normalization and a possible constant
Maslov phase) is fixed by the metaplectic requirement.
Since $F_{22}$ is linear in $P_\tau$ and $P_y$, the $P_\tau$-- and $P_y$--integrations
in \eqref{eq:K22_def} are elementary and collapse to delta distributions,
\be\label{eq:K22_delta}
K_{22}(\tau,y;t,x)
=
\delta\!\left(\tau-\frac{1}{\omega}R(\omega t)\right)\,
\delta\!\left(y-\sqrt{R'(\omega t)}\,x\right)\,
\exp\!\left[-\frac{i}{\hbar}\frac{m\omega}{4}\,\mathcal P_R(\omega t)\,x^{2}\right].
\ee
Thus the kernel is localized on the classical point transformation,
\[
\tau=\frac{1}{\omega}R(\omega t)\ \Longleftrightarrow\ t=t(\tau),
\qquad
y=\sqrt{R'(\omega t)}\,x=\frac{x}{\sqrt{\rho(\tau)}}\ \Longleftrightarrow\ x=\sqrt{\rho(\tau)}\,y,
\]
and the remaining quadratic term in $F_{22}$ produces the phase factor
$\exp\!\big(-\tfrac{i}{\hbar}S(\tau,y)\big)$ after substituting $t=t(\tau)$ and
$x=\sqrt{\rho(\tau)}\,y$ (using the identities relating $\mathcal{P}_{R}$ to
$\rho'/\rho$ and the definition of $S(\tau,y)$ given earlier).

To pass from this distributional collapse to a \emph{unitary} map between
configuration--space wave functions, one must still fix the overall amplitude
(metaplectic half--density).  Writing the most general pointwise form compatible
with the canonical map as
\[
\Phi(\tau,y)=A(\tau)\exp\!\left(-\frac{i}{\hbar}S(\tau,y)\right)\Psi(t(\tau),x),
\qquad x=\sqrt{\rho(\tau)}\,y,
\]
the metaplectic (unitarity) requirement on spatial slices,
\eqref{app:norm_invariance_lens},
together with $dx=\sqrt{\rho(\tau)}\,dy$, yields $|A(\tau)|^{2}=\sqrt{\rho(\tau)}$.
Hence $A(\tau)=\rho(\tau)^{1/4}$ (up to an irrelevant constant phase), and one
obtains the pointwise transformation
\begin{equation}
\Phi(\tau,y)=\rho(\tau)^{1/4}\exp\!\left(-\frac{i}{\hbar}S(\tau,y)\right)\Psi(t(\tau),x),
\qquad x=\sqrt{\rho(\tau)}\,y,
\end{equation}
which is equivalent to \eqref{app:lens_transform}.

The remaining generating functions $F_{23}$, $F_{32}$ and $F_{33}$ describe the
same classical canonical map, but their quantum operators are obtained from the
$F_{22}$ realization only after additional unitary Fourier transforms in the
corresponding conjugate pairs, and are therefore less economical for recovering
the configuration--space point transformation.

%%%%%%%%%%%%%%%%%%%%%%%%%%%%%%%%%%%%%%%%%%%%%%%%%%%%%%%%%%%%
%%%%%%%%%%%%%%%%%%%%%%%%%%%%%%%%%%%%%%%%%%

The quantum analog of the classical first--class constraint is obtained by the substitutions
$p_t\mapsto -i\hbar\,\partial_t$ and $p_x\mapsto -i\hbar\,\partial_x$.  For the free particle,
\be\label{eq:free_constraint_TDSE_selfcontained_v2}
\widehat C_{\rm free}\Psi=0,
\qquad
\widehat C_{\rm free}:=\hat p_t+\frac{\hat p_x^{\,2}}{2m}
\;\;\Longleftrightarrow\;\;
i\hbar\,\partial_t\Psi(t,x)=-\frac{\hbar^2}{2m}\,\partial_x^2\Psi(t,x)\, .
\ee
Under the unitary map \eqref{app:lens_transform} this is transformed into the constrained
Schr\"odinger equation associated with the oscillator--type Hamiltonian
$H_\Omega=\frac{1}{2m}P_y^2+\frac{m}{2}\Omega^2(\tau)y^2$,
\[
\widehat\Phi_\Omega\Phi=0,
\,\,
\widehat\Phi_\Omega:=\hat P_\tau+\frac{\hat P_y^{\,2}}{2m}+\frac{m}{2}\Omega^2(\tau)\,y^2
\;\Longleftrightarrow\;
i\hbar\,\partial_\tau\Phi(\tau,y)=
\left[-\frac{\hbar^2}{2m}\,\partial_y^2+\widetilde V(\tau,y)\right]\Phi(\tau,y),
\]
with the time--dependent harmonic--oscillator potential
\[
\widetilde V(\tau,y)=\frac{m}{2}\,\Omega^2(\tau)\,y^2,
\qquad
\Omega^2(\tau)=\frac12\,\omega^2\,\{F;\eta\},\qquad \eta=\omega\tau,
\]
cf.\ \eqref{OmwgaFreq}.  Thus the unitary operator induced by $F_{22}$ intertwines the
quantum constraints (equivalently, the corresponding TDSEs) in complete agreement with the
classical canonical equivalence established above. The boundary term in \eqref {eq:free_actiontilde}
 is the classical
precursor of the quantum ``chirp'' factor in \eqref{app:lens_transform}:
it is precisely the total derivative responsible for the metaplectic
phase (Maslov/Fresnel branch) bookkeeping when one crosses caustics.

%%%%%%%%%%%%%%%%%%%%%%%%%%%%%%%%%%%%%%%%%%%%%%%%%%%%%%%%%%%%%%%%%

\medskip
\noindent\textbf{Constant frequency.}
If we choose the projective time parametrization 
\be
t(\tau)=\frac{1}{\omega}\tan(\omega\tau)\,,
\qquad
-\frac{\pi}{2\omega}<\tau<\frac{\pi}{2\omega}\,,
\label{eq:Niederer_choice_section}
\ee
cf. Cayley map \eqref{Caypar}, 
then
\be
F(\eta)=\tan\eta\,, 
\qquad \rho(\tau)=\frac{dt}{d\tau}=\frac{1}{\cos^2(\omega\tau)}\,,
\qquad
\Schw{t}{\tau}=2\omega^2\,,
\label{eq:Niederer_rho_Schwarz}
\ee
so that $\Omega(\tau)=\omega$, and we obtain
the ordinary
harmonic oscillator. In this case \eqref{app:lens_transform} becomes 
the standard Niederer unitary transformation,
\be
\Psi\big(t(\tau),x\big)
=
(\cos\omega\tau)^{1/2}\,
\exp\!\Big(\frac{i m\omega}{2\hbar}\tan(\omega\tau)\,y^2\Big)\,
\Phi(\tau,y)\,,
\qquad
x=\frac{y}{\cos\omega\tau}\,.
\label{eq:Niederer_unitary_operator}
\ee

 The choice 
\be\label{FtanhIHO}
F(\eta)=\tanh\eta\,,\qquad \Schw{F}{\eta}=-2\,,
\ee
results in the inverted harmonic oscillator, which can be obtained from the case
\eqref{eq:Niederer_choice_section} by the obvious change $\omega\rightarrow i\omega$.

\medskip
\noindent\textbf{Factorised unitary operator in the Cayley case.}
In the extended (reparametrization-invariant) formulation, where kinematical wavefunctions depend on $(t,x)$ and the inner product is
$\int_{\R}\!dt\int_{\R}\!dx\,|\Psi(t,x)|^2$,
the Cayley--Niederer choice
$t(\tau)=\omega^{-1}\tan(\omega\tau)$, $x=\sec(\omega\tau)\,y$
admits an \emph{exact} unitary implementer on $L^2(\R_t\times\R_x)$ in a closed product form.
Writing $\eta:=\omega\tau$, one may express the corresponding unitary as
\begin{equation}
\mathcal U_{\tan}
=
\mathcal U^{(y)}_{\tan}(\tau)\,\bigl(\mathcal U^{(t)}_{\tan}(\tau)\otimes \mathrm{Id}\bigr),
\label{eq:Utan_factorised}
\end{equation}
where the \emph{time-sector} unitary can be written as an exponential in $(t,\hat p_t)$,
\[
\mathcal U^{(t)}_{\tan}(\tau)
=
\exp\!\left(-\frac{i}{\hbar}\,\eta\,\hat G_{\tan}\right),
\quad \,\,
\hat G_{\tan}
=
\frac{1}{2\omega}\Bigl((1+\omega^2 t^2)\hat p_t+\hat p_t(1+\omega^2 t^2)\Bigr),
\quad \,\,
\hat p_t=-i\hbar\,\partial_t,
\]
and implements the reparametrization together with the half-density factor on $L^2(\R,dt)$,
\begin{equation}
(\mathcal U^{(t)}_{\tan}\varphi)(\tau)
=
\sqrt{\frac{dt}{d\tau}}\;\varphi\bigl(t(\tau)\bigr)
=
\sec(\eta)\,\varphi\!\left(\frac{1}{\omega}\tan\eta\right).
\label{eq:Utan_time_action}
\end{equation}
The \emph{spatial} factor is the product of a quadratic ``chirp'' and a metaplectic dilation,
\begin{equation}
\mathcal U^{(y)}_{\tan}(\tau)
=
\exp\!\left(-\frac{i m\omega}{2\hbar}\tan\eta\,\hat y^{\,2}\right)
\exp\!\left(-\frac{i}{2\hbar}\ln(\sec\eta)\,(\hat y\hat p_y+\hat p_y\hat y)\right),
\quad
\hat p_y=-i\hbar\,\partial_y .
\label{eq:Utan_space_exponential}
\end{equation}
\paragraph{Remark (sign of the chirp).}
Our metaplectic convention is fixed by the projective implementation
$U(S)\,W(\xi)\,U(S)^{-1}=W(S\xi)$ (App.~\ref{app:metaplectic}), hence equivalently
$U(S)\,\hat\xi\,U(S)^{-1}=S\hat\xi$ with $\hat\xi=(\hat q,\hat p)^{T}$.
In particular, a lower shear $S=\bigl(\begin{smallmatrix}1&0\\ \gamma&1\end{smallmatrix}\bigr)$
is implemented by
$U(S)=\exp\!\left(-\frac{i}{2}\gamma\,\hat q^{2}\right)$, cf. \eqref{eq:metaplectic_b0}.
Accordingly, the quadratic chirp in \eqref{eq:Utan_space_exponential} appears with a minus sign.
When one writes instead the inverse pointwise relation for wavefunctions
(as in \eqref{app:lens_transform}), the corresponding quadratic phase appears with the opposite sign.

This closure on a finite set of quadratic generators (chirp and dilation, together with the finite-dimensional
projective time flow) is special to the constant-Schwarzian (projective) reparametrisations;
 for a generic reparametrization $t=\omega^{-1}F(\omega\tau)$ one typically loses such a closed factorization (time ordering enters), and recovering an operator product form requires additional structure (to be discussed in Sec.~\ref{SecGenF}).
\medskip

%%%%%%%%%%%%%%%%%%%%%%%%%%%%%%%%%%%%%%%%%%%%%%%%%%%%%%%%%%%%%%%%%%%%%%%%%%%%%%%%%%%%%%%%%%%%%%%%%%%%%

\medskip
\noindent\textbf{Propagators: free kernel, Maslov phases, and the Mehler kernel.}
The 
relation between propagators
of the free particle and harmonic oscillator (with time-dependent frequency in general case)  is obtained by inserting
\eqref{app:lens_transform} into the free evolution kernel:
\be
K_{\text{osc}}(y_2,\tau_2;y_1,\tau_1)
=
\rho_2^{1/4}\rho_1^{1/4}\,
\exp\!\Big(-\frac{i}{\hbar}S_2\Big)\,
K_{\text{free}}(x_2,t_2;x_1,t_1)\,
\exp\!\Big(\frac{i}{\hbar}S_1\Big)\,,
\label{eq:kernel_relation}
\ee
where
\(
t_j=t(\tau_j)\,,
\)
\(
\rho_j=\rho(\tau_j)\,,\)
\(
x_j=\sqrt{\rho_j}\,y_j\,,
\)
\(
S_j=S(\tau_j,y_j)\,.\)
For the free particle,
\be
K_{\text{free}}(x_2,t_2;x_1,t_1)
=
\Big(\frac{m}{2\pi i\hbar\,(t_2-t_1)}\Big)^{1/2}
\exp\!\Big[
\frac{i m}{2\hbar}\frac{(x_2-x_1)^2}{t_2-t_1}
\Big]\,,
\label{eq:free_kernel_section7}
\ee
where the square root is understood in the metaplectic sense (a choice of
branch of the Fresnel factor, see App. \ref {app:metaplectic}). In the oscillator variables this choice
becomes the usual Maslov rule: at a caustic (where the Van Vleck
determinant changes sign), the kernel picks up the universal phase jump
$\exp(-i\pi/2)$.

Combining \eqref{eq:kernel_relation} with \eqref{eq:free_kernel_section7}
and \eqref{eq:Niederer_choice_section} yields the Mehler kernel,
\be\label{Mehler}
K_{\mathrm{osc}}(y_2,\tau_2;y_1,\tau_1)
=\left(\frac{m\omega}{2\pi i\hbar\,\sin(\omega T)}\right)^{\!1/2}
\exp\!\left[
\frac{i m\omega}{2\hbar\,\sin(\omega T)}
\Big((y_2^2+y_1^2)\cos(\omega T)-2y_2y_1\Big)
\right]\,.
\ee
The square--root prefactor is understood in the metaplectic sense, and the caustics
$\sin(\omega T)=0$, $T:=\tau_2-\tau_1$,  produce the standard Maslov phase jumps.
The inverted oscillator follows by $\omega\rightarrow i\omega$.

\medskip
\noindent\textbf{Mehler kernel from the ABCD shear factorization.}
In conclusion of this Section we note that an alternative (and rather direct) way to recover the oscillator propagator uses the
$\Sp(2,\mathbb{R})$ shear factorization of the rotation matrix and its metaplectic lift
(App.~\ref{app:sp2r_abcd}, cf.\ \eqref{app:eq:rot}).  Writing
\(
T:=\tau_2-\tau_1
\)
and using that the classical oscillator flow over time~$T$ is the phase--space rotation by
angle~$\omega T$, one obtains the operator identity
\begin{equation}
\label{eq:Uho_shear_section7}
e^{-\frac{i}{\hbar}T\hat H_+}
=
\exp\!\Bigl(-\frac{i}{\hbar}\,\tan\!\frac{\omega T}{2}\,\frac{m\omega}{2}\,\hat y^{\,2}\Bigr)\,
\exp\!\Bigl(-\frac{i}{\hbar}\,\Delta t\,\hat H_0\Bigr)\,
\exp\!\Bigl(-\frac{i}{\hbar}\,\tan\!\frac{\omega T}{2}\,\frac{m\omega}{2}\,\hat y^{\,2}\Bigr)\,,
\end{equation}
where 
$\Delta t:=\frac{\sin(\omega T)}{\omega}$, and  $\hat H_0=\hat p_y^{\,2}/(2m)$ is the free Hamiltonian (in the $y$--coordinate).
Taking the $(y_2,y_1)$ matrix element and inserting a resolution of the identity between the
three factors gives
\begin{align}
K_{\mathrm{osc}}(y_2,\tau_2;y_1,\tau_1)
&=\big\langle y_2\big|e^{-\frac{i}{\hbar}T\hat H_+}\big|y_1\big\rangle
\nonumber\\
&=
\exp\!\Bigl(-\frac{i m\omega}{2\hbar}\tan\!\frac{\omega T}{2}\,y_2^2\Bigr)\,
K_{\text{free}}(y_2,\Delta t; y_1,0)\,
\exp\!\Bigl(-\frac{i m\omega}{2\hbar}\tan\!\frac{\omega T}{2}\,y_1^2\Bigr).
\nonumber
\end{align}
Using the explicit free kernel \eqref{eq:free_kernel_section7} with $x_j=y_j$ and
$t_2-t_1=\Delta t$, and the elementary identity
\[
\frac{(y_2-y_1)^2}{\Delta t}-\omega\tan\!\frac{\omega T}{2}\,(y_2^2+y_1^2)
=
\frac{\omega}{\sin(\omega T)}\Bigl((y_2^2+y_1^2)\cos(\omega T)-2y_2y_1\Bigr),
\]
one recovers exactly the Mehler kernel \eqref{Mehler} (with the same metaplectic
branch choice and the same Maslov phase jumps at $\sin(\omega T)=0$).

%%%%%%%%%%%%%%%%%%%%%%%%%%%%%%%%%%%%%%%%%%%%%%%%%%%%%%%%%%%%%%%%%%%%%%%%%%%%%%%%%

\section{General reparametrisations \texorpdfstring{$F$}{F} and a factorised metaplectic form: the Ermakov--Pinney amplitude}\label{SecGenF}

In the special constant-Schwarzian cases $F(\eta)=\eta$, $F(\eta)=\tan\eta$, and $F(\eta)=\tanh\eta$
(so $\{F;\eta\}=0,2,-2$, respectively), the time reparametrization is fractional-linear (projective),
i.e.\ it lies in $\mathrm{PSL}(2,\mathbb{R})$ acting on $\mathbb{RP}^{1}$; equivalently, on any interval
$I$ where it is smooth and strictly monotone it may be viewed as an element of $\mathrm{Diff}_{+}(I)$.
In these projective cases the quantum implementation closes on a finite set of quadratic generators,
and one can write an exact closed product of exponentials for the corresponding unitary operator
(including the time sector). In this section we turn to a general strictly monotone reparametrization
$t=\omega^{-1}F(\omega\tau)$ with $F'(\eta)>0$ and explain what survives of this picture in the extended
(reparametrization-invariant) formulation considered in the previous Section.
As will be seen, one can still write an explicit factorization into a time-reparametrization unitary,
a metaplectic dilation, and a quadratic phase (chirp), with the remaining dynamical content encoded in
Ermakov--Pinney data. This provides a convenient framework for the generic case; for complementary
remarks on the exact versus semiclassical implementation of canonical maps, see App. \ref{app:exact_vs_semiclassical_quantisation}.

\medskip
\noindent\textbf{Ermakov--Pinney amplitude and a closed product formula.}
As shown in the previous Sections, a strictly monotone reparametrization $t=t(\tau)$ induces, for the
free particle, a time-dependent oscillator with frequency
$\Omega(\tau)^2=\tfrac12\,\omega^2\,\{F;\eta\}$, where $\eta=\omega\tau$.
Here $\rho(\tau)=dt/d\tau>0$ is the Jacobian of the time change; in the lens/Cayley--Niederer map
the associated spatial scale factor is $b(\tau)=\sqrt{\rho(\tau)}$ (so $x=b(\tau)\,y$).
This should not be confused with the Ermakov--Pinney amplitude $r(\tau)>0$ defined below.

Introduce the standard $\mathfrak{sl}(2,\mathbb{R})$ quadratic generators on $L^{2}(\mathbb{R},dy)$,
\begin{equation}
\hat H:=\frac{1}{2m}\hat p_y^{\,2},\qquad
\hat K:=\frac{m}{2}\hat y^{\,2},\qquad
\hat D:=\frac{1}{2}\left(\hat y\hat p_y+\hat p_y\hat y\right),
\qquad
\hat p_y=-\,\mathrm{i}\hbar\,\partial_y,
\end{equation}
and the time-dependent oscillator Hamiltonian
\begin{equation}
\hat H_{\Omega}(\tau):=\hat H+\Omega(\tau)^2\hat K
=\frac{\hat p_y^{\,2}}{2m}+\frac{m}{2}\Omega(\tau)^2\hat y^{\,2}.
\end{equation}
The exact quantum evolution is the time-ordered exponential
\[
U_{\Omega}(\tau,\tau_{0})
=\mathcal{T}\exp\!\left(-\frac{\mathrm{i}}{\hbar}\int_{\tau_{0}}^{\tau}\hat H_{\Omega}(\sigma)\,d\sigma\right).
\]
Let $r(\tau)>0$ solve the Ermakov--Pinney equation \cite{ErmPin}
\begin{equation}
r''(\tau)+\Omega(\tau)^2r(\tau)=\frac{\omega_{0}^{2}}{r(\tau)^{3}},\qquad \omega_{0}>0\ \text{constant},
\end{equation}
and define the reparametrized time
\begin{equation}
\theta(\tau):=\int_{\tau_{0}}^{\tau}\frac{\omega_{0}}{r(\sigma)^{2}}\,d\sigma.
\end{equation}
Then one has the exact metaplectic factorization (up to an overall phase fixed by metaplectic
normalization),
\[
U_{\Omega}(\tau,\tau_{0})
=
\exp\!\Big(\frac{\mathrm{i}}{\hbar}\frac{r'(\tau)}{r(\tau)}\,\hat K\Big)\,
\exp\!\Big(-\frac{\mathrm{i}}{\hbar}\ln r(\tau)\,\hat D\Big)\,
\exp\!\Big(-\frac{\mathrm{i}}{\hbar}\theta(\tau)\,\hat H_{0}\Big),
\,\,
\hat H_{0}=\frac{1}{2m}\hat p_y^{\,2}+\frac{m}{2}\omega_{0}^{2}\hat y^{\,2}.
\]
Thus, once $r(\tau)$ is known, the time-dependent frequency problem is reduced to a constant-frequency
evolution in the new time $\theta$, together with an explicit chirp and a dilation. In the Cayley case
$F(\eta)=\tan\eta$ one may choose $\omega_{0}=\omega$ and the constant solution $r(\tau)=1$, so that
$\theta(\tau)=\omega(\tau-\tau_{0})$ and the above reduces to a product with constant coefficients.

\medskip
\noindent\textbf{Including the time sector in the extended formulation.}
In the extended (reparametrization-invariant) setting, the canonical map includes the time change
$t\mapsto\tau$. On $L^{2}(\mathbb{R},dt)$, the corresponding unitary is the pullback with the Jacobian:
\begin{equation}
\big(U_{F}^{(t)}\varphi\big)(\tau)=\sqrt{\frac{dt}{d\tau}}\ \varphi\big(t(\tau)\big)
=\sqrt{\rho(\tau)}\ \varphi\!\left(\frac{1}{\omega}F(\omega\tau)\right),
\qquad \rho(\tau):=\frac{dt}{d\tau}>0.
\end{equation}
Equivalently, if one represents the reparametrization as a (possibly non-autonomous) flow on the $t$-line,
its unitary action on half-densities is generated by the standard symmetrized operator built from
$\hat p_t=-\,\mathrm{i}\hbar\,\partial_t$,
\begin{equation}
\hat G_{\xi(\tau)}=\frac{1}{2}\big(\xi(\tau,t)\hat p_t+\hat p_t\,\xi(\tau,t)\big),
\qquad
\hat p_t=-\,\mathrm{i}\hbar\,\partial_t,
\end{equation}
so that the same time-sector unitary may be written (along the chosen flow) as the time-ordered exponential
\begin{equation}
U_{F}^{(t)}(\tau,\tau_{0})
=\mathcal{T}\exp\!\left(-\frac{\mathrm{i}}{\hbar}\int_{\tau_{0}}^{\tau}\hat G_{\xi(\sigma)}\,d\sigma\right).
\end{equation}
In the projective (constant-Schwarzian) cases the relevant time flow is generated within the
finite-dimensional span of $1,t,t^{2}$, and the corresponding exponential can be written without
ordering complications; for a generic $F$ one understands the exponential in the usual time-ordered sense.

Combining the time-sector unitary with the spatial factors (the slice identification from the previous
Section and/or the dynamical factorization $U_{\Omega}$ above, depending on the context) yields a compact
factorized operatorial description of the quantum implementer of the extended canonical map.

%%%%%%%%%%%%%%%%%%%%%%%%%%%%%%%%%%%%%%%%%%%%%%%%%%%%%%%%%%%%%%%%%%%%%%%%%%%%%%%%%%%%%%%%%%%%%%%%%%%%%%%%%%%%%%%%%%

\section{Summary, Discussion and Outlook}\label{Conclusion}
%%%%%%%%%%%%%%%%%%%%%%%%%%%%%%%%%%%%%%%%%%%%%%%%%%%%%%%%%%%%%%%%%%%%%%

\noindent
In this work we revisited the relation between the one--dimensional free particle (FP) and
the harmonic oscillator (HO) from a unified viewpoint that combines projective geometry,
Cayley transformations (in both phase space and time), and the Schwarzian derivative.
Although FP and HO represent the opposite intuitive extremes of \emph{freedom} and
\emph{confinement}, they share a common symmetry backbone: the semidirect product of the
Heisenberg group with the conformal group $SL(2,\R)$ (or, equivalently, $Sp(2,\R)$).
While the FP--HO correspondence has a long history (see e.g.\ \cite{Niederer1972,Niederer1973,Takagi}),
our aim was to make explicit a single organizing mechanism behind its various incarnations.
Our main point is that the \emph{projective} nature of the $SL(2,\R)$ action on time, together
with its metaplectic lift, provides an organizing principle that simultaneously explains the
FP--HO correspondence at the level of the TDSE (Cayley--Niederer/lens transforms) and at
the level of the SSE (conformal bridge transformation, CBT), and clarifies the role of the Schwarzian
as the universal cocycle inducing oscillator--type terms under time reparametrisations.

\medskip

The main results can be summarized as follows.

\begin{itemize}
\item
In Sec.~\ref{CTCCT} we emphasized the dual role of the (rotated) Cayley matrix: it acts
(i) as a linear canonical transformation between real and complexified symplectic bases in
phase space, and (ii) as a M\"obius transformation on the projectivized time line $\RP^1$,
linking naturally the upper half--plane and unit disc models of hyperbolic geometry.
This provides a convenient geometric language for complex structures compatible with
$\omega=dp\wedge dq$ (see Appendix~\ref{app:kahler_uplift}) and for the appearance of the
Cayley parameter as a global coordinate on $\RP^1$.

\item
In Secs.~\ref{Free particle section} and \ref{SymmetriesHO} we reviewed how the same
$\mathfrak{sl}(2,\R)$ skeleton is realized as a \emph{dynamical} symmetry of FP, and as the
natural symmetry algebra underlying HO evolution.  In particular, the global implementation
of conformal boosts and time inversion is most naturally formulated after projectivizing time,
$t\in\R \mapsto \RP^1=\R\cup\{\infty\}$, and using the corresponding Cayley parametrization.

\item
In Sec.~\ref{SchrodingerProjective} we separated two closely related, but conceptually distinct,
covariance principles: the projective--connection transformation law for the SSE potential
(the Liouville picture), and the half--density (unitary) transformation law for the TDSE.
This separation explains why the Schwarzian derivative appears in both settings, but with
different physical roles: as the projective cocycle controlling the SSE potential, and as the
universal term induced by \emph{time} reparametrisations in the TDSE.

\item
In Sec.~\ref{sec:free_to_osc_extended} we formulated the FP--HO relation in the \emph{extended},
reparametrization--invariant setting, where time is promoted to a dynamical variable and the
dynamics is encoded by a first--class constraint.  In this formulation, a time reparametrization
combined with the accompanying space scaling becomes a canonical transformation on the
extended phase space that maps the free constraint to an oscillator--type constraint with a
\emph{Schwarzian--induced} (generally time--dependent) frequency.  The standard
Cayley--Niederer (lens) transform arises as a distinguished choice of projective time map.

\item
In Sec.~\ref{QuantCayley} we identified the quantum analogue of the complexified Cayley map
with the Bargmann transform (metaplectic implementation of the same symplectic data).

\item
In Sec.~\ref{sec:QCayley_CBT} we showed how the same Cayley logic yields the CBT relating FP and HO at the level of the stationary problem.  In particular,
the CBT appears naturally as a (non--unitary) similarity transform generated by the inverted
oscillator evolution at a purely complex time, making manifest the algebraic unity between
the TDSE and SSE correspondences.

\item
Finally, Sec.~\ref{SecGenF} exhibited a factorized/metaplectic form for general time maps,
controlled by Ermakov--Pinney data.  This clarifies how the projective reparametrization
geometry governs not only the basic FP$\leftrightarrow$HO map but also its generalisations to
time--dependent oscillator profiles.
\end{itemize}

\medskip

A recurring theme is that the conformal sector $SL(2,\R)$ acts on time by M\"obius
transformations, hence the natural arena is $\RP^1$ rather than $\R$.
For the HO, this projective viewpoint dovetails with periodic evolution: the $SL(2,\R)$
rotation generator effectively sees the half--period $\pi/\omega$ as the fundamental cycle.
In the extended FP formulation, the projective time map
\(
t=\omega^{-1}\tan(\omega\tau)
\)
implements precisely this global completion of conformal boosts, while the requirement of
maintaining the standard kinetic form (and hence unitarity in the TDSE) forces the accompanying
space scaling and produces the quadratic term governed by the time Schwarzian
$\Schw{t}{\tau}$.
The Schwarzian thus plays the role of the canonical projective cocycle measuring the
departure from M\"obius reparametrisations (Appendix~\ref{Appendix Schwarzian}; see also the
original sources \cite{Cayley1880,Schwarz1869}).

\medskip

From the phase--space viewpoint, the same $Sp(2,\R)\cong SL(2,\R)$ can be generated by
elementary canonical transformations (shears and ``thin--lens'' chirps).  Their Lie algebra
generators span $\mathfrak{sl}(2,\R)$ and interpolate continuously between free evolution,
oscillator evolution (compact rotation), and inverted--oscillator evolution (hyperbolic rotation).
This algebraic fact underlies the various manifestations of the FP--HO correspondence:
in classical Hamiltonian flows, in the TDSE lens transform, and in the quantum metaplectic
operators that implement the same symplectic data.

\medskip

The framework developed here suggests several natural extensions.

\begin{itemize}
\item
\emph{Inverted oscillator, resonances, and analyticity.}
Since HO and inverted HO are related by $\omega\mapsto \ii\omega$, the present constructions
have a direct analytic continuation to the inverted case.  In particular, it would be interesting
to sharpen the relation between FP Jordan states (Appendix~\ref{app:Jordan-analytic}) and
resonant (Gamow) states of the inverted oscillator, and to clarify how the corresponding ladder
structures are transported by the Cayley/CBT operators.

\item
\emph{Linear potential and mixed generators.}
Another natural testbed is the FP versus a particle in a linear potential, where the Hamiltonian
mixes quadratic and linear generators of the Schr\"odinger/Jacobi symmetry.
Understanding its projective--geometric organization may illuminate new ``bridge'' maps beyond
the purely quadratic realm (cf.\ the Schr\"odinger-group perspective in \cite{BacryLevyLeblond1968,Hagen1972,Jackiw1972,DuvalHorvathy2009}).

\item
\emph{Integrable hierarchies and projective connections.}
The projective--connection viewpoint on the SSE operator is closely related to integrable
isospectral deformations: KdV and its hierarchy can be viewed as flows on Schr\"odinger
operators/potentials preserving spectral data (Lax/Gelfand--Dikii picture).
It would be valuable to explore systematically how the Schwarzian/projective cocycle language
used here interfaces with KdV--type evolutions, Miura maps, and dressing/Darboux--Crum constructions
(see e.g.\ \cite{GGKM1967,Lax1968,FadTakh,MatveevSalle1991}).

\item
\emph{AdS$_2$, SYK, and hyperbolic geometry.}
Because our central mechanism is the projective action of $SL(2,\R)$ and the Schwarzian
cocycle, it is natural to connect it more directly with the low--energy dynamics of SYK and
Jackiw--Teitelboim gravity, governed by ${\rm Diff}(S^1)/SL(2,\R)$ and the Schwarzian action.
In particular, the hyperbolic--geometric interpretation of the Cayley map suggests revisiting
Schwarzian boundary dynamics from the perspective of motion on (or moduli of) hyperbolic
structures (see e.g.\ \cite{SachdevYe1993,Kitaev2015,MaldacenaStanford2016,KitaevSuh2018,SaadShenkerStanford2019,StanfordWitten2019}).
\end{itemize}

%\medskip
%%%%%%%%%%%%%%%%%%%%%
\paragraph{Acknowledgments.}
The work was partially supported by the FONDECYT Project 1242046.

%%%%%%%%%%%%%%%%%%%%%%%%%%%%%%%%%%%%%%%%%%%%%%%%%%%%%%%%%%%%%%%%%%%%%%%%%%%%%%%%%%%%%%%%%%%%

%%%%%%%%%%%%%%%%%%%%%%%%%%%%%%%%%%%%%%%%%%%%%%%%%%%%%%%%%%%%%%%%%%%%%%%%%%%%%%%%%%%%%%%%%%%%

\appendix
\section{Compatible complex structures on \texorpdfstring{$(\R^2,\omega)$}{(R2, omega)} and the upper half-plane}
\label{app:kahler_uplift}

\noindent\textbf{Symplectic conventions.}
We consider the real symplectic plane $V\cong\R^2$ with coordinates
\(\xi_\alpha=(q,p)\), $\alpha=1,2$, endowed with the symplectic form
$\omega = dp\wedge dq.$
In matrix notation,
\( 
\omega(u,v)=u^{\mathsf T}\,\breve{\Omega}\,v,
\)
\(
\breve{\Omega}:=-J,\) 
\(
J:=\big(\begin{smallmatrix}
0 & 1 \\
-1 & 0\end{smallmatrix} \big)\),
so that the Poisson brackets are $\{\xi_\alpha,\xi_\beta\}=J_{\alpha\beta}$.
The linear symplectic group is
$Sp(2,\R)=\{S\in GL(2,\R)\mid S^{\mathsf T} \breve{\Omega}S= \breve{\Omega}\}$, and in dimension $2$
this coincides with $SL(2,\R)$.

\medskip
\noindent\textbf{
Compatible complex structures and $\H_+$.}
A (linear) complex structure on $V$ is an endomorphism $\mathcal{J}:V\to V$ with
$\mathcal{J}^2=-\mathbf{1}$. It is called \emph{compatible} with $\omega$ if
\begin{equation}
\mathcal{J}^{\mathsf T} \breve{\Omega}\mathcal{J}= \breve{\Omega}
\qquad\mbox{and}\qquad
g_{\mathcal{J}}(u,v):=\omega(u,\mathcal{J}v)\ \ \mbox{is positive definite.}
\end{equation}
In real dimension $2$ the space of compatible complex structures is naturally
parametrized by the upper half-plane
\(
\H_+:=\{\tau=x+\ii y\in\C\mid y>0\}.
\)
A convenient explicit representative is
\begin{equation}
\mathcal{J}_\tau
=\frac{1}{\Im\tau}
\left(\begin{array}{cc}
\Re\tau & |\tau|^2\\[2pt]
-1 & -\Re\tau
\end{array}\right)
=\frac{1}{y}
\left(\begin{array}{cc}
 x & x^2+y^2\\[2pt]
-1 & -x
\end{array}\right), 
\label{eq:Jtau_compact_v2}
\end{equation}
One checks directly that $\mathcal{J}_\tau^2=-\mathbf{1}$ and
$\mathcal{J}_\tau^{\mathsf T} \breve{\Omega}\mathcal{J}_\tau=
\breve{\Omega}$.
Moreover, the associated K\"ahler metric is
\begin{equation}
 g_\tau(u,u)=\omega(u,\mathcal{J}_\tau u)=\frac{1}{\Im \tau}\,|q+\tau p|^2
 =\frac{(q+xp)^2+y^2p^2}{y}>0\qquad (u=(q,p)\neq 0),
\label{eq:gtau_compact_v2}
\end{equation}
which is positive definite precisely for $\Im\tau>0$.
Consequently,
\begin{equation}\label{A.6}
\{\mbox{compatible }\mathcal{J}\}\cong Sp(2,\R)/U(1)\cong SL(2,\R)/SO(2)\cong \H_+.
\end{equation}
The induced invariant metric on this 
parameter space (often loosely called a moduli space)
 is the Poincar\'e metric (\ref{H+metic}), 
$ds^2_{\H_+}=(dx^2+dy^2)/y^2$.
Indeed, denote this space by $\mathcal{M}$; by \eqref{A.6} it is the Riemannian symmetric space
$Sp(2,\mathbb{R})/U(1)$, hence it carries a canonical $Sp(2,\mathbb{R})$--invariant
Riemannian metric, unique up to an overall constant factor.
 One convenient way to write it is via the trace pairing on endomorphisms: for a smooth curve
$s\mapsto \mathcal{J}(s)$ in $\mathcal{M}$ (so $\mathcal{J}(s)^2=-\mathbf{1}$), its tangent vector
$\dot{\mathcal{J}}=\frac{d}{ds}\mathcal{J}(s)$ satisfies $\dot{\mathcal{J}}\mathcal{J}+\mathcal{J}\dot{\mathcal{J}}=0$, and one sets
\(
\langle \dot{\mathcal{J}}_1,\dot{\mathcal{J}}_2\rangle_{\mathcal{J}}
:=\frac{1}{2}\,\mathrm{tr}\,\!\big(\dot{\mathcal{J}}_1\,\dot{\mathcal{J}}_2\big),
\)
\(
ds^2=\frac{1}{2}\,\mathrm{tr}\,\!\big(d\mathcal{J}\,d\mathcal{J}\big).
\)
This expression is invariant under the natural $Sp(2,\mathbb{R})$ action $\mathcal{J}\mapsto S\mathcal{J}S^{-1}$ because
$\mathrm{tr}\,\!\big((SXS^{-1})(SYS^{-1})\big)=\mathrm{tr}\,(XY)$.
Using the explicit parametrization $\tau=x+iy\in\H_+$ and \eqref{eq:Jtau_compact_v2}, one checks directly that
\[
\frac{1}{2}\,\mathrm{tr}\,\!\big(d\mathcal{J}_\tau\,d\mathcal{J}_\tau\big)
=\frac{dx^2+dy^2}{y^2},
\]
i.e. the Poincar\'e metric on the upper half-plane.

A useful viewpoint is the homogeneous-space identification $\H_+\cong PSL(2,\R)/SO(2)$ (cf.~\eqref{A.6}).
The induced action is the standard M\"obius transformation
\(
\tau\mapsto\tau'=\frac{a\tau+b}{c\tau+d},\)
\(\left(\begin{smallmatrix}a&b\\ c&d\end{smallmatrix}\right)\in SL(2,\R),
\)
whose stabiliser at $\tau=i$ is $SO(2)$.

\medskip
\noindent\textbf{Holomorphic polarization and complex coordinate.}
The complex structure $\mathcal{J}_\tau$ determines a holomorphic polarization.
A convenient $(1,0)$ one-form is
\(
\theta_\tau := dq+\tau\,dp,
\)
and the associated complex coordinate
\begin{equation}
 z_\tau:=\frac{q+\tau p}{\sqrt{2\,\Im\tau}},
 \qquad
 \bar z_\tau:=\frac{q+\bar\tau\,p}{\sqrt{2\,\Im\tau}}.
\end{equation}
With our convention $\omega=dp\wedge dq$, one verifies
\(
\omega = \ii\,dz_\tau\wedge d\bar z_\tau,
\) 
\( g_\tau = 2\,dz_\tau\,d\bar z_\tau.
\)
Thus the same upper half-plane label $\tau\in\H_+$ that appears in hyperbolic geometry
also parametrises complex polarisations in geometric quantization 
\cite{GeomQuant1}--\cite{GeomQuant3}.

\medskip
\noindent\textbf{$SL(2,\R)$ action and the Cayley map.}
For
$S=\left(\begin{smallmatrix}
a&b\\ c&d
\end{smallmatrix}
\right)\in SL(2,\R)$ acting linearly as
$\xi\mapsto\xi'=S\xi$, the induced action on $\tau\in\H_+$ is the fractional
linear transformation
\(
\tau\longmapsto \tau' = \frac{a\tau+b}{c\tau+d},
\) 
\( \mathcal{J}_{\tau'} = S\,\mathcal{J}_\tau\,S^{-1}.\) 
The Cayley map
\(
 w = i\, \frac{\tau-\ii}{\tau+\ii}
\)
identifies $\H_+$ with the unit disk
$\D:=\{w\in\C\mid |w|<1\}$ and conjugates $PSL(2,\R)$ to $PSU(1,1)$.
In the quantum theory, this same Cayley change of polarization underlies the
passage from the real (Schr\"odinger) representation to the holomorphic
(Bargmann--Fock) one; in the holomorphic basis the induced linear action on
creation/annihilation operators takes the Bogoliubov ($SU(1,1)$) form.

%%%%%%%%%%%%%%%%%%%%%%%%%%%%%%%%%%%%%%%%%%%%%%%%%%%%%%%%%%%%%%%%%%%%%%%%%%%%%%%%%%%%%%%

%========================================================
\section{Metaplectic representation in quantum mechanics}
\label{app:metaplectic}
%========================================================

\noindent\textbf{Heisenberg group, central extension, and the Weyl system.}
We work with the symplectic plane $(\R^2,\omega)$, $\omega = dp\wedge dq$, and
$\hat q\,\psi(q)=q\psi(q)$, $\hat p\,\psi(q)=-\ii\,\partial_q\psi(q)$ so that
$[\hat q,\hat p]=\ii$.

A convenient ``exponentiated'' form of the Schr\"odinger representation is the
Weyl system of unitary operators
\begin{equation}
  W(\xi)\;:=\;\exp\!\bigl(\ii(p_0\hat q-q_0\hat p)\bigr),
  \qquad \xi=(q_0,p_0)\in\R^2 .
  \label{eq:Weyl_def}
\end{equation}
The map $\xi\mapsto W(\xi)$ is \emph{projective} as a representation of the
abelian group $\R^2$ (phase-space translations). Indeed, using the
Baker--Campbell--Hausdorff formula and the commutator
$[p_0\hat q-q_0\hat p,\;p_1\hat q-q_1\hat p]=-\ii\,\omega(\xi_0,\xi_1)$ with
\(
  \omega(\xi_0,\xi_1)=p_0q_1-q_0p_1\,,
\)
one finds the Weyl relations
\begin{equation}
  W(\xi_0)\,W(\xi_1)
  \;=\;
  e^{\frac{\ii}{2}\omega(\xi_0,\xi_1)}\,W(\xi_0+\xi_1).
  \label{eq:Weyl_relations}
\end{equation}
The phase factor in \eqref{eq:Weyl_relations} is a (nontrivial) $2$-cocycle on
$\R^2$; it obstructs $\xi\mapsto W(\xi)$ from being a true representation of
$\R^2$.

This obstruction is removed by passing to the \emph{Heisenberg group}
$\mathrm{Heis}_3$, the central extension of $\R^2$ by $\R$:
\begin{equation}
  \mathrm{Heis}_3 \cong \R^2\times \R,
  \qquad
  (\xi,s)\cdot(\eta,t)
  =
  \Bigl(\xi+\eta,\;s+t+\tfrac12\omega(\xi,\eta)\Bigr).
  \label{eq:Heis_group_law}
\end{equation}
Then
\begin{equation}
  U(\xi,s)\;:=\;e^{\ii s}\,W(\xi)
  \label{eq:Heis_rep}
\end{equation}
is a genuine unitary representation of $\mathrm{Heis}_3$.
(Equivalently, \eqref{eq:Heis_group_law} is precisely engineered so that
\eqref{eq:Weyl_relations} becomes associativity of $U(\xi,s)$.)  The
Stone--von Neumann theorem states that, once the action of the center is fixed,
this irreducible unitary representation is unique up to unitary equivalence.

\medskip
\noindent\textbf{Symplectic automorphisms, projective implementation, and $\mathrm{Mp}(2,\R)$.}
The group $Sp(2,\R)\cong SL(2,\R)$ consists of real $2\times 2$ matrices
\(
  S=\left(\begin{smallmatrix} a & b \\ c & d \end{smallmatrix}\right)\),
 \( ad-bc=1,\)
preserving $\omega$, i.e. $\omega(S\xi,S\eta)=\omega(\xi,\eta)$.
Therefore $S$ acts by automorphisms of $\mathrm{Heis}_3$:
\begin{equation}
  \alpha_S:\ (\xi,s)\mapsto(S\xi,s),
  \qquad
  W(\xi)\mapsto W(S\xi).
  \label{eq:symp_auto_Heis}
\end{equation}
By Stone--von Neumann, for each $S\in Sp(2,\R)$ there exists a unitary operator
$\mathcal{U}(S)$ on $L^2(\R,dq)$ (unique up to an overall phase) such that
\begin{equation}
  \mathcal{U}(S)\,W(\xi)\,\mathcal{U}(S)^{-1}=W(S\xi),
  \qquad \forall\,\xi\in\R^2 .
  \label{eq:metaplectic_intertwining}
\end{equation}
Because $\mathcal{U}(S)$ is only determined up to phase, the map
$S\mapsto\mathcal{U}(S)$ is generally only a \emph{projective} representation:
\begin{equation}
  \mathcal{U}(S_1)\,\mathcal{U}(S_2)
  \;=\;
  e^{\ii\,\sigma(S_1,S_2)}\,\mathcal{U}(S_1S_2),
  \label{eq:Sp2_cocycle}
\end{equation}
where $\sigma$ is a $2$-cocycle (the ``multiplier'').

The \emph{metaplectic group} $\mathrm{Mp}(2,\R)$ is the double cover of $\Sp(2,\R)$
that resolves this cocycle. Concretely, there exists a canonical projection
$\pi:\mathrm{Mp}(2,\R)\to \Sp(2,\R)$ with $\ker\pi=\{\pm\mathbf 1\}$, and one can choose a
continuous lift $S\mapsto \widetilde S$ together with operators
$\widetilde S\mapsto \mathcal U(\widetilde S)$ giving a \emph{true} (not merely
projective) representation,
\begin{equation}
  \mathcal{U}(\widetilde{S}_1)\,\mathcal{U}(\widetilde{S}_2)
  =\mathcal{U}(\widetilde{S}_1\widetilde{S}_2),
  \qquad
  \pi(\widetilde{S})=S\in \Sp(2,\R).
  \label{eq:Mp_true_rep}
\end{equation}
For a fixed $S\in \Sp(2,\R)$ there are exactly two lifts $\widetilde S$ and
$-\widetilde S$ in the fiber $\pi^{-1}(S)$, so $\mathcal U(\widetilde S)$ is
defined only up to an overall sign. Imposing continuity along paths in
$ \Sp(2,\R)$ makes this sign jump in a controlled way when one crosses caustics
(see below); equivalently, the relevant square-root branch changes by a phase
quantized in multiples of $\pi/2$, encoded by the Maslov (Fresnel) index.

\medskip
\noindent\textbf{Generating function and the metaplectic kernel for $b\neq 0$.}
Consider the linear canonical transformation
\(
  (Q,P)^T = S\,(q,p)^T,
  \)
  \(
  Q=a q + b p,\) \(P=c q + d p,
  \)
with $S=\left(\begin{smallmatrix}
a&b\\ c&d
\end{smallmatrix}
\right)\in SL(2,\R)$.  For $b\neq 0$ one can solve
$p=(Q-aq)/b$ and $P=(dQ-q)/b$, and a (type-1) generating function
$G_S(q,Q)$ satisfying
\(
  p=\frac{\partial G_S}{\partial q},
  \)
  \(
  P=-\frac{\partial G_S}{\partial Q},
  \)
is
\begin{equation}
  G_S(q,Q)
  =
  \frac{1}{2b}\Bigl(2qQ-aq^2-dQ^2\Bigr),
  \qquad (b\neq 0).
  \label{eq:genfun_GS}
\end{equation}
(Equivalently, $p\,dq-P\,dQ=dG_S$.)

Treating the phase $\exp(-\ii G_S)$ as an integral kernel leads to the standard
metaplectic operator (one of the two lifts; the other differs by an overall
minus sign):
\begin{equation}
  (\mathcal{U}(S)\psi)(Q)
  =
  \frac{1}{\sqrt{2\pi\,|b|}}\,
  e^{-\ii\frac{\pi}{4}\,\mathrm{sgn}(b)}
  \int_{\R}
  \exp\!\left[
    \frac{\ii}{2b}\Bigl(aq^2-2qQ+dQ^2\Bigr)
  \right]\psi(q)\,dq ,
  \qquad (b\neq 0).
  \label{eq:metaplectic_kernel_bneq0}
\end{equation}
Here the \emph{square-root factor} $|b|^{-1/2}$ and the Fresnel phase
$e^{-\ii\pi\,\mathrm{sgn}(b)/4}$ arise from enforcing unitarity and the correct
composition law under Gaussian (Fresnel) integrals. The same phenomenon is
already visible in the elementary Fresnel integral:
\[
  \int_{\R} \exp\!\left(\frac{\ii}{2}\,a\,x^2\right)\,dx
  =
  \sqrt{\frac{2\pi}{|a|}}\,
  \exp\!\left(\ii\,\frac{\pi}{4}\,\mathrm{sgn}(a)\right),
  \qquad a\in\R\setminus\{0\}.
\]
It exhibits the square-root/branch effect: the factor $a^{-1/2}$ is not
single-valued globally, and its continuation produces phase changes by
multiples of $\pi/4$. Setting $a=1/b$ shows that the prefactor in
\eqref{eq:metaplectic_kernel_bneq0} is precisely the corresponding normalization,
i.e.\ a branch choice for $(\ii b)^{-1/2}$.
They are also the simplest manifestation of the Maslov-type ambiguity: the
choice of branch for $\sqrt{b}$ (or $\sqrt{|b|}$ together with the Fresnel phase)
cannot be made globally on $Sp(2,\R)$ without passing to the double cover
$\mathrm{Mp}(2,\R)$.

\medskip
\noindent\textbf{Half-density viewpoint.}
The Jacobian-type factor in \eqref{eq:metaplectic_kernel_bneq0} is the
one-dimensional version of the statement that, under changes of polarization,
wavefunctions naturally transform as \emph{half-densities}.  In the present
linear setting this shows up precisely through the square-root of the relevant
determinant (here $|b|$; see also \eqref{eq:metaplectic_b0} and the
$\det B$ factor in the $ Sp(2n,\R)$ remark below).

\medskip
\noindent\textbf{The case $b=0$ and caustics.}
When $b=0$, the matrix has the form
\(
  S=
  \left(\begin{smallmatrix}
    a & 0\\
    c & a^{-1}
  \end{smallmatrix}\right),\)
  \( a\neq 0\),
and \eqref{eq:metaplectic_kernel_bneq0} is not applicable (this is a ``caustic''
for the $b$-chart).  In this case one uses the factorization into a shear and a
dilation,
\begin{equation}
  \begin{pmatrix}
    a & 0\\
    c & a^{-1}
  \end{pmatrix}
  =
  \begin{pmatrix}
    1 & 0\\
    c/a & 1
  \end{pmatrix}
  \begin{pmatrix}
    a & 0\\
    0 & a^{-1}
  \end{pmatrix},
  \label{eq:b0_factorization}
\end{equation}
whose metaplectic lifts are elementary:
\begin{equation}
  \mathcal{U}\!\left(\begin{pmatrix}1&0\\ \gamma&1\end{pmatrix}\right)
  =\exp\!\left(-\ii\frac{\gamma}{2}\hat q^{\,2}\right),
  \qquad
  \bigl(\mathcal{U}(\mathrm{diag}(a,a^{-1}))\psi\bigr)(Q)
  =\sqrt{|a|}\,\psi(aQ).
  \label{eq:shear_dilation_lifts}
\end{equation}
Combining them gives, up to the global sign $\pm$,
\begin{equation}
  (\mathcal{U}(S)\psi)(Q)
  =
  \sqrt{|a|}\,
  \exp\!\left(-\ii\,\frac{c}{2a}\,Q^2\right)\,
  \psi(aQ),
  \qquad (b=0).
  \label{eq:metaplectic_b0}
\end{equation}

Crossing from $b\neq 0$ to $b=0$ (or, more generally, changing charts in
$ \Sp(2,\R)$) forces a choice of branch for the square root factor and produces
a quantized phase jump.  Tracking these jumps along a closed loop yields an
integer Maslov index; the associated phase $e^{-\ii\pi\mu/2}$ is exactly the
Fresnel/Maslov correction needed for a globally consistent metaplectic lift.

\medskip
\noindent\textbf{Metaplectic group as a double cover via square--root automorphy factors.}
A recurring theme in the main text and in App.~\ref{app:sp2r_abcd} below  is that linear canonical maps
$S\in\Sp(2,\R)$ have a perfectly honest action on classical phase space, but their quantum
implementation (as unitary operators on wavefunctions) is only \emph{projective}: one meets a
$\pm$ sign (more generally, a phase) in composition.  The conceptual reason is always the same:
unitarity forces a \emph{square root} of a Jacobian (equivalently, an action on half--densities),
and square roots carry an unavoidable sign/branch ambiguity.  The metaplectic group
$\mathrm{Mp}(2,\R)$ is the canonical device that resolves this ambiguity by passing to a double cover.

\medskip
\noindent
\textbf{(i) Automorphy factors and the double cover.}
Let
\(
S=\left(\begin{smallmatrix}a&b\\ c&d\end{smallmatrix}\right)\in SL(2,\R)\cong \Sp(2,\R),
\)
\( \tau\in \H_+:=\{\tau\in\C\, |\,\Im\tau>0\},\)
\(
S\cdot\tau=\frac{a\tau+b}{c\tau+d}.
\)
The standard automorphy factor is
\[
j(S,\tau):=c\tau+d.
\]
It satisfies the chain rule
\begin{equation}
  j(S_1S_2,\tau)=j(S_1,S_2\cdot\tau)\,j(S_2,\tau),
  \label{eq:j_chain_rule}
\end{equation}
Indeed, writing $S_k=\bigl(\begin{smallmatrix}a_k&b_k\\ c_k&d_k\end{smallmatrix}\bigr)$, one has
\[
  j(S_1,S_2\cdot\tau)\,j(S_2,\tau)
  =
  \left(c_1\frac{a_2\tau+b_2}{c_2\tau+d_2}+d_1\right)(c_2\tau+d_2)
  =
  (c_{12}\tau+d_{12})
  =
  j(S_1S_2,\tau),
\]
where $\bigl(\begin{smallmatrix}a_{12}&b_{12}\\ c_{12}&d_{12}\end{smallmatrix}\bigr)=S_1S_2$.

Many constructions (metaplectic kernels, modular weight $1/2$, etc.) require a choice of
$\sqrt{j(S,\tau)}$.  If one tries to choose $\sqrt{j(S,\tau)}$ globally and multiplicatively in $S$,
one encounters the $\{\pm1\}$ multiplier
\begin{equation}
\label{app:eq:mp_cocycle}
\sigma(S_1,S_2;\tau):=
\frac{\sqrt{j(S_1,S_2\cdot\tau)}\,\sqrt{j(S_2,\tau)}}{\sqrt{j(S_1S_2,\tau)}}\ \in\ \{\pm1\},
\end{equation}
which is a $2$--cocycle (it satisfies the cocycle identity because it is built from the chain rule
for $j$).  The metaplectic double cover can be described as the set of pairs
\[
\mathrm{Mp}(2,\R)=\Bigl\{(S,\phi(\tau)):\ S\in SL(2,\R),\ \phi(\tau)^2=j(S,\tau)\Bigr\},
\]
with multiplication
\begin{equation}
\label{app:eq:mp_law}
(S_1,\phi_1)\,(S_2,\phi_2)=\bigl(S_1S_2,\ \phi_1(S_2\cdot\tau)\,\phi_2(\tau)\bigr).
\end{equation}
Projectively, $\phi$ is precisely the missing square root: once the choice is made in
$\mathrm{Mp}(2,\R)$, compositions become strictly consistent.

\medskip
\noindent
\textbf{(ii) Half--densities and Liouville transformations.}
The same ``square--root mechanism'' underlies the ubiquitous $\pm\frac12$ exponent in
Liouville transformations of second--order equations, as used in
\eqref{app:SSE_Liouville} (stationary Schr\"odinger) and \eqref{app:TDSE} (time--dependent case).
Concretely, under a reparametrization $x=x(y)$ a wavefunction treated as a \emph{half--density}
transforms as
\(
\psi(x)\,|dx|^{1/2}=\widetilde\psi(y)\,|dy|^{1/2},
\)
\(\Longleftrightarrow
\widetilde\psi(y)=\Bigl(\frac{dx}{dy}\Bigr)^{1/2}\psi(x(y)).\)
Exactly this $\bigl(dx/dy\bigr)^{1/2}$ factor removes the first--derivative term that otherwise
appears when rewriting a second--order differential operator in the new variable; in the Schr\"odinger
context it is the familiar Liouville rescaling.  The sign/phase ambiguity of the square root is the
same phenomenon that, for linear canonical maps, manifests itself as the metaplectic sign (Maslov-type
phases along paths).

\medskip
\noindent
\textbf{(iii) Brief remark on $\Sp(2n,\R)$.}
For $n>1$ the same structure persists.
On the open chart where $B$ is invertible for
$S=\bigl(\begin{smallmatrix}A&B\\ C&D\end{smallmatrix}\bigr)$,
the metaplectic operator is an oscillatory integral whose prefactor contains the
half--density factor $|\det B|^{-1/2}$; choosing a continuous branch of $\sqrt{\det B}$
along paths produces the familiar Maslov-type phase and is globally resolved by the
double cover $\mathrm{Mp}(2n,\R)$.
Equivalently, $\Sp(2n,\R)$ acts on the Siegel upper half space $\mathfrak H_n$ by
$\Omega\mapsto (A\Omega+B)(C\Omega+D)^{-1}$ with automorphy factor
$j(S,\Omega)=\det(C\Omega+D)$, and unitarity requires a consistent choice of the square root
of $j(S,\Omega)$, again leading to $\mathrm{Mp}(2n,\R)$.

%%%%%%%%%%%%%%%%%%%%%%%%%%%%%%%%%%%%%%%%%%%%%%%%%%%%%%%

\section{Jordan states from analyticity in the energy}
\label{app:Jordan-analytic}

\noindent\textbf{General construction.}
Consider the one--dimensional stationary Schr\"odinger problem
\begin{equation}
  \hat H\,\psi(x;E)=E\,\psi(x;E),
  \qquad
  \hat H=-\frac{d^2}{dx^2}+V(x),
  \label{app:SchroE}
\end{equation}
and fix an energy value $E_0$.
Assume that for $E$ near $E_0$ there exists a (nontrivial) solution $\psi(x;E)$ analytic in $E$.
Then
\begin{equation}
  \psi(x;E)
  =
  \psi_0(x)
  +
  \sum_{n=1}^{\infty}\frac{(E-E_0)^n}{n!}\,\chi_n(x),
  \quad
  \psi_0(x):=\psi(x;E_0),
  \quad
  \chi_n(x):=\left.\partial_E^n\psi(x;E)\right|_{E=E_0}.
  \label{app:TaylorPsi}
\end{equation}
Substituting \eqref{app:TaylorPsi} into \eqref{app:SchroE} and comparing powers of $(E-E_0)$ gives
\begin{equation}
  (\hat H-E_0)\psi_0=0,
  \qquad
  (\hat H-E_0)\chi_1=\psi_0,
  \qquad
  (\hat H-E_0)\chi_n = n\,\chi_{n-1},
  \quad n\ge2,
  \label{app:JordanChainRec}
\end{equation}
hence
\begin{equation}
  (\hat H-E_0)^{n}\chi_n = n!\,\psi_0\neq0,
  \qquad
  (\hat H-E_0)^{n+1}\chi_n = 0,
  \qquad n=1,2,\dots
  \label{app:JordanNilpotent}
\end{equation}
so $\chi_n$ is a generalized eigenfunction (Jordan state) of rank $n+1$ at $E_0$.
(Jordan chains may also be generated by confluent Darboux/Darboux--Crum transformations
\cite{CJP2015,CarPlyu}, but in this paper we only use the analyticity construction above.)

\medskip
\noindent\textbf{Free particle at $E_0=0$.}
For the free Hamiltonian $\hat H_0=-\frac12\,\frac{d^2}{dq^2}$ the stationary equation is
$\psi''(q)+2E\,\psi(q)=0$.
At $E_0=0$ we may take the basis
\begin{equation}
  \psi^{(+)}_0(q)=1,
  \qquad
  \psi^{(-)}_0(q)=q.
  \label{app:E0basis}
\end{equation}
A convenient choice of analytic families extending these solutions away from $E=0$ is
\begin{equation}
  \psi^{(+)}(q;E)
  :=
  \sum_{n=0}^{\infty}(-1)^n\,\frac{(2E)^n\,q^{2n}}{(2n)!},
  \qquad
  \psi^{(-)}(q;E)
  :=
  \sum_{n=0}^{\infty}(-1)^n\,\frac{(2E)^n\,q^{2n+1}}{(2n+1)!},
  \label{app:FreeAnalyticFamilies}
\end{equation}
(so $\psi^{(+)}=\cos(\sqrt{2E}\,q)$ and $\psi^{(-)}=\sin(\sqrt{2E}\,q)/\sqrt{2E}$ via their $E$--Taylor series).
Define Jordan states by energy derivatives,
$\chi^{(\pm)}_n(q):=\left.\partial_E^n\psi^{(\pm)}(q;E)\right|_{E=0}$.
From \eqref{app:FreeAnalyticFamilies} one finds explicitly
\begin{equation}
  \chi^{(+)}_n(q)
  =
  (-1)^n\,2^n\,n!\,\frac{q^{2n}}{(2n)!},
  \qquad
  \chi^{(-)}_n(q)
  =
  (-1)^n\,2^n\,n!\,\frac{q^{2n+1}}{(2n+1)!}.
  \label{app:FreeJordanExplicit}
\end{equation}

\medskip
\noindent\textbf{Remark.}
The two chains $\{\chi^{(+)}_n\}$ and $\{\chi^{(-)}_n\}$ span the even and odd polynomial sectors at $E=0$.
In the main text we work with the normalized monomials
\begin{equation}
  \chi_n(q):=\frac{q^n}{\sqrt{n!}},
  \label{app:MonomialsMainText}
\end{equation}
which are obtained from \eqref{app:FreeJordanExplicit} by simple rescalings and re-indexing.

%%%%%%%%%%%%%%%%%%%%%%%%%%%%%%%%%%%%%%%%%%%%%%%%%%%%%%%
%%%%%%%%%%%%%%%%%%%%%%%%%%%%%%%%%%%%%%%%%%%%%%%%%%%%%%%%
\section{ABCD optics and \texorpdfstring{$\Sp(2,\R)$}{Sp(2,R)} factorisations: shears, oscillators, and metaplectic lifts}
\label{app:sp2r_abcd}

A linear canonical map $\xi=(q,p)^{\top}\mapsto \xi'=S\,\xi$ with
$S=\left(\begin{smallmatrix}A&B\\ C&D\end{smallmatrix}\right)$ and $AD-BC=1$
lies in $\Sp(2,\R)\cong SL(2,\R)$ and preserves $\omega=dp\wedge dq$.
Equivalently, with
$J=\left(\begin{smallmatrix}0&1\\ -1&0\end{smallmatrix}\right)$
(the matrix $\Omega$ of Sec.~\ref{CTCCT}),
$S^{\top}JS=J$, and the bilinear form
$\Omega(\xi_1,\xi_2):=\xi_1^{\top}J\xi_2=q_1p_2-q_2p_1$
is invariant:
$\Omega(\xi_1',\xi_2')=\Omega(\xi_1,\xi_2)$.

In paraxial optics \cite{GuilSter}  the same $S$ is the \emph{ABCD} matrix.  Writing $q:=x$ and
$p:=\kappa\,\theta$ for transverse position $x$ and slope $\theta$ gives
$(q',p')^{\mathsf T}=S\,(q,p)^{\mathsf T}$ and the Lagrange invariant
$\mathcal I=\kappa(x_1\theta_2-x_2\theta_1)=x_1p_2-x_2p_1$ (independent of the choice of $\kappa$).

\medskip
\noindent
\textbf{Elementary shears (drift and thin lens).}
Introduce
\[
\mathcal T(a):=\begin{pmatrix}1&a\\0&1\end{pmatrix}=\exp\,(a\,\sigma_+),\qquad
\mathcal L(b):=\begin{pmatrix}1&0\\ b&1\end{pmatrix}=\exp\,(b\,\sigma_-),
\qquad a,b\in\R,
\]
so that $\mathcal T(a):(q,p)\mapsto(q+ap,p)$ and $\mathcal L(b):(q,p)\mapsto(q,p+bq)$.
In the conventions of Sec.~\ref{CTCCT},
$M_{H_0}(\tau)=\mathcal T(-\tau)$ for $H_0=\tfrac12p^2$ and
$M_{K}(\tau)=\mathcal L(\tau)$ for $K=\tfrac12q^2$; in optics,
$\mathcal T(L)$ is free propagation and $\mathcal L(-1/f)$ a thin lens.

\medskip
\noindent
\textbf{Generation of $\Sp(2,\R)$ and Gauss factorization.}
Let $\bigl(\begin{smallmatrix}A&B\\C&D\end{smallmatrix}\bigr)\in \Sp(2,\R)$.
If $B\neq0$, then
\begin{equation}
\label{app:eq:gauss}
\begin{pmatrix}A&B\\ C&D\end{pmatrix}
=
\mathcal L\!\left(\frac{D-1}{B}\right)\,\mathcal T(B)\,\mathcal L\!\left(\frac{A-1}{B}\right),
\qquad (B\neq0),
\end{equation}
i.e.\ \emph{lens--propagation--lens}.  If $B=0$, symplecticity forces the matrix to be of
the form $\bigl(\begin{smallmatrix}A&0\\C&A^{-1}\end{smallmatrix}\bigr)$ with $A\neq0$.
Introduce the dilation/squeeze matrix
\begin{equation}
\label{app:eq:Ddef}
\mathcal D(d):=\begin{pmatrix}d^{-1}&0\\0&d\end{pmatrix}=\exp\,(-(\ln d)\,\sigma_3)\,, \qquad(d\neq0),
\end{equation}
so that, for $B=0$,
\begin{equation}
\label{app:eq:B0}
\begin{pmatrix}A&0\\ C&A^{-1}\end{pmatrix}
=
\mathcal D(1/A)\,\mathcal L(AC),
\qquad
\mathcal D(1/A)=\begin{pmatrix}A&0\\0&A^{-1}\end{pmatrix}.
\end{equation}

\medskip
\noindent
\textbf{Key examples.}
\emph{(i) Harmonic oscillator rotation.}
Define
\[
\mathcal R(\theta):=
\begin{pmatrix}\cos\theta&-\sin\theta\\ \sin\theta&\cos\theta\end{pmatrix}.
\]
For $\sin\theta\neq0$ one has the shear decomposition
\begin{equation}
\label{app:eq:rot}
\mathcal R(\theta)=\mathcal L\!\left(\tan\frac{\theta}{2}\right)\,
\mathcal T(-\sin\theta)\,\mathcal L\!\left(\tan\frac{\theta}{2}\right).
\end{equation}

\smallskip
\emph{(ii) Quarter turn, Fourier transform, and inversion.}
The symplectic matrix
\[
J=\begin{pmatrix}0&1\\-1&0\end{pmatrix}=\mathcal R(-\pi/2)\in\Sp(2,\R)
\]
acts as $(q,p)\mapsto(p,-q)$; in the metaplectic lift it is (up to phase) the Fourier transform
and induces $\tau\mapsto-1/\tau$.  Conjugation by $J$ swaps the two shears:
\begin{equation}
\label{app:eq:conj}
J\,\mathcal T(a)\,J^{-1}=\mathcal L(-a),
\qquad\text{equivalently}\qquad
\mathcal L(b)=J\,\mathcal T(-b)\,J^{-1}.
\end{equation}

\smallskip
\emph{(iii) Dilation/squeeze.}
A shear product is
\begin{equation}
\label{app:eq:dil}
\mathcal D(d)
=
\mathcal T(1)\,\mathcal L(d^{-1}-1)\,\mathcal T(-d)\,
\mathcal L\!\left(\frac{d-1}{d^{2}}\right),
\qquad (d\neq0).
\end{equation}

\smallskip
\emph{(iv) Hyperbolic rotation (inverted oscillator).}
Define
\[
\mathcal H(\tau):=
\begin{pmatrix}\cosh\tau&-\sinh\tau\\ -\sinh\tau&\cosh\tau\end{pmatrix}.
\]
For $\tau\neq0$ it factorizes as
\begin{equation}
\label{app:eq:boost}
\mathcal H(\tau)=
\mathcal L\!\left(-\tanh\frac{\tau}{2}\right)\,
\mathcal T(-\sinh\tau)\,
\mathcal L\!\left(-\tanh\frac{\tau}{2}\right).
\end{equation}

\medskip
\noindent
\textbf{Metaplectic lifts and quadratic propagators.}
At the quantum level one uses the metaplectic cover $\Mp(2,\R)$: for each $S\in\Sp(2,\R)$ there are two lifts
$\pm\widehat S$ implementing $\widehat S\,\hat\xi\,\widehat S^{-1}=S\hat\xi$ on
$\hat\xi=(\hat q,\hat p)^{\mathsf T}$ (products are defined up to the Maslov sign).
For the elementary shears one may take the standard quadratic operators
\[
\widehat{\mathcal T}(a):=\exp\!\Big(+\ii a\,\frac{\hat p^{2}}{2}\Big),\qquad
\widehat{\mathcal L}(b):=\exp\!\Big(-\ii b\,\frac{\hat q^{2}}{2}\Big),
\qquad
\widehat{\mathcal D}(d):=\exp\!\big(-\ii (\ln d)\,\hat D\big),
\]
where $\hat D=\tfrac12(\hat q\hat p+\hat p\hat q)$.  These implement the same linear actions on
$(\hat q,\hat p)$ as $\mathcal T(a)$, $\mathcal L(b)$ and $\mathcal D(d)$ on $(q,p)$.
Thus classical factorisations immediately yield quantum ones.  For example, \eqref{app:eq:rot} gives
\begin{equation}
\label{app:eq:Urot}
\widehat{\mathcal R}(\theta)
=
\widehat{\mathcal L}\!\left(\tan\frac{\theta}{2}\right)\,
\widehat{\mathcal T}(-\sin\theta)\,
\widehat{\mathcal L}\!\left(\tan\frac{\theta}{2}\right),
\qquad(\sin\theta\neq0),
\end{equation}
and \eqref{app:eq:boost} yields
\begin{equation}
\label{app:eq:Uboost}
\widehat{\mathcal H}(\tau)
=
\widehat{\mathcal L}\!\left(-\tanh\frac{\tau}{2}\right)\,
\widehat{\mathcal T}(-\sinh\tau)\,
\widehat{\mathcal L}\!\left(-\tanh\frac{\tau}{2}\right),
\qquad(\tau\neq0).
\end{equation}
In the oscillator normalization of Sec.~\ref{CTCCT}, $\widehat{\mathcal R}(\theta)=\exp(-\ii\theta\,\hat H_+)$ with
$\hat H_+=\tfrac12(\hat p^2+\hat q^2)$, while $\widehat{\mathcal H}(\tau)=\exp(-\ii\tau\,\hat H_-)$ with
$\hat H_-=\tfrac12(\hat p^2-\hat q^2)$ (analytic continuation $\theta\mapsto\ii\tau$ relates the two).

\smallskip
\noindent
\emph{Cayley operator.}
The bridge operator $U(\pi/4)=\exp(\frac{\pi}{4}\hat H_-)$ used in the main text is the metaplectic lift of the
complex--parameter flow $M_{H_-}(\tau)$ at $\tau=\ii\pi/4$.  Its factorization used in Sec.~\ref{QuantCayley} is
\begin{equation}
\label{app:eq:Uc_factor}
U(\pi/4)
=
\exp\!\Big(-\frac12\,\hat q^{2}\Big)\,
\exp\!\Big(\frac12\,\hat H_0\Big)\,
\exp\!\Big(-\frac{\ii}{2}\ln 2\,\hat D\Big)\,,
\qquad \hat H_0=\frac12\,\hat p^{2},\quad \hat D=\frac12(\hat q\hat p+\hat p\hat q).
\end{equation}
At the level of matrices this corresponds to the Gauss decomposition (consistent with $\widehat S\,\hat\xi\,\widehat S^{-1}=S\hat\xi$)
\begin{equation}
\label{app:eq:C_gauss}
C
=
\mathcal L(-\ii)\,\mathcal T\!\Big(-\frac{\ii}{2}\Big)\,
\mathcal D\big({2}^{1/2}\big)\,,
\qquad
C=\frac{1}{\sqrt2}\begin{pmatrix}1&-\ii\\ -\ii&1\end{pmatrix}
=\exp\!\Big(-\ii\frac{\pi}{4}\sigma_1\Big)\in \Sp(2,\C).
\end{equation}

\medskip
\noindent
\textbf{Link to $SU(1,1)$ disentangling.}
Under $\Sp(2,\R)\cong SL(2,\R)\cong SU(1,1)$, Gauss decompositions lift to the standard disentangling identity
for the generators \eqref{K+K-K0},
$\hat{K}_\pm=\frac12(\hat a^\pm)^2$ and $\hat{K}_0=\frac12\hat H_+$:
\begin{equation}
\label{app:eq:su11}
\exp\!\big[-s\,(\hat{K}_+ + \hat{K}_-)\big]
=
\exp\!\big[-\tan s\,\hat{K}_+\big]
\exp\!\big[-2\ln(\cos s)\,\hat{K}_0\big]
\exp\!\big[-\tan s\,\hat{K}_-\big],
\quad(\cos s\neq0)\,,
\end{equation}
whose special case $s=\pi/4$ yields the factorization \eqref{Uc_factorised},  that is equivalent to \eqref{app:eq:Uc_factor}.

%%%%%%%%%%%%%%%%%%%%%%%%%%%%%%%%%%%%%%%%%%%%%%%%%%%%%%%%%%%%%%%%%%%%%%%%%%%%%%%%%%%%

%%%%%%%%%%%%%%%%%%%%%%%%%%%%%%%%%%%%%%%%%%%%%%%%%%%%%%%%%%%%%%%%%%%%%%%%%%%%%%%%%%%%%%%%%%%%%%%%%%%%%%%%%%%%%%%

%%%%%%%%%%%%%%%%%%%%%%%%%%%%%%%%%%%%%%%%%%%

%========================================================
% Appendix: Demonstration of Eq. (\ref{psi0zeta})
%========================================================
\section{Demonstration of Eq.~(\ref{psi0zeta})}
\label{app:demo_psi0zeta}

In this Appendix we demonstrate Eq.~(\ref{psi0zeta}), i.e.\ the closed Gaussian
form of the squeezed-vacuum wavefunction
\(\psi_{0,\zeta}(q)=\langle q|S(\zeta)|0\rangle \).
Starting from the number-basis expansion (\ref{eq:sqvac_number_23}), we have
\begin{equation}
\psi_{0,\zeta}(q)
= \langle q|S(\zeta)|0\rangle
= (1-|\kappa|^2)^{1/4}\sum_{n=0}^{\infty}\frac{\sqrt{(2n)!}}{2^n\,n!}\,\kappa^n\,
\psi_{2n}(q),
\label{eq:psi0zeta_series_start}
\end{equation}
where \(\kappa=e^{i\phi}\tanh r\) with \(|\kappa|<1\), and \(\psi_{n}(q)\) are the
normalized harmonic-oscillator eigenfunctions,
\begin{equation}
\psi_n(q)=\pi^{-1/4}\frac{1}{\sqrt{2^n n!}}\,e^{-q^2/2}H_n(q).
\label{eq:psi_n_def_demo}
\end{equation}
Substituting (\ref{eq:psi_n_def_demo}) with \(n\mapsto 2n\) into
(\ref{eq:psi0zeta_series_start}) gives
\begin{align}
\psi_{0,\zeta}(q)
&=(1-|\kappa|^2)^{1/4}\sum_{n=0}^{\infty}\frac{\sqrt{(2n)!}}{2^n n!}\,\kappa^n\,
\left[
\pi^{-1/4}\frac{1}{\sqrt{2^{2n}(2n)!}}\,e^{-q^2/2}H_{2n}(q)
\right]\nonumber\\
&=\pi^{-1/4}(1-|\kappa|^2)^{1/4}\,e^{-q^2/2}\sum_{n=0}^{\infty}
\frac{\kappa^n}{4^n\,n!}\,H_{2n}(q).
\label{eq:psi0zeta_reduced_Hermite}
\end{align}
Thus it remains to evaluate the even-Hermite series
\(
\sum_{n\ge0}\frac{\kappa^n}{4^n n!}H_{2n}(q).
\)

We use the standard identity relating even Hermite polynomials to generalized
Laguerre polynomials,
\begin{equation}
H_{2n}(q)=(-1)^n\,2^{2n}\,n!\,L_n^{-1/2}(q^2)
= (-1)^n\,4^n\,n!\,L_n^{-1/2}(q^2).
\label{eq:Hermite_to_Laguerre_demo}
\end{equation}
Inserting (\ref{eq:Hermite_to_Laguerre_demo}) into the series in
(\ref{eq:psi0zeta_reduced_Hermite}) yields
\begin{equation}
\sum_{n=0}^{\infty}\frac{\kappa^n}{4^n\,n!}\,H_{2n}(q)
=\sum_{n=0}^{\infty}(-\kappa)^n\,L_n^{-1/2}(q^2).
\label{eq:Laguerre_series_demo}
\end{equation}
Now apply the generating function for generalized Laguerre polynomials,
\begin{equation}
\sum_{n=0}^{\infty}L_n^{\alpha}(x)\,t^n
=(1-t)^{-\alpha-1}\exp\!\Bigl(-\frac{t}{1-t}\,x\Bigr),
\qquad |t|<1,
\label{eq:Laguerre_generating_demo}
\end{equation}
with \(\alpha=-\tfrac12\), \(x=q^2\), \(t=-\kappa\). Since \(|\kappa|<1\), the sum
converges and we obtain
\begin{equation}
\sum_{n=0}^{\infty}(-\kappa)^n\,L_n^{-1/2}(q^2)
=(1+\kappa)^{-1/2}\exp\!\Bigl(\frac{\kappa}{1+\kappa}\,q^2\Bigr).
\label{eq:Laguerre_summed_demo}
\end{equation}
Substituting (\ref{eq:Laguerre_summed_demo}) into
(\ref{eq:psi0zeta_reduced_Hermite}) gives
\begin{align}
\psi_{0,\zeta}(q)
&=\pi^{-1/4}(1-|\kappa|^2)^{1/4}\,e^{-q^2/2}\,
(1+\kappa)^{-1/2}\exp\!\Bigl(\frac{\kappa}{1+\kappa}\,q^2\Bigr)\nonumber\\
&=\pi^{-1/4}(1-|\kappa|^2)^{1/4}(1+\kappa)^{-1/2}
\exp\!\Bigl(-\frac12\,\frac{1-\kappa}{1+\kappa}\,q^2\Bigr),
\label{eq:psi0zeta_final_demo}
\end{align}
which is exactly Eq.~(\ref{psi0zeta}).

%%%%%%%%%%%%%%%%%%%%%%%%%%%%%%%%%%%%%%%%%%%%%%%%%%%%%%%%

%%%%%%%%%%%%%%%%%%%%%%%%%%%%%%%%%%%%%%%%%%%%%%%%%%%%%%%%%%%%%%%%%%%%%%
\section{Schr\"odinger-group Gaussian packets, covariance, and the upper half-plane}
\label{app:Sch_free_coherent}
%%%%%%%%%%%%%%%%%%%%%%%%%%%%%%%%%%%%%%%%%%%%%%%%%%%%%%%%%%%%%%%%%%%%%%

This Appendix collects the basic facts about Gaussian
wavepackets of the free particle that are needed in the discussion of the
Cayley/conformal bridge.  The essential point is that the natural parameter space
of (pure) Gaussian covariances is the upper half-plane $\H_+$, the same space that
appears as the moduli of compatible complex structures in
App.~\ref{app:kahler_uplift}.

\medskip
\noindent\textbf{Robertson--Schr\"odinger uncertainty relation and covariance matrix.}
We work with $[\hat q,\hat p]=i$ ($\hbar=1$).  For a normalized state define
\[
\Delta \hat q:=\hat q-\langle \hat q\rangle,\qquad
\Delta \hat p:=\hat p-\langle \hat p\rangle,
\]
and the real symmetric covariance matrix
\be
\Sigma:=
\begin{pmatrix}
\langle (\Delta \hat q)^2\rangle &
\frac12\langle \Delta \hat q\,\Delta \hat p+\Delta \hat p\,\Delta \hat q\rangle\\[4pt]
\frac12\langle \Delta \hat q\,\Delta \hat p+\Delta \hat p\,\Delta \hat q\rangle &
\langle (\Delta \hat p)^2\rangle
\end{pmatrix}
=
\begin{pmatrix}
(\Delta q)^2 & \operatorname{Cov}(q,p)\\[2pt]
\operatorname{Cov}(q,p) & (\Delta p)^2
\end{pmatrix}.
\label{eq:app_Sigma_def}
\ee
\medskip
\noindent\textbf{Remark: Meaning of $ \operatorname{Cov}(q,p)$\label{rem:covariance_qp}.}
The off--diagonal entry
\begin{equation}
   \operatorname{Cov}(q,p)=\frac12\langle \Delta\hat q\,\Delta\hat p+\Delta\hat p\,\Delta\hat q\rangle
  =\Re\, \langle \Delta\hat q\,\Delta\hat p\rangle
  \label{eq:cov_qp_def}
\end{equation}
is the (symmetrized) covariance of the noncommuting observables $q$ and $p$.
It is real and quantifies the \emph{linear correlation} between the fluctuations
$\Delta\hat q$ and $\Delta\hat p$ in the given state.  Introducing the
dimensionless correlation coefficient
\begin{equation}
  \rho_{qp}:=\frac{ \operatorname{Cov}(q,p)}{\Delta q\,\Delta p}\,,\qquad |\rho_{qp}|\le 1,
  \label{eq:rho_qp}
\end{equation}
one may regard $ \operatorname{Cov}(q,p)\neq 0$ as encoding a \emph{tilt} of the uncertainty ellipse
in phase space, while
$ \operatorname{Cov}(q,p)=0$ corresponds to an ``untilted'' (often called \emph{unchirped})
situation.  In particular, for a Gaussian wave packet this tilt is precisely the
quadratic $qp$ (``chirp'') term in the exponent and is preserved under linear
canonical transformations.

The sharp Robertson--Schr\"odinger (RS) uncertainty relation
\cite{GutierrezVega2021}
 is equivalently
\be
\det\Sigma\ \ge\ \frac14,
\qquad\text{i.e.}\qquad
(\Delta q)^2(\Delta p)^2-\operatorname{Cov}(q,p)^2\ \ge\ \frac14.
\label{eq:app_RS_det_form}
\ee
Dropping the nonnegative $\operatorname{Cov}(q,p)^2$ term yields the familiar
Heisenberg product bound $\Delta q\,\Delta p\ge \tfrac12$.

\medskip
\noindent\textbf{Phase-space ellipse and ``minimal area'' interpretation.}
The matrix $\Sigma$ defines the uncertainty ellipse in the classical $(Q,P)$-plane
by
\be
E_1:=\left\{(Q,P)\in\R^2\ \Big|\ 
\begin{pmatrix}Q\\ P\end{pmatrix}^{\!\mathsf T}
\Sigma^{-1}
\begin{pmatrix}Q\\ P\end{pmatrix}\le 1
\right\}.
\label{eq:app_uncertainty_ellipse}
\ee
Its Euclidean area is
\be
\operatorname{Area}(E_1)=\pi\,\sqrt{\det\Sigma}.
\label{eq:app_area_ellipse}
\ee
Hence \eqref{eq:app_RS_det_form} implies $\operatorname{Area}(E_1)\ge \pi/2$.
In this precise sense, \emph{saturating} the RS inequality means \emph{minimal
phase-space area} of the covariance ellipse (for $\hbar=1$).

\medskip
\noindent\textbf{A convenient Gaussian parametrization by $\tau\in\H_+$.}
Introduce the upper half-plane parameter
\(
\tau=u+iv\in\C, \) \( v>0,
\)
and the normalized centered Gaussian, cf. \eqref{eq:free_gauss_tau_23},
\be
\psi_{\tau}(q)
:=
\left(\frac{\Im\tau}{\pi|\tau|^2}\right)^{\!1/4}
\exp\!\left(-\,\frac{i}{2\tau}\,q^2\right).
\label{eq:app_psi_tau_def}
\ee
More general packets are obtained by Heisenberg translation; up to an overall phase,
a convenient explicit form is
\be
\psi_{q_0,p_0;\tau}(q)
:=
e^{\,i p_0\,(q-\frac12 q_0)}\,
\psi_{\tau}(q-q_0),
\qquad (q_0,p_0)\in\R^2,\ \ \tau\in\H_+.
\label{eq:app_psi_shifted}
\ee
For these Gaussians one finds the second moments
\be
(\Delta q)^2=\frac{|\tau|^2}{2\,\Im\tau},\qquad
(\Delta p)^2=\frac{1}{2\,\Im\tau},\qquad
\operatorname{Cov}(q,p)=-\,\frac{\Re\tau}{2\,\Im\tau}.
\label{eq:app_covariances_tau}
\ee
Consequently,
\[
(\Delta q)^2(\Delta p)^2-\operatorname{Cov}(q,p)^2=\frac14,
\]
so the RS bound is saturated for all $\tau\in\H_+$.

\medskip
The Heisenberg product is
\[
\Delta q\,\Delta p=\frac{|\tau|}{2\,\Im\tau}\ge \frac12,
\]
and equality holds iff $\Re\tau=0$. Thus ``unchirped'' means $\Re\tau=0$, equivalently $\operatorname{Cov}(q,p)=0$.

\medskip
\noindent\textbf{Stability under free evolution and evolution of moments.}
For the free Hamiltonian $\hat H_0=\frac12\hat p^{\,2}$, Heisenberg evolution gives
\(
\hat q(t)=e^{it\hat H_0}\hat q\,e^{-it\hat H_0}=\hat q+t\hat p,
\)
\(
\hat p(t)=e^{it\hat H_0}\hat p\,e^{-it\hat H_0}=\hat p.
\)
Hence the centered operators satisfy $\Delta \hat q(t)=\Delta\hat q+t\,\Delta\hat p$,
$\Delta\hat p(t)=\Delta\hat p$, and the second moments evolve exactly as
\[
(\Delta p(t))^2 = (\Delta p(0))^2, \label{eq:app_Dp_t}\qquad
\operatorname{Cov}(q,p)(t)
=\operatorname{Cov}(q,p)(0)+t\,(\Delta p(0))^2, 
\]
\[
(\Delta q(t))^2
=(\Delta q(0))^2+2t\,\operatorname{Cov}(q,p)(0)+t^2(\Delta p(0))^2. 
\]
For pure Gaussians \eqref{eq:app_psi_tau_def}--\eqref{eq:app_psi_shifted} one has
$\det\Sigma=\frac14$ at $t=0$, and the RS-saturating character is preserved by the
free evolution (the ellipse is sheared/tilted, but its symplectic area remains minimal).

\medskip
Equivalently, on the parameter $\tau$ one has the simple flow (up to phase):
\be
e^{-it\hat H_0}\,\psi_{q_0,p_0;\tau}
\ \propto\
\psi_{q_0+t p_0,\,p_0;\,\tau-t}.
\label{eq:app_tau_flow}
\ee
This expresses the ``stability under free time evolution'' as closure of the Gaussian
family under the one-parameter subgroup $e^{-it\hat H_0}$.
%%%%%%%%%%%%%%%%%%%%%%%%%%%%%%%%%%%%%%%%%%%%%%%%%%%%%%%%%%%%%%%%%%%%%%

%%%%%%%%%%%%%%%%%%%%%%%%%%%%%%%%%%%%%%%%%%%%%%%%%%%%%%%%%%%%%%%%%%%%%%%%%%%%%%%%%%%%%%%%%%%%%%%%%%%%%%%%%%%%%%%%%%%%%%%%%%

%%%%%%%%%%%%%%%%%%%%%%%%%%%%%%%%%%%%%%%%%%%%%%%%%%%%%%%%%%%%%%%%%%%%%%
\section{Schwarzian derivative: cocycle property and inverse-map identity}
\label{Appendix Schwarzian}

\providecommand{\Schw}[2]{\left\{#1\,;\,#2\right\}}

For a locally univalent function $f=f(x)$, its Schwarzian derivative \cite{OTbook,Osgood,OTams} is
\be
\Schw{f}{x}
:=\frac{f'''(x)}{f'(x)}-\frac{3}{2}\left(\frac{f''(x)}{f'(x)}\right)^{\!2}.
\label{app:Schwarz_def}
\ee
It vanishes iff $f$ is a M\"obius (projective) transformation,
\be
\Schw{f}{x}=0 \quad \Longleftrightarrow \quad
f(x)=\frac{ax+b}{cx+d}\,,\qquad ad-bc\neq0.
\label{app:Schwarz_zero_mobius}
\ee
In terms of the \emph{pre-Schwarzian}
\be
\mathcal P_f(x):=\frac{f''(x)}{f'(x)}=\bigl(\ln f'(x)\bigr)'\,,
\label{app:preSchwarz_def}
\ee
one may also write
\be
\Schw{f}{x}=\mathcal P_f'(x)-\frac12\bigl(\mathcal P_f(x)\bigr)^2\,.
\label{app:Schwarz_logform}
\ee

\medskip
\noindent
\textbf{Cocycle (chain) property.}
As a quadratic differential $S(f):=\Schw{f}{x}(dx)^2$, the Schwarzian is a (nontrivial) $1$-cocycle on $\mathrm{Diff}(I)$:
\be
\Schw{f\circ g}{x}=\bigl(g'(x)\bigr)^2\,\Schw{f}{g(x)}+\Schw{g}{x}\,,
\label{app:Schwarz_chain}
\ee
equivalently $S(f\circ g)=g^{*}S(f)+S(g)$.
In particular, $\Schw{M}{z}=0$ for any M\"obius map $M$, hence
\be
\Schw{M\circ f}{x}=\Schw{f}{x}.
\label{app:Schwarz_postmobius}
\ee
(Under pre-composition $x\mapsto M(x)$ one uses \eqref{app:Schwarz_chain}.)

\medskip
\noindent
\textbf{Inverse-map identity.}
If $y=y(x)$ is locally invertible with inverse $x=x(y)$, then
\be
\Schw{x}{y}
=-\bigl(x'(y)\bigr)^2\,\Schw{y}{x}\Big|_{x=x(y)}.
\label{app:Schwarz_inverse}
\ee
This is the form used in the main text (cf.\ \eqref{app:Q_transform_SSE_xy} and \eqref{app:Q_transform_SSE_yx}).
%%%%%%%%%%%%%%%%%%%%%%%%%%%%%%%%%%%%%%%%%%%%%%%%%%%%%%%%%%%%%%%%%%%%%%

\noindent
\paragraph{Historical note.}
The Schwarzian derivative was isolated in conformal mapping by Schwarz (1869, 1873) \cite{Schwarz1869,Schwarz1873}
and the terminology became standard after Cayley (1880) \cite{Cayley1880}; related third-order expressions occur
(\emph{implicitly}) earlier in transformation theory \cite{Lagrange,Jacobi,Kummer1836}.
As a simple corollary, for $Q\equiv k^2$ one has $\{ \tan(kx);x\}=2k^2$ (up to post-composition by a M\"obius map).

%%%%%%%%%%%%%%%%%%%%%%%%%%%%%%%%%%%%%%%%%%%%%%%%%%%%%%%%%%%%%%%%%%%%%%

\section{Exact versus semiclassical quantization of canonical transformations}
\label{app:exact_vs_semiclassical_quantisation}

This Appendix clarifies the relation between classical canonical transformations
and their quantum counterparts.  The key point is that a generating function
controls the \emph{oscillatory phase} of an integral kernel, but in general does
not determine a full unitary operator: one must also include a nontrivial
amplitude (Van Vleck/half-density factor) and a Maslov phase.  Exact
quantization in the strong sense of an exact Egorov property \cite{EgoTeor} on a
large algebra of observables is exceptional, although there are important exact
classes strictly larger than the linear/metaplectic one.

\medskip
\noindent\textbf{Two notions of ``quantum analogue''.}
Let $\kappa$ be a canonical transformation (symplectomorphism) on phase space,
\(
\kappa:\ (q,p)\mapsto(Q,P),\)
\( dp\wedge dq = dP\wedge dQ.
\)
There are two distinct (and often conflated) requirements:

\medskip
\noindent
\textbf{(A) Exact implementation on observables (exact Egorov).}
One asks for a unitary $U_\kappa$ such that for a large class of symbols $a$,
\be
U_\kappa^{-1}\,\mathrm{Op}^{\rm W}(a)\,U_\kappa=\mathrm{Op}^{\rm W}(a\circ\kappa),
\label{app:eq:exact_egorov_goal}
\ee
where $\mathrm{Op}^{\rm W}$ denotes Weyl quantization.  This is a very strong
property; it holds, for example, for the metaplectic representation of linear
symplectic maps.

\medskip
\noindent
\textbf{(B) Exact intertwining of a \emph{specific} differential operator.}
Often one only needs an exact intertwiner for one operator (or a small family),
for instance a Schr\"odinger/constraint operator $\mathcal{S}$:
\be
U\,\mathcal{S}_{\rm old}=\mathcal{S}_{\rm new}\,U\,.
\label{app:eq:intertwiner_goal}
\ee
This can hold exactly even when \eqref{app:eq:exact_egorov_goal} fails for the
full observable algebra.  In particular, the free$\leftrightarrow$oscillator
unitary map in the extended formulation is naturally viewed as an exact
intertwiner of Schr\"odinger operators.

\medskip
\noindent\textbf{Generating functions determine the phase, not the full kernel.}
Assume that $\kappa$ admits locally a generating function of type $(q,Q)$,
\(
p=\frac{\partial S(q,Q)}{\partial q},\)
\(
P=-\frac{\partial S(q,Q)}{\partial Q}.
\)
A standard semiclassical ansatz for an implementer is a Fourier integral
operator (FIO) with kernel
\be
K(Q,q)\ \sim\ (2\pi\ii\hbar)^{-1/2}\,A(Q,q;\hbar)\,
\exp\!\left(\frac{\ii}{\hbar}S(q,Q)\right),
\label{app:eq:FIO_ansatz_1D}
\ee
where $A$ is an amplitude.  The classical function $S$ supplies the eikonal
phase $\exp(\ii S/\hbar)$, while $A$ is fixed by unitarity and composition
(transport equations).  In $1D$ and away from caustics
($\partial^2 S/\partial q\,\partial Q\neq0$), the leading amplitude is the Van
Vleck factor
\be
A_{\rm VV}(Q,q)=\left|\frac{\partial^2 S(q,Q)}{\partial q\,\partial Q}\right|^{1/2}
\label{app:eq:VanVleck_1D}
\ee
(up to an overall phase); higher orders in $\hbar$ generally correct it.

\medskip
\noindent\textbf{Egorov theorem: generic nonlinear symplectomorphisms are only
asymptotically exact.}
For a general (nonlinear) symplectomorphism $\kappa$, suitable FIO implementers
satisfy Egorov's theorem in the form of an \emph{asymptotic} expansion:
\be
U_\kappa^{-1}\,\mathrm{Op}^{\rm W}(a)\,U_\kappa
=\mathrm{Op}^{\rm W}(a\circ\kappa)
+\hbar\,\mathrm{Op}^{\rm W}(a_1)+\hbar^2\,\mathrm{Op}^{\rm W}(a_2)+\cdots.
\label{app:eq:egorov_asymptotic}
\ee
The correction symbols $a_j$ depend on the \emph{full} quantization data
(including amplitudes), not only on $S$; hence exactness in the strong sense
\eqref{app:eq:exact_egorov_goal} is not generic.

\medskip
\noindent\textbf{Exact classes beyond the linear/metaplectic case.}

\noindent
\textbf{(i) Linear symplectic maps and the metaplectic representation.}
If $\kappa$ is linear symplectic ($\kappa(z)=Mz$, $M\in\Sp(2,\R)$), then $S$ is
quadratic and the metaplectic operator $U_M\in\Mp(2,\R)$ is exact.  In this case
the Van Vleck factor is constant (a determinant of a constant block), and the
remaining global datum is the Maslov sign/phase (the $\Mp\to\Sp$ double cover).

\medskip
\noindent
\textbf{(ii) Polarisation-preserving canonical transformations on $T^*\R$.}
A nonlinear exact class is given by the canonical maps that preserve the
Schr\"odinger (vertical) polarization: cotangent lifts of diffeomorphisms of
configuration space, possibly composed with an exact one-form shift.  Given a
diffeomorphism $Q=Q(q)$ and a function $G(q)$, consider
\be
Q=Q(q),\qquad P=\frac{p+G'(q)}{Q'(q)}.
\label{app:eq:pol_preserving_classical}
\ee
One checks that $ p\,dq - P\,dQ = dG(q)$, so
\eqref{app:eq:pol_preserving_classical} is canonical.  Its exact unitary action
on wavefunctions is
\be
\bigl(U\psi\bigr)(Q)
=\left(\frac{dQ}{dq}\right)^{-1/2}
\exp\!\left(\frac{\ii}{\hbar}G(q)\right)
\psi(q)\Big|_{q=q(Q)}.
\label{app:eq:pol_preserving_unitary}
\ee
The Jacobian factor expresses that $\psi$ is naturally a half-density:
$\psi(q)\,(dq)^{1/2}=\Psi(Q)\,(dQ)^{1/2}$.

\medskip
\noindent
\textbf{(iii) Exactness by truncation of the Moyal commutator.}
A useful sufficient condition for exact Egorov on the Weyl/Moyal algebra is
that the Moyal commutator truncates for the generator Hamiltonian.  This
includes quadratic Hamiltonians (metaplectic case), and also Hamiltonians at
most affine in momentum,
\be
H(q,p)=a(q)\,p+b(q),
\label{app:eq:affine_in_p}
\ee
whose flows are precisely of the polarization-preserving form
\eqref{app:eq:pol_preserving_classical}.  In such cases the quantum evolution
$U(t)=\exp(-\ii\hat H t/\hbar)$ produces an exact automorphism of the Weyl
algebra (no infinite $\hbar$-series is generated).

\medskip
\noindent\textbf{Maslov phase and the metaplectic double cover.}
Whenever the kernel involves a square root of a determinant (or, in $1D$, a
square root of $\partial^2 S/\partial q\,\partial Q$), one must choose a branch.
Along continuous families this branch can jump when passing through caustics
(where the chosen projection becomes singular).  The standard way to encode
this is a Maslov phase.  In $1D$ the leading kernel is
\be
K(Q,q)\ \approx\ (2\pi\ii\hbar)^{-1/2}
\left|\frac{\partial^2 S}{\partial q\,\partial Q}\right|^{1/2}
\exp\!\left(\frac{\ii}{\hbar}S(q,Q)-\frac{\ii\pi}{2}\,\mu\right),
\label{app:eq:FIO_with_maslov_1D}
\ee
where $\mu\in\Z$ changes by $\pm1$ upon crossing a caustic.  In the linear
symplectic case, $\mu$ reduces to the discrete datum distinguishing the two
lifts in $\Mp(2,\R)$ of a given element of $\Sp(2,\R)$.

\medskip
\noindent\textbf{Remark.}
The exact classes \eqref{app:eq:pol_preserving_classical}--\eqref{app:eq:pol_preserving_unitary}
are the natural nonlinear extensions of ``phase from a generating function'' in
the Schr\"odinger polarization: the phase is supplied by the exact differential
$dG$ in the Liouville form, while the half-density Jacobian is forced by
unitarity (and is the same mechanism behind the familiar Schwarzian terms
appearing when one transforms second-order operators).

%%%%%%%%%%%%%%%%%%%%%%%%%%%%%%%%%%%%%%%%%%%%%%%%%%%%%%%%%%%%%%%%%%%%%%%%%%%%%%%%%%%%%%%%%%%%%%%%%%%%%%%%%%%%%%%%

% ======================================================================
% Appendix: Exact vs semiclassical quantization of canonical transformations
% ======================================================================

%%%%%%%%%%%%%%%%%%%%%%%%%%%%%%%%%%%%%%%%%%%%%%%%%%%%%%%%%%%%%%%%%%%%%%%%%%%%%%%%%%%%%%%%%%%%%%%%%%%%%%%%%%%%%%%%

\end{document}